\documentclass[twocolumn,showpacs,preprintnumbers,prc,superscriptaddress]{revtex4-2}
\usepackage{graphicx,color}
\usepackage{float}
\usepackage{amsmath,amssymb}
\usepackage{bm}
\usepackage{tkz-berge,tikz}
\usepackage[qm,braket]{qcircuit}

\usepackage[pdftex,colorlinks=true, linkcolor = blue, citecolor=blue,urlcolor=blue, bookmarksnumbered=true, bookmarksopen=true]{hyperref}
\usepackage{appendix}
\usepackage{soul}

\usepackage[braket]{qcircuit}

\usepackage[normalem]{ulem}

\newcommand{\bs}[1]{\ensuremath{\boldsymbol{#1}}}
\newcommand{\be}{\begin{equation}}
\newcommand{\ee}{\end{equation}}
\newcommand{\bea}{\begin{align}}
\newcommand{\eea}{\end{align}}

\newcommand{\smblksquare}{\scalebox{0.6}{$\blacksquare$}}

\begin{document}

\title{Solving reaction dynamics with quantum computing algorithms}

\author{R. Weiss}
\affiliation{
  Los Alamos National Laboratory, Theoretical Division, Los Alamos, New Mexico 87545, USA}
\affiliation{
Department of Physics, Washington University in St. Louis, St. Louis, Missouri 63130, USA}
\author{A. Baroni}
\affiliation{National Center for Computational Sciences, Oak Ridge National Laboratory, TN 37831, USA}
\author{J. Carlson}
\affiliation{
  Los Alamos National Laboratory, Theoretical Division, Los Alamos, New Mexico 87545, USA}
\author{I. Stetcu}
\affiliation{
  Los Alamos National Laboratory, Theoretical Division, Los Alamos, New Mexico 87545, USA}
  
\date{\today}
\preprint{LA-UR-24-20333, LA-UR-25-20324}

\begin{abstract}
    The description of quantum many-body dynamics is extremely challenging on classical computers, as it can involve many degrees of freedom. On the other hand, the time evolution of quantum states is a natural application for quantum computers that are designed to efficiently perform unitary transformations. In this paper, we study quantum algorithms for response functions, relevant for describing different reactions governed by linear response. We focus on nuclear-physics applications and consider a qubit-efficient mapping on the lattice, which can efficiently represent the large volumes required for realistic scattering simulations. 
    For the case of a contact interaction, we develop an algorithm for time evolution based on the Trotter approximation that scales logarithmically with the lattice size, and is combined with quantum phase estimation. We eventually focus on the nuclear two-body system and a typical response function relevant for electron scattering as an example. We also investigate ground-state preparation and examine the total circuit depth required for a realistic calculation and the hardware noise level required to 
    interpret the signal. 
\end{abstract}

\maketitle


\section{Introduction}

Quantum computing has the potential to become one of the most powerful methods for calculating the dynamics of many-body quantum systems. As hardware continues to improve, it is vital to develop appropriate quantum algorithms. In various studies, relevant aspects of quantum computation have been examined. Various methods for mapping Hamiltonians to qubits have been investigated \cite{JW,BRAVYI2002210,Tavernelli:2016ckh,Bravyi:2017eoo,Babbush:2017oum,Steudtner:2018ujo,DiMatteo:2020dhe,Kirby:2021vkt,PhysRevResearch.4.023154}, considering trade-offs with respect to the number of required qubits and the number of Pauli strings in the Hamiltonian, or the circuit depth for different algorithms. The preparation of specific states on a quantum computer is a necessary step in the description of many-body systems. Indeed, different ground-state preparation methods, including variational algorithms \cite{Kandala:2017HE,Tilly:2021}, adiabatic evolution \cite{Farhi2000,Albash2018}, and projection-based methods \cite{Motta:2020,PhysRevResearch.4.033121,Turro2022,Jouzdani2022,Ge2018,Dong2022,Keen2021,Choi2021, Stetcu:2022nhy}, have been developed, as well as algorithms for initialization to a known state \cite{Bergholm:2005ibb,Plesch:2011vwn,Zhang:2021uwi,Zhang:2021bue,Rosenthal:2021rcb,Zhang:2022pue,Sun:2023bwt}.


In scientific computing for nuclear physics, quantum computers should be especially useful for calculating reaction cross sections. For example, exact calculations of exclusive electron- and neutrino-scattering are not accessible with classical computers, even for relatively light nuclei. Several algorithms for response functions, required for calculating such reaction cross sections, have been suggested \cite{Terhal:1998yh,Lidar:1998mf,Ortiz:2000gc,Soma-SpectrumState,Kassal_2008,Rogerro2019,Somma:2019rmm,Roggero:2020qoz,PhysRevA.102.022408}. This might be an early application of quantum computers, but significant algorithmic progress is still needed \cite{Roggero:2019myu}. In many cases, the steps requiring the most resources in such algorithms involve time evolution of a given state. Trotter-based time evolution and the associated uncertainty estimation have been studied in different works \cite{Trotter1959,Suzuki:1985wzj,Lloyd1996,Heyl_2019,Kivlichan_2020,Childs:2019hts,Tran_2020,Sieberer:2019htd,Layden:2021ols}, as well as other approaches \cite{Berry_2015,Berry:2014ivo,Low:2016sck,Low:2016znh}. 

For such applications, the system is commonly discretized on a lattice, where a three-dimensional lattice with about $10\times10\times10$ sites, or more, is needed, even if a small number of particles is involved \cite{Meissner:2023cvo}. In such cases, mappings like the Jordan-Wigner (JW) method \cite{JW} will require thousands of logical qubits. Expected early fault-tolerant devices will not be able to support such calculations. Therefore, it is important to investigate the integration of qubit-efficient mappings with relevant quantum algorithms. Different such mappings are available in the literature, including both the first-quantization mapping and second-quantization methods \cite{DiMatteo:2020dhe,PhysRevResearch.4.023154}, and will require only a few tens of logical qubits for applications that go beyond the abilities of classical computers.

In this work, we study reaction dynamics in quantum many-body systems on a lattice using quantum computing. We focus on systems with small number of particles, which are expected to be most relevant for early applications on hardware.  Our main goal is to investigate the combination of qubit-efficient mappings with reaction algorithms and, especially, the calculation of response functions. We study the optimization of the required quantum circuits to provide reliable resource estimation and explicit quantum circuits. 

In Sec. \ref{sec:Ham_Map}, we discuss the qubit-efficient mapping used in this work and the general form of the Hamiltonian.
In Sec. \ref{sec:2body}, we start analyzing the two-body system and present the contact interaction model, which we focus on in this work. 
Such an interaction model is justified in different cases, including in nuclear systems and condensed matter physics. This also opens the path for studies of more complex interactions and systems with more particles using the same mapping. 
Time evolution with optimized quantum circuits using the Trotter approximation is discussed in Sec. \ref{sec:time_evo} for the two-body system. We present two methods for the time evolution and show that with a small number of ancilla qubits, gate complexity, which scales logarithmically with the lattice size, can be obtained. An extension of this time-evolution algorithm to the many-body case is discussed in Sec. \ref{sec:time_evo_many_body}. We consider different approaches for ground-state preparation in Sec. \ref{sec:GS_prep}. We analyze an energy-filter method and another approach for state initialization, assuming the ground state is known.
Finally, combining these different ingredients, we study a response function algorithm relevant for nuclear systems and analyze the required circuit depth in Sec. \ref{sec:res_func}, investigate the impact of noise in Sec. \ref{sec:noise}, and summarize in Sec. \ref{sec:summary}.

\section{Hamiltonian and Mapping}
\label{sec:Ham_Map}

We consider a fermionic non-relativistic many-body system with a two-body interaction. We will describe it using a cubic three-dimensional lattice with periodic boundary conditions. The side lengths of the lattice are denoted by $L_x$, $L_y$, and $L_z$, with $N_x$, $N_y$, and $N_z$ sites along each direction, all assumed to be even. The total number of sites is  $N = N_x N_y N_z$ and the lattice volume is $\Omega=L_x L_y L_z$. We will use momentum basis states denoted by $|\bs{p}\rangle = |p_x,p_y,p_z\rangle$. On the lattice, $p_x$ can take the values $\{0,\pm \frac{2\pi}{L_x},\pm 2\frac{2\pi}{L_x},...,\pm (\frac{N_x}{2}-1) \frac{2\pi}{L_x}, -\frac{N_x}{2} \frac{2\pi}{L_x} \}$, and similarly for $y$ and $z$.

We assume a local radial two-body interaction $V$, such that the Hamiltonian is given by the sum of kinetic and potential energy
\be \label{eq:Ham}
{\cal H} = \sum_{i=1}^A \frac{p_i^2}{2m}+\sum_{i<j}V(|\bs{r}_i-\bs{r}_j|).
\ee
Here, $A$ is the number of particles, $p_i$ is the momentum operator of particle $i$ and $m$ is the mass of the particles. To be specific, we consider the case of nuclear systems, where each particle can be either a proton or a neutron, corresponding to isospin-half particles ($t=1/2$) with isospin projection $t_z=\pm 1/2$. They also contain a spin-half ($s=1/2$, $s_z=\pm1/2$) degree of freedom. For simplicity, we use a spin-isospin independent interaction, i.e. the interaction is the same whether the particles are protons or neutrons or whether $s_z=+1/2$ or $s_z=-1/2$. Extensions to other interactions will be considered in future works.

Different nuclear properties, including energies and radii, have been studied using classical computing of nuclei on lattices. See, for example, Refs. \cite{Lee:2008fa,Drut:2012md,Lee:2016fhn,Lahde:2019npb,Elhatisari:2022zrb} for the Nuclear Lattice EFT approach. In this approach, realistic interaction models from chiral effective field theory are combined with quantum Monte Carlo simulations using auxiliary fields.

The first step in describing such a system on a quantum computer is mapping the Hamiltonian to qubits. As mentioned above, we focus in this work on a mapping in which the number of required qubits is relatively small. Specifically, we will consider a qubit-efficient mapping of the Hamiltonian, following the mapping discussed in Ref. \cite{PhysRevResearch.4.023154} (which, in the case studied here, is also very similar to the first quantization mapping \cite{AbramsLloyd1997}). In this mapping, we start by choosing a relevant set of basis functions. Working with momentum states, and for the case of $A$ particles, we construct the basis states by choosing the momentum of the first $A-1$ particles, while the momentum of the last particle is fixed by requiring zero total CM momentum. This allows us to consider only the space of zero-CM $A$-particle states, unlike the JW mapping, where the number of particles and the CM momentum are not restricted.

We note that for small systems of up to 2 protons and 2 neutrons, we do not need to consider explicitly the spin-isospin degrees of freedom. The four particles (proton-up, proton-down, neutron-up, neutron-down) can be considered as distinguishable particles. 
In this case there are $N^{A-1}$ basis states, and the number of required qubits is 
\be \label{eq:n_q}
n_q=\lceil \log_2(N^{A-1}) \rceil
=  \lceil (A-1) \log_2(N) \rceil.
\ee
The number of qubits scales logarithmically with the number of sites, unlike the linear scaling that is obtained using the JW and similar mappings (where the number of qubits is equal to the number of orbitals). But notice that here the number of qubits depends also on the number of particles. For simplicity we will assume that $N$ is a power of $2$, and, therefore, the ceiling function is not needed in Eq. \eqref{eq:n_q}.

Following Ref. \cite{PhysRevResearch.4.023154}, the mapping is defined by associating each of the many-body basis states with a different bit-string vector of $n_q$ digits (0 or 1). The latter are basis states of the $n_q$-qubit system. The Hamiltonian can be written using the physical basis states $\{|v_i\rangle\}$
\be \label{eq:H_with_basis}
{\cal H} = \sum_{i,j} \langle v_j | {\cal H} | v_i \rangle |v_j\rangle \langle v_i |.
\ee
In its mapping, the operators $|v_j\rangle \langle v_i |$ should be consistently mapped. Since each basis function is mapped to bit-string, $|v_j\rangle \langle v_i |$ is mapped to a tensor product of the form $\prod_{a=0}^{n_q-1} | \tilde{v}_j^a \rangle \langle \tilde{v}_i^a|$, where each $|\tilde{v}_k^a\rangle$ is either $|0\rangle$ or $|1\rangle$. 
This product can be written as a sum of Pauli strings using the following single-qubit relations \cite{PhysRevResearch.4.023154}:
\begin{align} \label{eq:mapping}
|0\rangle \langle 0| &= \frac{1}{2}(I+Z) 
\; \; ; \; \;
|1\rangle \langle 1| = \frac{1}{2}(I-Z)
\nonumber \\
|1\rangle \langle 0| &= \frac{1}{2}(X-iY)
\; \; ; \; \;
|0\rangle \langle 1| = \frac{1}{2}(X+iY),
\end{align}
where $X$, $Y$, and $Z$ are the Pauli $x$, $y$, and $z$ matrices, and $I$ is the two-by-two identity matrix. These relations are obtained under the convention that $|0\rangle = \begin{pmatrix} 1 \\ 0 \end{pmatrix}$ and $|1\rangle = \begin{pmatrix} 0 \\ 1 \end{pmatrix}$. Pauli strings and their numerical coefficients coming from different $i,j$ pairs in Eq. \eqref{eq:H_with_basis} should be combined together to obtain the full mapping of the Hamiltonian, which is eventually written as a sum of Pauli strings $\sum_i h_i S_i$. Here, $S_i$ are the Pauli strings and $h_i$ are numerical coefficients.

As mentioned above, for the case considered here, this mapping is very similar to the first quantization mapping \cite{AbramsLloyd1997}.
For a larger number of particles, or in the case of a more complex interaction, spin-isospin degrees of freedom and fermionic anti-symmetrization should be accounted for explicitly. Following Ref. \cite{PhysRevResearch.4.023154}, this can be done using anti-symmetrized many-body states $\{|v_i\rangle\}$ that are mapped to bit-strings. This guarantees an antisymmetric wave function (see more details in Appendix \ref{sec:app_anti_symmetrization}). Alternatively, one can use the first quantization mapping combined with known and efficient anti-symmetrization algorithms \cite{AbramsLloyd1997,Berry2018a}. A detailed study of these alternatives will be the focus of future studies. 

In the next section we will discuss the application of this mapping for the two-body system. The many-body case will be discussed in Sec. \ref{sec:time_evo_many_body}, where a very similar mapping using coordinate-space basis is used.

\section{The two-body system}
\label{sec:2body}

We can start analyzing the use of the above mapping and the number of Pauli string in the Hamiltonian by considering the two-body system $A=2$. For two particles, the zero-CM-momentum basis states are of the form $|\bs{k},-\bs{k}\rangle$, $N$ states in total and $n_q=\log_2(N)$. 
Since we work with momentum basis states, the kinetic energy matrix is diagonal. Therefore, only $I$ and $Z$ matrices can be obtained in its mapping, i.e. at most $2^{n_q}=N$ Pauli strings. All these strings commute with each other. We will see soon in an explicit example that eventually much fewer strings are obtained for the kinetic energy due to symmetries of the matrix. Mapping the potential energy matrix results in at most $4^{\log_2(N)}=N^2$, due to the 4 possible matrices ($X,Y,Z,I$) for each qubit. 

We note for comparison that, using JW mapping with momentum-space orbitals, the number of Pauli strings generally scales as $N^3$ (due to the contribution from the two-body interaction involving creation and annihilation operators in the form $a^\dagger b^\dagger d c$ together with CM momentum conservation). If coordinate-space orbitals corresponding to the lattice sites are used in JW mapping, $O(N^2)$ Pauli strings are obtained for a local potential, and $O(N^3)$ for a non-local potential. For short-range potentials, if only a limited neighborhood of each lattice site is affected, the scaling with $N$ is further improved. The number of qubits scales linearly with $N$.

\subsection{Contact interaction}

In the rest of the paper we will focus on a specific interaction model. We model the interaction between the two particles using a contact interaction. This type of interaction acts only between two particles that are on the same site, i.e. $V=V_0 \delta(\bs{r})$. Despite its seeming simplicity, this is the interaction obtained in the pionless effective field theory at the leading order for nuclear systems \cite{Bedaque:2002mn,Hammer:2019poc,VANKOLCK1999273,CHEN1999386}. Therefore, performing calculations with such an interaction model is relevant for different nuclear observables. The contact interaction is also relevant for condensed matter physics, as it is used in the Hubbard model.   Finite-range interactions will be considered in future works.

When discussing the two-body system, to be specific, we will consider the deuteron, i.e. the bound state of a proton and a neutron, but the general ideas are relevant for other non-relativistic quantum systems. For this purpose, we will consider an $8\times 8\times 8$ lattice, i.e. $N_x=N_y=N_z=8$, $N=512$, with a distance of $a=1$ fm between two adjacent sites, i.e. $L_x=L_y=L_z=8$ fm. We note that these parameter values are very close to realistic parameter values that are needed to perform accurate calculations of nuclear observables for light systems. For this lattice size, the number of qubits using the above qubit-efficient mapping is $n_q=9$. Using JW mapping would require $2048$ qubits.

$V_0$, the strength of the contact interaction, can be tuned to obtain the binding energy of the system, for example. In our case, we choose $V_0=-235$ MeV. For fixed $a=1$ fm, and in the infinite lattice limit, this leads to binding energy of approximately $2.2$ MeV, in agreement with the experimental value for the deuteron. For the finite value of $L_x=L_y=L_z=8$ fm, the binding energy is $4.375$ MeV. The exact value used for $V_0$ is not important for the general results and conclusion of this work. 

The Hamiltonian matrix for this interaction model can now be constructed using two-body momentum basis states of the form $|\bs{k},-\bs{k}\rangle$, and mapped to Pauli strings. Following the relations in Eq. \eqref{eq:mapping}, and using a specific order of the basis states (see more details in Appendix \ref{sec:app_order_states}), the kinetic part translates to $19$ Pauli strings involving no more than two $Z$ matrices per string, and they all commute with each other. Notice that this is indeed much smaller than the number of all strings with $I$ and $Z$, $2^9=512$, as mentioned before. 
The contact two-body interaction is a constant in momentum space, i.e. proportional to the all-ones $N\times N$ matrix. Since,
\be
X+I = \begin{bmatrix} 1 1 \\ 1 1 \end{bmatrix},
\ee
the all-ones $N\times N$ matrix is obtained as $(X+I)^{\otimes n_q}$. Therefore, the two-body contact interaction translates to all $N$ Pauli strings with $X$ and $I$ matrices. They all have the same coefficient of $V_0/N$, and they all commute with each other. Notice that in this case, the number of Pauli strings is linear with $N$, as opposed to the $N^2$ scaling of a general interaction in this mapping. We have here $512$ strings due to the interaction and a total of $530$ Pauli strings for the whole Hamiltonian (the all-$I$ string appears in both the kinetic and potential contributions). A JW mapping of a Hamiltonian with a contact interaction also results in a number of Pauli strings that scale linearly with the lattice size, but the number of groups of commuting strings is greater than two \cite{Roggero:2019myu}.

We will continue investigating this mapping in the next sections. One possible disadvantage of this mapping is the possibly large entanglement between qubits, due to Pauli strings that include many non-I Pauli matrices. This, in principle, can lead to deep circuits in the execution of time evolution. We will discuss how such circuit length can be significantly shortened by canceling gates originating from different Pauli strings (Sec. \ref{subsec:evo_with_Pauli_strings}). We will also present an efficient circuit construction using ancilla qubits that does not involve a separate evolution of each Pauli string (Sec. \ref{sec:evo_with_mcp}). We will start by discussing time evolution for the two-body system with a contact interaction in the next section, and later present an efficient extension to systems with more than two particles in Sec. \ref{sec:time_evo_many_body}.

We note that the extension of some of the algorithms presented in the next sections to other interaction models is not trivial. One possible approach to deal with more general interaction models is to utilize time propagation of the two-body system to simulate time evolution in the many-body system. Additional work is needed in this direction, including the consideration of other mappings to qubits. 
In this context, a recent work \cite{Watson:2023oov} should be mentioned, in which specific second-quantization mappings were used to describe similar systems with different interaction models (contact interaction and others), leading to parallelization and reduction in circuit depth (but not necessarily in the total number of one-body and two-body gates, and requiring more qubits).

\section{Time evolution: the two-body system} \label{sec:time_evo}

Time evolution is an important part of different algorithms relevant for describing quantum systems, including ground-state preparation and cross section calculations. We will consider here time evolution based on the Trotter approximation \cite{Trotter1959,Suzuki:1985wzj,Lloyd1996}. In this approach, time evolution over time $t$, i.e., the operator $\exp(-i{\cal H}t)$, is divided to short time steps $dt=t/r$ involving $r$ Trotter steps. Specifically, we can split ${\cal H}$ into the sum of the kinetic energy $T$ and the potential energy $V$. 
First-order Trotter approximation with $r$ Trotter steps then gives
\be \label{eq:1st_trotter}
e^{-i{\cal H}t} \approx \left(e^{-iV dt} e^{-iT dt} \right)^r \equiv U_1(dt)^r.
\ee
Second-order Trotter approximation leads to 
\be \label{eq:2nd_trotter}
e^{-i{\cal H}t} \approx \left(e^{-iT \frac{dt}{2}} e^{-iV dt} e^{-iT \frac{dt}{2}} \right)^r
\equiv U_2(dt)^r.
\ee
$T$ and $V$ can be exchanged in both Eq. \eqref{eq:1st_trotter} and Eq. \eqref{eq:2nd_trotter}.
Notice that the first-order and second-order expressions differ only at the very beginning and end of the circuit because $\exp(-iTdt/2)\exp(-iTdt/2) = \exp(-iTdt)$ \cite{Layden:2021ols}. Therefore, the circuit for full time evolution using the second-order Trotter approximation will generally involve only a negligible number of additional gates, compared to using the first-order approximation. Nevertheless, one can obtain non-negligible improvement using the second-order approximation \cite{Layden:2021ols}. Therefore, we will mostly use the second-order formula.

We now need to separately consider the operators $\exp(-iT dt)$ and $\exp(-iV dt)$. We will start with the kinetic term.
Since in our case the kinetic energy $T$ translates to a set of commuting Pauli strings, we can write
\be \label{eq:T_evol}
e^{-iT dt}=\prod_{S_j \in T} e^{-i \alpha_j S_j dt},
\ee
where $S_j$ are the strings in the mapping of $T$, and $\alpha_j$ are the corresponding numerical coefficients. Each operator $\exp(-i\alpha S dt)$, with $k$ non-$I$ elements in the string $S$, can be performed using a simple quantum circuit, involving a ladder of $2(k-1)$ controlled-NOT (CNOT) gates and a single $Z$-rotation \cite{Wecker:2014vsy}. For the lattice dimensions chosen in this work, it results in a circuit with $18$ CNOTs and $18$ $Z$-rotations for the exact application of the operator $\exp(-iT dt)$. The circuit can be organized as
\be \label{eq:T_evo_circuit}
\Qcircuit @C=1em @R=1.2em {
& \lstick{q_0} & \targ & \gate{R_1}  & \targ & \targ & \gate{R_2} & \targ & \qw & \gate{R_4} & \qw & \qw \\
& \lstick{q_1} &   \qw &    \qw   &  \qw   & \ctrl{-1} & \gate{R_2} & \ctrl{-1} & \ctrl{1} & \qw & \ctrl{1} & \qw\\
& \lstick{q_2} & \ctrl{-2} & \gate{R_1} & \ctrl{-2} & \qw &\qw &\qw & \targ & \gate{R_3} & \targ & \qw \\
& \lstick{q_3} & \targ & \gate{R_1}  & \targ & \targ & \gate{R_2} & \targ & \qw & \gate{R_4} & \qw & \qw \\
& \lstick{q_4} &   \qw &    \qw   &  \qw   & \ctrl{-1} & \gate{R_2} & \ctrl{-1} & \ctrl{1} & \qw & \ctrl{1} & \qw \\
& \lstick{q_5}  & \ctrl{-2} & \gate{R_1} & \ctrl{-2} & \qw &\qw &\qw & \targ & \gate{R_3} & \targ & \qw   \\
& \lstick{q_6} & \targ & \gate{R_1}  & \targ  & \targ & \gate{R_2} & \targ & \qw & \gate{R_4} & \qw & \qw \\
& \lstick{q_7} &   \qw &    \qw   &  \qw   & \ctrl{-1} & \gate{R_2} & \ctrl{-1} & \ctrl{1} & \qw & \ctrl{1} & \qw  \\
& \lstick{q_8} & \ctrl{-2} & \gate{R_1} & \ctrl{-2} & \qw &\qw &\qw & \targ & \gate{R_3} & \targ & \qw  \\
}
\ee
Here, we employ the notation $R_j=R_Z(2\varphi_j dt)$ for $Z$-rotations. $\varphi_j$ are determined according to the numerical coefficients in the mapping of $T$ to Pauli strings: $\varphi_1=-51.163$, $\varphi_2=25.581$, $\varphi_3=-102.326$, and $\varphi_4=12.791$ (all in units of MeV). To build this circuit we used the fact that $Z$-rotation and CNOT that is controlled on the same qubit of the $Z$-rotation commute. 
The above circuit does not include the impact of the all-$I$ Pauli string as it is just a global phase (its numerical coefficient is $\alpha=422.093$ MeV). All other $18$ Pauli strings and their numerical coefficients can be read from the above circuit.
We can see that the circuit is built from $3$ identical and disconnected $3$-qubit circuits. The state of each group of three qubits describes the value of the momentum in one axis ($k_x$, $k_y$, or $k_z$) and the kinetic energy operator acts separately on each axis. This leads to such three disconnected circuits.
Notice that such a construction of the circuit for $\exp(-iT dt)$ is general and does not depend on the interaction model.

In the above circuit there are $4$ groups of same-angle parallel rotations (two groups with $6$ rotations and two groups with $3$ rotations). The Hamming-weight phasing method \cite{Gidney2018halvingcostof,Kivlichan_2020} can be used to reduce the number of arbitrary rotations. For $n$ parallel rotations with the same angle, it provides implementation with $\lfloor{\log_2{n} +1 }\rfloor$ rotations at the cost of at most $4n-4$ explicit T gates and $n-1$ ancilla qubits. This is relevant for error-corrected implementations where arbitrary $Z$-rotations should be translated to $T$ gates, and are, thus, expensive. For us, $6$ parallel rotations can be replaced by  $\lfloor{\log_2{6} +1 }\rfloor = 3$ rotations, using $20$ T gates and $5$ ancilla qubits. And $3$ parallel rotations can be replaced by $\lfloor{\log_2{3} +1 }\rfloor = 2$ rotations using $8$ T gates and $2$ ancilla qubits. In total we can have $10$ rotations instead of the original $18$. The ancilla qubits are used only for the time of execution of the parallel rotations, so we need at most $5$ ancilla at a given time. The total number of explicit T gates that are added to the circuit is $56$.
Notice that the number of $T$ gates required for a $Z$-rotation with an arbitrary angle with error $\epsilon$ is approximately $1.15 \log_2{(1/\epsilon)}+9.2$ \cite{Bocharov:2015tpl}. Therefore, reducing $6$ rotations to $3$ with the cost of $20$ T gates is beneficial even if low accuracy rotations are used. The same is true for reducing $3$ rotations to $2$ with the cost of $8$ T gates. We note that there are suggestions for partially fault-tolerant protocols where arbitrary $Z$-rotations can be directly implemented \cite{Akahoshi:2023xck}, for which the original circuit would probably be better. 
In this work, we will consider the circuit given in Eq. \eqref{eq:T_evo_circuit}.

Moving to the interaction term $\exp(-iV dt)$, we will present here two approaches to constructing an appropriate quantum circuit for a two-body system with a contact interaction. In the first approach, we follow a similar idea of time evolution of each Pauli string, but with significant cancellation of gates from adjacent strings.
In the second approach, we rely on the specific structure of the contact interaction to identify an efficient time evolution algorithm with gate complexity linear with the number of qubits, i.e. logarithmic with the lattice size. It requires the use of ancilla qubits. The total number of qubits is still logarithmic with the lattice size. 

\subsection{Evolution with Pauli strings}
\label{subsec:evo_with_Pauli_strings}

As mentioned above, the operator $V$ translates to all $2^{n_q}=512$ $9$-qubit strings built from the $X$ and $I$ matrices with the same numerical coefficient. We can therefore write
\be \label{Eq:VtoD}
e^{-iV dt} = H^{\otimes n_q} e^{-iD dt} H^{\otimes n_q},
\ee
where $H^{\otimes n_q}$ is the tensor product of Hadamard gates, and $D$ is a diagonal matrix in momentum space. Thanks to the identity $HZH=X$, the mapping of $D$ to Pauli strings is identical to $V$, only with $Z$ matrices instead of $X$. 
$\exp(-iD dt)$ can be written similarly to Eq. \eqref{eq:T_evol} since all the Pauli strings commute. 
Following naively the same idea of CNOT ladders used for the kinetic term results in $2-2^{n_q+1}+n_q2^{n_q}=3,586$ CNOT gates and $2^{n_q}-1=511$ rotations \cite{Welch_2014}. But CNOTs coming from neighboring Pauli strings can cancel as CNOT$^2=1$. This depends on the order of Pauli strings used in the implementation of time evolution. Following Ref. \cite{Welch_2014}, the Pauli strings can be ordered based on the Gray Code. Then, using commutation properties of CNOT gates, we obtain a circuit that includes only a single CNOT between two adjacent $Z$-rotations, with a total of $2^{n_q}-2=510$ CNOTs and $2^{n_q}-1=511$ $Z$-rotations in the exact application of $\exp(-iV dt)$ (together with 9 Hadamard gates at the beginning and at the end of the circuit, Eq. \eqref{Eq:VtoD}). We can see that despite the fact that $V$ includes Pauli strings with many non-$I$ elements (e.g., the all-$X$ string), the final number of CNOTs per Pauli string is relatively small due to CNOT cancellations.

Using first-order Trotter approximation, Eq. \eqref{eq:1st_trotter}, we get 528 CNOTs and 529 $Z$-rotations per trotter step. Implementing second-order Trotter approximation, Eq. \eqref{eq:2nd_trotter}, requires only $18$ additional CNOTs and $18$ additional $Z$-rotations for the total time evolution, independent of the number of Trotter steps.

\subsection{Evolution with multicontrolled gate}
\label{sec:evo_with_mcp}

The above approach of CNOT cancellation is applicable for any diagonal matrix. However, the diagonal operator $D$ has a special structure. It is the sum of all $2^{n_q}$ Pauli string with $Z$ and $I$ matrices with the same numerical coefficient. As a result, with the help of ancilla qubits, 
the gate complexity for implementing $\exp(-iVdt)$ can be significantly improved. Explicitly,
\be \label{eq:D_def}
D = \frac{V_0}{N} (Z+I)^{\otimes n_q}.
\ee
Since 
\be
(Z+I)/2 =
\begin{pmatrix}
 1 & 0 \\
0 & 0
\end{pmatrix},
\ee
$D$ has a single non-zero entry: the first term on the diagonal is $V_0$ (because $N=2^{n_q}$). Therefore, for the diagonal operator $e^{-i D dt}$, the first element on the diagonal is $e^{-i V_0 dt}$ and all others are simply $1$. For convenience, we denote $\theta =  V_0 dt$. We can see that $e^{-i D dt}$ is a multi-controlled operator. It acts on the first qubit if, and only if, all the remaining $n_q-1$ qubits are in the $|0\rangle$ state. In that case, the operation that is applied on the first qubit is given by 
\be \label{eq:U_def}
U = \begin{pmatrix}
 e^{-i\theta} & 0 \\
0 & 1
\end{pmatrix}.
\ee
This is a phase gate, where the phase is applied on state $|0\rangle$ (as opposed to the common definition of a phase gate).
Explicitly, for our $9$-qubit case, we can implement $\exp(-iVdt)$ as
\be \label{eq:exp_V_with_mcp}
\Qcircuit @C=1em @R=1.2em {
& \lstick{q_0} & \gate{H}& \qw      & \gate{U}  & \qw       & \gate{H}& \qw \\
& \lstick{q_1} & \gate{H}& \gate{X} & \ctrl{-1} & \gate{X}  & \gate{H}& \qw \\
& \lstick{q_2} & \gate{H}& \gate{X} & \ctrl{-1} & \gate{X}  & \gate{H}& \qw \\
& \lstick{q_3} & \gate{H}& \gate{X} & \ctrl{-1} & \gate{X}  & \gate{H}& \qw \\
& \lstick{q_4} & \gate{H}& \gate{X} & \ctrl{-1} & \gate{X}  & \gate{H}& \qw \\
& \lstick{q_5} & \gate{H}& \gate{X} & \ctrl{-1} & \gate{X}  & \gate{H}& \qw \\
& \lstick{q_6} & \gate{H}& \gate{X} & \ctrl{-1} & \gate{X}  & \gate{H}& \qw \\
& \lstick{q_7} & \gate{H}& \gate{X} & \ctrl{-1} & \gate{X}  & \gate{H}& \qw \\
& \lstick{q_8} & \gate{H}& \gate{X} & \ctrl{-1} & \gate{X}  & \gate{H}& \qw \\
}
\ee
The Hadamard gates are due to Eq. \eqref{Eq:VtoD} and the $X$ gates are to ensure that the gate $U$ is only applied to qubit $q_0$ if qubits $q_1,q_2,...,q_8$ are in the $\ket{0}$ state.

\begin{figure*}
    \centering
    \includegraphics[scale=1]{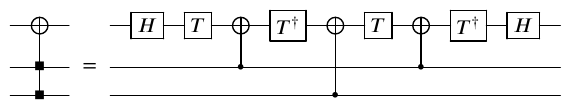}
    \caption{Circuit for a $3$-qubit gate that is equivalent to the Toffoli gate up to phases. Based on a circuit presented in Ref. \cite{Kivlichan_2020} (page 22). The squares here represent the control qubits.}
    \label{fig:almost_Toffoli}
\end{figure*}

\begin{figure*}
    \centering

\includegraphics{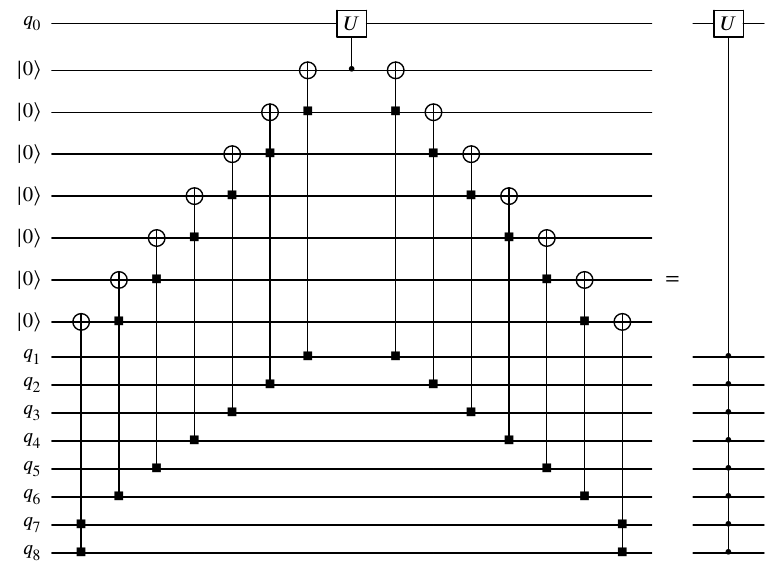}
    \caption{Circuit for a multicontrolled-U gate $c^{n_q-1}U$ with $n_q-2$ ancilla qubits initialized to the state $\ket{0}$, for $n_q=9$. }
    \label{fig:mcU_with_almost_Toffoli}
\end{figure*}

With this realization, it is left to efficiently construct the multi-controlled gate $c^{n_q-1}U$ using more basic quantum gates. Following Refs. \cite{Nielsen_Chuang_2010,Selinger:2013ksm}, this can be done with the help of $n_q-2$ ancilla qubits and a ladder of Toffoli gates. In fact, one can use $3$-qubit gates that are equivalent to a Toffoli gate except for possible phases \cite{Selinger:2013ksm}. We provide an explicit construction of such a gate in Fig. \ref{fig:almost_Toffoli}. This gate flips the target qubit only if both control qubits, denoted here by squares, are in state $\ket{1}$. However, unlike the Toffoli gate, it also changes the phase of some states. Notice that this gate is its own inverse. It is built using $4$ T gates and $3$ CNOTs (and $2$ Hadamard gates). This can be compared to the Toffoli gate, which requires $7$ T gates and $6$ CNOTs. 
Now, this gate can be utilized to construct a circuit for $c^{n_q-1}U$ using $n_q-2$ ancilla qubits \cite{Nielsen_Chuang_2010,Selinger:2013ksm}. We provide such a circuit for the case of $n_q=9$ in Fig. \ref{fig:mcU_with_almost_Toffoli}. Modifications for other values of $n_q$ are straightforward. The circuit includes the gate from Fig. \ref{fig:almost_Toffoli} $2(n_q-2)$ times, together with a single $cU$ gate. We note that the circuit can be reorganized so that some of the $3$-qubits gates in Fig. \ref{fig:mcU_with_almost_Toffoli} can be executed in parallel \cite{Selinger:2013ksm,He:2017wow}.

\begin{figure*}
    \centering

\includegraphics{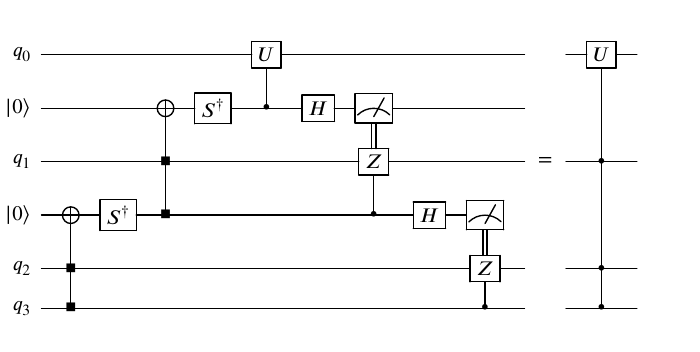}
    \caption{Circuit for $c^3U$ with $2$ ancilla qubits initialized to the state $\ket{0}$, following Ref. \cite{Jones:2013gpb}.}
    \label{fig:c3U_mid_circuit}
\end{figure*}

If we allow for mid-circuit measurements and classical feed-forward, the gate complexity for executing $c^{n_q-1}U$ can be further reduced, as was shown in Ref. \cite{Jones:2013gpb}. Starting with $c^2U$, it can be executed as \cite{Jones:2013gpb,Kivlichan_2020}
\be \label{eq:c2U_mid_circuit}
\Qcircuit @C=1em @R=1.2em {
\lstick{q_0}     & \qw                   & \qw              &\gate{U}  &\qw      &\qw           &\qw & && \gate{U}  & \qw \\
\lstick{\ket{0}} & \targ                 & \gate{S^\dagger} &\ctrl{-1} &\gate{H} &\meter        &    &=&&           &     \\
\lstick{q_1}     & \qw \qwx \smblksquare & \qw              &\qw       &\qw      &\gate{Z} \cwx &\qw & && \ctrl{-2} & \qw \\
\lstick{q_2}     & \qw \qwx \smblksquare & \qw              &\qw       &\qw      &\ctrl{-1}     &\qw & && \ctrl{-1} & \qw  \\
}
\ee
The above circuit includes a measurement of an ancilla qubit and a $cZ$ gate that is applied only if the result of the measurement is $\ket{1}$.
This approach can be extended to a general $c^{n_q-1}U$ gate \cite{Jones:2013gpb}. In Fig. \ref{fig:c3U_mid_circuit} we provide an example for $c^3U$. It is based on the fact that Eq. \eqref{eq:c2U_mid_circuit} is valid also for operators that act on multiple qubits instead of the gate $U$. For the general case, $c^{n_q-1}U$ can be executed using $n_q-2$ ancilla qubits (as before). The $3$-qubit gate of Fig. \ref{fig:almost_Toffoli} appears $n_q-2$ times, and there are also $n_q-2$ measurements, once for each of the ancilla qubits, followed by a classically-conditioned $cZ$ gate.

Finally, the controlled-phase gate $cU$ can be applied using $3$ rotations and 2 CNOTs. Explicitly,
\be \label{eq:cU_circuit}
\Qcircuit @C=1em @R=0.6em {
& \gate{U}  &\qw  & &&  \targ   & \gate{R_Z\left(-\frac{\theta}{2}\right)} & \targ & \gate{R_Z\left(\frac{\theta}{2}\right)} & \qw &\\ 
&           &     &=&&          &      &           &   & &\\ 
& \ctrl{-2} &\qw  & && \ctrl{-2}& \qw  & \ctrl{-2} & \gate{R_Z\left(-\frac{\theta}{2}\right)}& \qw &\\
}
\ee
$\theta$ here is the same angle used in the definition of U, Eq. \eqref{eq:U_def}.
We note that, in this construction, the circuit on the right comes with a global phase of $\theta/4$ (i.e., with an overall factor of $\exp(i\theta/4)$) compared to the definition of $cU$. We discuss the implications of this phase difference in Appendix \ref{sec:app_global_phase}.

Combining all these components, a single application of $\exp(-iVdt)$ using $n_q-2$ ancilla qubits requires $5n_q-6$ Hadamard gates, $2(n_q-1)$ $X$ gates, $n_q-2$ $S$ gates, $4(n_q-2)$ $T$ gates, $3n_q-4$ CNOTs, $n_q-2$ mid-circuit measurements followed by a measurement-conditioned $cZ$ gate, and $3$ $Z$-rotations (we do not separate here between a gate and its inverse, e.g. $T$ and $T^\dagger$). Notice the linear dependence of all gates with respect to the number of qubits, i.e. logarithmic with respect to the lattice size $N$. Notice also that the number of arbitrary $Z$-rotations is fixed to $3$ (independent of the number of qubits) and all other non-Clifford gates are explicit $T$-gates. This is important for error-corrected implementations, where arbitrary $Z$-rotations should be translated to (possibly many) $T$ gates.
We used here the latter approach (involving mid-circuit measurements) for the $C^{n_q-1}U$ gate. Using the approach of Fig. \ref{fig:mcU_with_almost_Toffoli}, not requiring mid-circuit measurements and feed-forward,  will more-or-less double the number of $T$ gates and CNOTs, but still with a linear dependence on the number of qubits and only $3$ arbitrary $Z$-rotations. 

We obtained here a significant improvement compared to Section \ref{subsec:evo_with_Pauli_strings}, where an exponential dependence on the number of qubits is obtained for both CNOTs and arbitrary Z-rotations. Specifically, for $n_q=9$ we have here $23$ CNOTs and $7$ conditioned $cZ$, $28$ $T$ gates, and $3$ Z-rotations (together with $39$ Hadamard gates, $16$ $X$ gates, and $7$ $S$ gates). The approach of Section \ref{subsec:evo_with_Pauli_strings} requires $510$ CNOTs and $511$ $Z$-rotations.
In addition, increasing the lattice size by doubling the number of sites in each dimension to $N_x=N_y=N_z=16$ (requiring $n_q=12$) will lead to only modest increase in the number of gates in the current approach, e.g., $32$ CNOTs, $11$ conditioned $cZ$, and $40$ $T$ gates (and still $3$ Z-rotations). On the other hand, following the approach of Section \ref{subsec:evo_with_Pauli_strings} will result in an exponential increase in the number of gates, leading to $4094$ CNOTs and $4095$ $Z$-rotations. 

\section{Time evolution: the many-body case} \label{sec:time_evo_many_body}

The efficient time-evolution algorithm of Sec. \ref{sec:evo_with_mcp} can be extended beyond two particles to the many-body case. For this purpose, it is simpler to use coordinate-space basis states instead of the momentum-space states used before. For $A$ particles, each many-body basis state is given by the location of all particles on the lattice. Thus, we have $N^A$ states, and the number of qubits is given by $n_q=A\log_2(N)$. As before, in the mapping to qubits, each basis state is assigned a bitstring of length $n_q$. We organize the basis states such that this mapping coincides with the well-known first-quantization mapping \cite{AbramsLloyd1997}. In this mapping, the lattice sites are ordered in some way, labeled from $0$ to $N-1$, and each label is translated to its binary representation using $\log_2(N)$ bits. Then, the full bitstring of a given many-body basis state is chosen such that the first $\log_2(N)$ bits describe the location of the first particle, the second $\log_2(N)$ bits describe the location of the second particle, etc.

We can now discuss the algorithm for the time evolution with the potential energy term $\exp(-iVdt)$. Here, $V=\sum_{i<j} V_{ij}$, where $V_{ij}$ is acting on the $ij$ pair. Since $\exp(-iVdt) = \prod_{i<j} \exp(-iV_{ij}dt)$, we can focus on a single $\exp(-iV_{ij}dt)$ term.

\begin{figure*}

\centering
\includegraphics[scale=0.88]{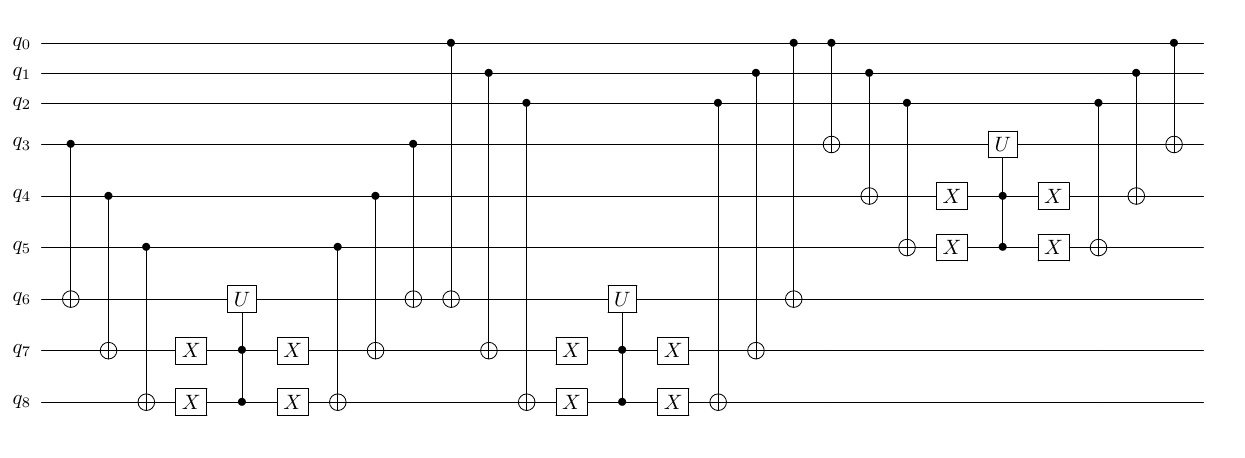}

    \caption{An example of a circuit for $\exp(-iVdt)$ for $A=3$ and $N=8$, and contact two-body interaction. The first $3$ qubits represent the location of particle $1$, the next $3$ qubits are for particle $2$ and the last $3$ qubits are for particle $3$. $\exp(-iVdt)$ is decomposed as $\exp(-iV_{12}dt)\exp(-iV_{13}dt)\exp(-iV_{23}dt)$. Extensions to more particles and a larger lattice is straightforward as explained in the text.}    \label{fig:Aequal3_time_evo}
\end{figure*}

In coordinate space, $V_{ij}$ is diagonal and the elements on the diagonal are non-zero (and equal to $V_0$) only if particles $i$ and $j$ are on the same lattice site, as we still consider the contact interaction model. 
Therefore, $\exp(-iV_{ij}dt)$ is diagonal, with the value $\exp(-iV_0dt)$ for the above elements, and $1$ for the others. 
In the first-quantization mapping, the non-zero values correspond to the bitstrings in which the $i$'s and $j$'s $\log_2(N)$ bits are the same. Thus, we can apply $\log_2(N)$ CNOTs controlled on the qubits corresponding to the location of particle $i$ and targeted on those corresponding to particle $j$ (first to first, second to second, etc.). This will change all the $j$'s $\log_2(N)$ digits to zero for all the above bitstrings. Now, applying the phase $\exp(-iV_0dt)$ to these bitstrings can be done using the multicontrolled phase gate $c^{\log_2(N)-1}U$ gate with $\log_2(N)-1$ $X$ gates before and after it, as in Sec. \ref{sec:evo_with_mcp}. Finally, the same structure of $\log_2(N)$ CNOTs should be applied to return the bitstrings to their original form. Following the construction of $c^{\log_2(N)-1}U$ using $\log_2(N)-2$ ancilla qubits discussed in Sec. \ref{sec:time_evo_many_body} leads to total gate count that scales logarithmically with the lattice size $N$ for the application of $\exp(-iV_{ij}dt)$. A similar circuit is required for each $ij$ pair, leading to $A \choose 2$ scaling with respect to the number of particles for the time evolution with the full interaction $V$. 
See Fig. \ref{fig:Aequal3_time_evo} for a simple example with $A=3$. 
This algorithm is similar to the general ideas discussed for simulating the Hubbard model in the first quantization in Ref. \cite{AbramsLloyd1997}, but the construction of the actual quantum circuit was not discussed. 

Time evolution with the kinetic term can be applied based on its mapping to Pauli strings. The kinetic term can be expressed in its common form based on finite-differences expression of the derivative. A relatively small number of terms is obtained, resulting in a small number of gates, scaling linearly with the number of particles $A$ for the full kinetic energy. One can also follow the ideas of Ref. \cite{AbramsLloyd1997} for the kinetic energy, or use quantum Fourier transform to diagonalize it.   

\section{Ground-state preparation}
\label{sec:GS_prep}

As mentioned before, the first step of many quantum algorithms is the preparation of a desired initial state on the quantum computer. Here we will focus on the ground state of the Hamiltonian. We will discuss and analyze two methods. First, we analyze the complexity of the projection-based energy-filtering approach of Ref. \cite{Stetcu:2022nhy}, where only limited information regarding the ground state and energy spectrum is assumed to be known. Next, assuming that the ground-state wave function is known, we study a method to prepare this state on a quantum computer. We analyze the success probability and gate complexity. In this section, we continue to consider the two-body case with the contact interaction and momentum-space representation. 

\subsection{Measurement-based energy filter}
\label{subsec:energy_filter}

We will first consider here the projection-based approach of Ref. \cite{Stetcu:2022nhy}. The first step is to initialize the system in a state $|\psi_i\rangle$, preferably with significant overlap with the exact ground state $|\Psi_0\rangle$. We assume $\langle \Psi_0 | \Psi_0 \rangle=\langle \psi_i | \psi_i \rangle=1$. In our case, we consider the state in which the two particles have zero momentum. It simply corresponds to the all-zero $9$-qubit state with the mapping used here, and obeys $|\langle \psi_i | \Psi_0 \rangle|^2 \approx 0.75$. Using ancillary qubit $a$ in state $|0\rangle$, the initial state is $|\psi_i\rangle \otimes |0\rangle$. Then, a time evolution operator of the form $\exp[-i({\cal H}-E_0 I_N) t \otimes Y_a]$ is applied, where $E_0$ is the ground-state energy, $I_N$ is the $N \times N$ identity matrix, and $Y_a$ is the Y Pauli matrix acting on the ancillary qubit. In principle, a phase can also be included \cite{Stetcu:2022nhy}, but we will not utilize this freedom here. If we then read that qubit $a$ is in state $|0\rangle$, the system is projected to the state ${\cal N}\cos[({\cal H}-E_0 I_N)t]|\psi_i \rangle$, where ${\cal N}$ is a normalization factor. This happens with probability $\langle \psi_i | \cos^2[({\cal H}-E_0 I_N)t]|\psi_i \rangle$. We can try to choose the time $t$ to get close to the ground state $|\Psi_0\rangle$. We assume to know the energy gap $\Delta$ between the ground state and the first excited state of the Hamiltonian, and choose the time $t_\Delta\equiv\frac{\pi}{2\Delta}$. As a result, the first excited state is exactly projected out from the initial state. In our case $\Delta=13.5$ MeV. The average energy of the initial state is $\langle \psi_i | {\cal H} | \psi_i \rangle \approx -0.46$. After projection with time $t_\Delta$, the average energy of the state, denoted by $\psi_\Delta$, is $-4.19$ MeV, less than $5\%$ difference compared to the exact ground-state energy. The success probability for producing $\psi_\Delta$, i.e. measuring the ancillary qubit in state $|0\rangle$, is approximately $75\%$, and $|\langle \psi_\Delta | \Psi_0 \rangle|^2 \approx 0.9988$. We can see that, by projecting out the first excited state, we obtain a good approximation for the ground state. This can be improved by performing additional projections with different values for the time $t$ \cite{Stetcu:2022nhy}. 

\begin{figure} \begin{center} 
\includegraphics[width=\linewidth]{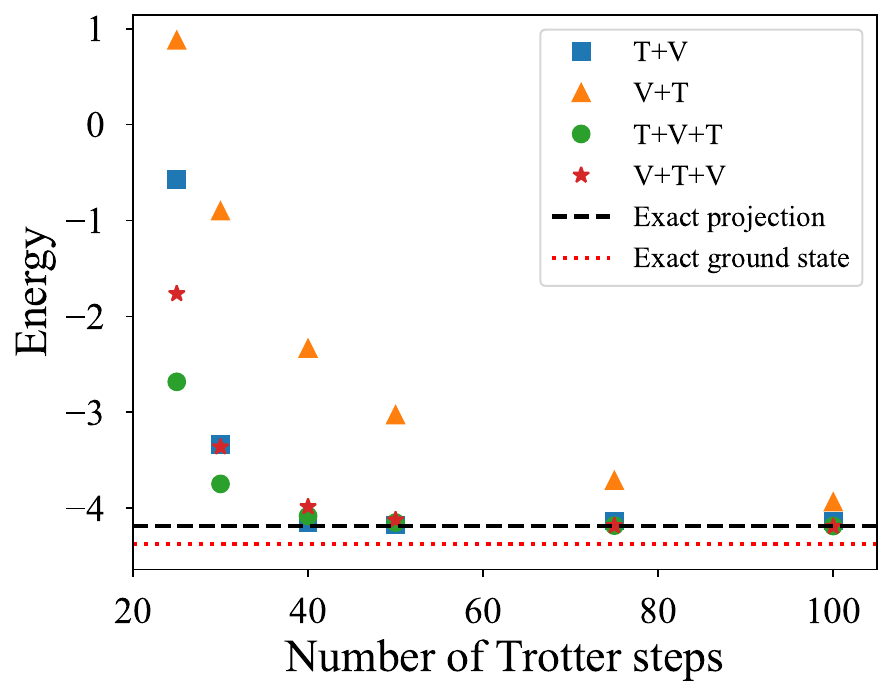}
\caption{\label{fig:gap_projection}
The energy of the state obtained after the projection-based approach described in the text, as a function of the number of Trotter steps used in the time evolution. The initial state is the zero-momentum state, and the time used in the algorithm is $t_\Delta$. Results using different Trotter approximations are shown. The label 'V+T' corresponds to Eq. \eqref{eq:1st_trotter}, with the operators $V \otimes Y_a$ and $(T-E_0 I_N)\otimes Y_a$ instead of $V$ and $T$, respectively. The label 'T+V' corresponds to the opposite order. 'T+V+T' and 'V+T+V' similarly correspond to the second-order Trotter approximation (Eq. \eqref{eq:2nd_trotter}). 
The energy obtained using exact time evolution and the exact ground-state energy are also shown (dashed and dotted lines, respectively).
}
\end{center}\end{figure}

To apply this algorithm on a quantum computer, a Trotter approximation can be used, similar to the discussion in Section \ref{sec:time_evo} (see Appendix \ref{sec:app_energy_filter} for more technical details). 
We study the convergence of such a calculation with respect to the number of Trotter steps in Fig. \ref{fig:gap_projection}. We can see that, as the number of Trotter steps increases, the energy of the resulting state converges to the value obtained using the exact projection ($-4.19$ MeV). We compare the use of first-order and second-order Trotter approximations, as well as the order of the operators, i.e. as in Eqs. \eqref{eq:1st_trotter} and \eqref{eq:2nd_trotter}, or if $V$ and $T$ are replaced. We can generally see that the use of second-order approximation leads to faster convergence compared to first order. Moreover, we see that one order of operators in the first-order approximation (orange triangles in the plot) leads to especially slow convergence. It would be interesting to try and understand the origin of this difference and its relevance to other studies, systems and algorithms. Except for this case, about $40$ Trotter steps are needed to obtain a good approximation of the time evolution, with less than $5\%$ error in the energy of the resulting state. 

Following the same ideas of the Gray-Code order discussed in Section \ref{subsec:evo_with_Pauli_strings}, for the first-order Trotter approximation, each step involves $536$ CNOTs and $531$ $Z$-rotations for $n_q=9$. For $40$ step, this is $21440$ CNOTs and $21240$ $Z$-rotations. The second-order approximation results in only $24$ more CNOTs and 19 more $Z$-rotations for the whole time evolution. If a larger lattice is required, the scaling of the number of gates in this approach is exponential with respect to the number of qubits, which is linear with respect to the lattice size.

Alternatively, we can build a shorter circuit utilizing a few ancilla qubits and the ideas of Section \ref{sec:evo_with_mcp}, 
adapted to the operator $\exp(-iV \otimes Y_a dt)$. This leads to only $6$ additional CNOTs compared to the circuit for $\exp(-iV dt)$. 
In total, the circuit will include $53$ CNOTs, $7$ conditioned $cZ$, $28$ $T$ gates, and $22$ $Z$-rotations per first-order Trotter step. For $40$ steps we get $2120$ CNOTS, $280$ conditioned $cZ$, $1120$ T gates, and $880$ $Z$-rotations (together with additional single-qubit Clifford gates). 
As before, second-order Trotter approximation results in only $24$ more CNOTs and 19 more $Z$-rotations for the whole time evolution. See technical details in Appendix \ref{sec:app_energy_filter}. This approach, compared to the Gray-Code order approach, leads to an order of magnitude reduction in the number of gates and even a larger reduction in the number of arbitrary $Z$-rotations. In total, here we require an order of $4\times10^3$ gates to obtain a good approximation of the ground state. As the scaling of the number of gates in the current approach is linear with the number of qubits (logarithmic in the lattice size), compared to the exponential scaling of the Gray-Code order approach, this advantage will be even more significant for larger lattice sizes.

\subsection{Measurement-based state initialization}
\label{subsec:initialization}

We can also consider a case in which we know classically the exact ground state and would like to initialize a quantum computer to this state. Then, for example, reaction that cannot be calculated on a classical computer could be studied using the quantum computer. We will provide here an algorithm with similarities to the above measurement-based approach. 

We assume that we know the expansion of the desired state using the basis states $\{|v_j\rangle \}$ defined in Section \ref{sec:Ham_Map}
\be \label{eq:gs_expan}
|\Psi_0 \rangle = \sum_j g_j |v_j \rangle,
\ee
where $g_j$ are complex numbers.
We start by initializing the qubits to a state that includes all basis states with {\it real} non-zero coefficients $\{b_j\}$
\be \label{eq:ini_state_bj}
|\psi_i \rangle = \sum_j b_j |v_j \rangle.
\ee
This can be obtained, for example, by initializing all qubits to the $|0\rangle$ state and applying a Hadamard gate to each qubit.
Now, with a single ancillary qubit $a$, the initial state is assumed to be $|\psi_i\rangle \otimes |0\rangle$. Then, an operator $\exp[-i Q \otimes Y_a]$ can be applied, where 
\be \label{eq:Q_op}
Q = \sum_j d_j |v_j\rangle \langle v_j|,
\ee
for real numbers $\{d_j\}$.
If we measure the ancillary qubit in state $|0\rangle$, the system collapses to the state 
$|\psi_Q \rangle = {\cal N} \sum_j \cos(d_j) b_j |v_j \rangle$, where ${\cal N}$ is an appropriate normalization factor. The values of $d_j$ can be chosen such that we obtain the state $|\psi_Q \rangle = \sum_j |g_j| |v_j \rangle$. To get the right phase of $g_j$, we can apply the operator $\exp(i\Theta)$, where $\Theta = \sum_j \theta_j |v_j\rangle \langle v_j|$ and $g_j=|g_j|\exp(i\theta_j)$. This finally results in the desired state $|\Psi_0\rangle$.
We provide more technical details on this algorithm in Appendix \ref{sec:app_state_ini}, including a discussion about the success probability of this approach. 

Since $Q$ and $\Theta$ are diagonal, the corresponding quantum circuit for this algorithm is relatively simple to construct in the mapping discussed in this work. No Trotter approximation is required. 
Applying $\exp[-i Q \otimes Y_a]$ involves at most $2^{n_q}$ CNOTs and $2^{n_q}$ $Z$-rotations. If the coefficients $\{g_j\}$ are not all real, additional $2^{n_q}-2$ CNOTs and $2^{n_q}-1$ $Z$-rotations are needed to apply the operator $\exp(i\Theta)$.
Since we use a qubit-efficient mapping, the number of gates scales linearly with the lattice size for the two-body system; see Eq. \eqref{eq:n_q}.

We also note that it was shown in Ref. \cite{Welch_2014}, that operators like $\exp[-i Q \otimes Y_a]$ and $\exp(i\Theta)$ can be approximated using a polynomial number of one-body and two-body gates. Therefore, if we are only interested in a good approximation of the desired state, the number of gates in our algorithm will scale polynomially with the number of qubits. 
Thus, it might provide an advantage over other algorithms in some cases, especially if a low success probability can be avoided. This includes algorithms that require exponential circuit depth, without any ancilla qubits or measurements \cite{Sun:2023bwt}, or 
algorithms with polynomial depth, but with possibly exponential number of ancilla qubits
\cite{Zhang:2021uwi,Zhang:2021bue,Rosenthal:2021rcb,Zhang:2022pue,Sun:2023bwt}.

In our specific $9$-qubit two-body problem, $512$ $Z$-rotations and CNOTs are needed for exact initialization to the ground state (all $g_j$ are real). 
If a $10\%$ accuracy on the energy is required, only $300$ $Z$-rotations and a similar number of CNOTs \cite{Welch_2014} are needed. This is a shallower circuit compared to the final method discussed in Section \ref{subsec:energy_filter}, where $2400$ two-qubit entangling gates, $1120$ T gates and $880$ $Z$-rotations are needed. The main difference is the need for multiple Trotter steps in the approach of Section \ref{subsec:energy_filter}. Therefore, the method discussed in the current section could be useful for the description of reactions in relatively small systems, where the ground state can be solved on a classical computer. Nevertheless, if a larger lattice is used, the linear scaling of the number of gates with respect to the number of qubits can lead to a shallower circuit when using the method of Section \ref{subsec:energy_filter} compared to the current method.

We have analyzed here two methods for ground-state preparation to compare their advantages. The first method, based on time evolution with the Hamiltonian, has a linear gate-complexity scaling with respect to the number of qubits but requires multiple Trotter steps and leads to an approximate ground state. The second method requires knowing the ground-state wave function and scales exponentially for exact state preparation or polynomially for approximated preparation, but requires no Trotter approximation. Although there are other available state preparation methods that we did not analyze here, this analysis shows possible trade-offs of different approaches.

In the calculation of the response function in the next section, we assume that the ground state has been prepared with a relatively short circuit, as suggested in the current section, such that this step has a negligible contribution to the total complexity of the calculation.

\section{Response function}
\label{sec:res_func}

After discussing time evolution and ground-state preparation, we are set to discuss the calculation of response functions.
Response functions are needed to calculate the cross sections of reactions that are governed by linear response. In nuclear physics, this includes electron- and neutrino-scattering reactions. Accurate calculations of these reactions are crucial for extracting nuclear structure information as well as fundamental properties of neutrinos from experiments.

In some cases, response functions can be calculated using classical computers. Utilizing integral transforms, methods like Green’s function Monte Carlo and coupled cluster can be used to calculate inclusive lepton-nucleus response functions for kinematics dominated by quasielastic reactions for light or medium-mass nuclei \cite{Carlson:2001mp,Lovato:2015qka,Lovato:2016gkq,Lovato:2017cux,Sobczyk:2021dwm,Sobczyk:2023sxh}. In addition, the short-time approximation \cite{Pastore:2019urn,Andreoli:2021cxo,Andreoli:2024ovl} and spectral function approaches \cite{Benhar:1994hw,Rocco:2015cil,Rocco:2018mwt}, which are based on factorization of the final hadronic states, can enable calculations of quasielastic response functions, including exclusive reactions. However, they are limited in their description of final-state interaction, including effects like rescattering of a knocked-out nucleon off spectator nucleons. Exact calculations of exclusive response functions are beyond reach even for light nuclei.

We generally follow here the algorithm presented in Ref. \cite{Rogerro2019}, with some modifications. The resources required for the calculation of neutrino-nucleus scattering using these algorithms were estimated in Ref. \cite{Roggero:2019myu} with the JW mapping. We focus here on its integration with the qubit-efficient mapping in momentum space, to provide optimized circuits and resource estimation, with an end-to-end calculation for the two-body system. We will repeat the main steps of the algorithm in the following.

The response function is defined as
\be
S(\omega) = \sum_\nu |\langle \Psi_\nu | \hat{O} | \Psi_0 \rangle |^2 \delta(E_\nu - E_0 - \omega),
\ee
where $\hat{O}$ is a relevant transition operator, $\omega$ is the energy transfer, and $|\Psi_\nu\rangle$ is an eigenstate of the Hamiltonian with energy $E_\nu$.
We will continue to analyze here the two-body deuteron system, and consider a transition operator of the form
\be \label{Eq:O_op}
\hat{O} = e^{i \bs{q} \cdot \bs{r}_p},
\ee
transferring momentum $\bs{q}$ to the proton. $\bs{r}_p$ is the proton coordinate. Such an operator is relevant, for example, for the longitudinal response function in electron-nucleus scattering. We can separate the action of this operator to the relative and CM motion
\be
\hat{O} = e^{i \frac{\bs{q}}{2} \cdot \bs{r}} e^{i \bs{q} \cdot \bs{R}},
\ee
where $\bs{r}$ and $\bs{R}$ are the relative and CM coordinates of the pair. The response function can then be written as
\be
S(\omega) = \sum_\nu |\langle \Psi_\nu^{int} | e^{i \frac{\bs{q}}{2} \cdot \bs{r}} | \Psi_0 \rangle |^2 \delta(E_\nu^{int} + E_{CM} - E_0 - \omega),
\ee
where $E_{CM}=q^2/4m$ is the energy of the CM motion, and $|\Psi_\nu^{int}\rangle$ and $E_\nu^{int}$ are the internal wave function and energy, corresponding to the relative coordinate. This allows us to remain in the space of zero-CM states. We note that, on a lattice, the relevant values of $\bs{q}/2$ are quantized. The lowest non-zero momentum transfer in the $z$ direction obeys $q/2 = 2\pi/L_z$.

Similar to Ref. \cite{Rogerro2019}, we define a shifted and scaled Hamiltonian
\be
\bar{{\cal H}} = \frac{{\cal H}-E_0}{\Delta {\cal H}}
\ee
and energy transfer $\bar{\omega}=(\omega-E_{CM})/\Delta {\cal H}$, where $\Delta {\cal H}$ is the difference between the largest zero-CM eigenstate of ${\cal H}$ and the ground-state energy. The response function corresponding to $\bar{{\cal H}}$ obeys \cite{Rogerro2019}
\be
\bar{S}(\bar{\omega}) = \Delta {\cal H} S(\omega).
\ee
Notice that $\omega \in [E_{CM},E_{CM}+\Delta {\cal H}]$ and $\bar{\omega}\in [0,1]$.
We can, therefore, focus on calculating $\bar{S}(\bar{\omega})$.

After preparing the ground state on a quantum computer, one should apply the transition operator \cite{Rogerro2019}. In our case, $\exp{(i \frac{\bs{q}}{2} \cdot \bs{r})}$ is a unitary operator. It can be applied by mapping the Hermitian operator $\frac{\bs{q}}{2} \cdot \bs{r}$ to Pauli strings, combined with Trotter approximation. Using the mapping discussed in this work, this operator, for the lowest possible non-zero momentum transfer in the $\hat{z}$ direction, translates to a small number of only $16$ strings.  As a result, $15$ $Z$-rotations and $14$ CNOTs are needed per Trotter step. About $30$ first-order Trotter steps produce an accurate result, and therefore the state $\exp{(i \frac{\bs{q}}{2} \cdot \bs{r})} |\Psi_0\rangle$ can be prepared with $450$ $Z$-rotations and $420$ CNOTs. These are relatively small resources compared to the next step.

At this point, the quantum phase estimation (QPE) algorithm \cite{Abrams1999,Cleve:1997dh} can be applied to obtain the response function. In this approach,
$W$ ancilla qubits are initialized to $|0\rangle$ and a Hadamard gate is applied to each of them. Then,
a series of controlled time evolution operators 
\be \label{eq:U2k}
U^{2^k} = e^{i2\pi 2^k \bar{{\cal H}}}
\ee
are applied, where $k=0,1,...,W-1$, each controlled on the corresponding $k$-th ancilla qubit. The inverse quantum Fourier transform is then applied to the ancilla qubits, and their state is measured. Repetition of this process, the probability $P(a)$ of measuring a binary string corresponding to an integer number $a\in [0,2^W-1]$ provides an estimation for $\bar{S}(\bar{\omega})$, where $\bar{\omega}=a/2^W$. In total, this results in $2^W$ evenly-spaced energy values in the range $\bar{\omega}\in[0,1]$ in which $\bar{S}(\bar{\omega})$ is evaluated, i.e. with energy resolution of $2^{-W}$.

The response function corresponding to the momentum transfer operator of Eq. \eqref{Eq:O_op} is expected to have a clear quasi-elastic (QE) peak around $\omega=q^2/2m-E_0$. The strength of the response function should be negligible far enough from this peak. Therefore, we suggest here some modifications to this algorithm. Instead of the operators in Eq. \eqref{eq:U2k}, one can apply the operators
\be \label{eq:QPE_mod_op}
U^{2^k}_{\alpha,\beta} = e^{i2\pi 2^k \alpha (\bar{{\cal H}}-\beta)},
\ee
involving two real parameters $\alpha$ and $\beta$. With this definition, the probability $P(a)$ corresponds to $\bar{S}(\bar{\omega})$, with a modified relation
\be
\bar{\omega}=\frac{a}{2^W \alpha}+\beta.
\ee
Since $a\in [0,2^W-1]$, we get values of $\bar{\omega}$ in the range $\beta \leq \bar{\omega} < 1/\alpha + \beta$. Therefore, we can use $\alpha>1$ to limit the calculation to a smaller range of energies. It is important that the response function is negligible outside this energy range to get a correct result. Compared to the original algorithm (i.e. $\alpha=1$ and $\beta=0$), this can reduce the number of ancilla qubits for a given required energy resolution, and also shorten somewhat the total time evolution (see details in Appendix \ref{sec:app_modified_QPE}). $\beta$ also allows us to shift the energy bins and control the energy values for which the response function is evaluated.

\begin{figure} \begin{center}
\includegraphics[width=\linewidth]{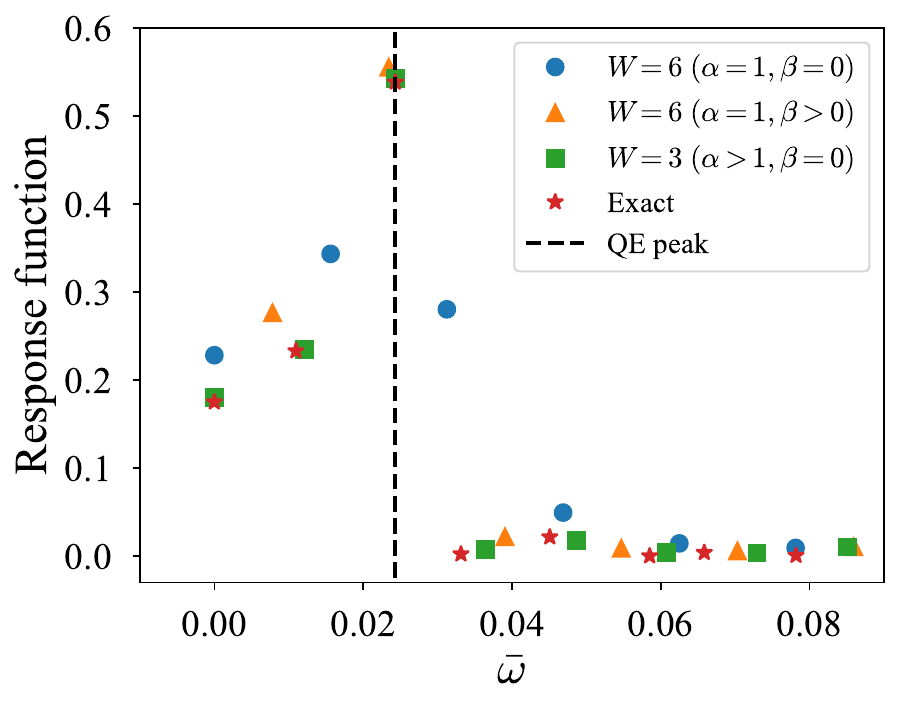}
\caption{\label{fig:response_exact} The response function $\bar{S}(\bar{\omega})$ for the lowest non-zero momentum transfer $\bs{q}= 4\pi/L_z \hat{z} = \pi/2$ fm$^{-1} \hat{z} = 310$ MeV/c $\hat{z}$. The exact calculation (red stars) is compared to the results of the quantum algorithm described in the text, assuming exact implementation (negligible Trotter errors) and exact ground-state initialization. The blue circles correspond to a $6$-ancilla-qubit calculation with the original algorithm ($\alpha=1$, $\beta=0$). The orange triangles correspond to a $6$-ancilla-qubit calculation with translated energy grid using $\beta>0$. The green squares correspond to a 3-ancilla-qubit calculation limited to the QE peak using $\alpha>1$. The expected center of the QE peak is shown by the dashed line. Both $W=6$ calculations extend to $\bar{\omega}=1$.
}
\end{center}\end{figure}

We compare these different possibilities in Fig. \ref{fig:response_exact}, where the response function $\bar{S}(\bar{\omega})$ for the lowest non-zero momentum transfer $\bs{q}= 4\pi/L_z \hat{z}$ is shown. We see that the original algorithm with $W=6$ (blue circles) provides reasonable results compared to the exact calculation, but with a slightly larger width. The calculation for $W=6$ with $\beta>0$ (orange triangles) results in a translated energy grid which allows us to identify the maximum of the QE peak. The green squares show the result of a $W=3$ calculation with $\alpha>1$, such that the energy grid is limited to the QE peak. These results include $8$ points, all in the relevant energy range and with slightly better resolution than the $W=6$ calculations, showing a good agreement with the exact calculations. We note that the total time evolution of the $W=3$ calculation is similar to the $W=6$ calculations, and, therefore, they should all require similar circuit depth.

We now consider the quantum circuit required to implement QPE. We can focus on the step of multiple time evolutions, which is the most resource-demanding part of the response function calculation. It was shown in Ref. \cite{Wecker:2015fib} that instead of executing the controlled time evolution discussed above, it is more efficient to consider a time evolution in which the "direction" is controlled by the ancilla qubit. In essence, instead of performing time evolution $\exp(iHt)$ if the ancilla qubit is in state $\ket{1}$ (and no time evolution if it is in state $\ket{0})$, we want to apply $\exp(-iHt/2)$ if the ancilla qubit is in state $\ket{0}$ and $\exp(+iHt/2)$ it is in state $\ket{1}$. This does not impact the final result of QPE as the difference between the two components remains $\exp(iHt)$ \cite{Wecker:2015fib}. 
When considering the symmetric Trotter approximation for the time evolution, Eq. \eqref{eq:2nd_trotter}, such a change in the direction of the time evolution is simply obtained by $dt\to -dt$ \cite{Kivlichan_2020}, which allows efficient circuit implementation. As a result, due to the factor of $1/2$ in the time evolution, half the number of Trotter steps are needed, compared to the case of controlled time evolution, with exactly the same final result \cite{Kivlichan_2020,Wecker:2015fib}. The number of rotations involved in each Trotter step is also reduced by another factor of $1/2$ \cite{Wecker:2015fib}.

We can now provide more explicit details on the construction of the appropriate circuit. The only gates that depend explicitly on the time step $dt$ of the Trotter approximation are the $Z$-rotations. To perform the above "directionally-controlled" time evolution, the sign of the angle in these $Z$-rotations should be determined by the state of the ancilla qubit. This can be done by adding a CNOT gate before and after each rotation, controlled on the ancilla qubit, and targeted on the rotated qubit (this is where there is an advantage compared to controlled-rotation, for which another $Z$-rotation is needed).

To obtain an explicit gate count, we can consider the two approaches for time evolution discussed in Sections \ref{subsec:evo_with_Pauli_strings} and \ref{sec:evo_with_mcp}. In the approach of Section \ref{subsec:evo_with_Pauli_strings} we have about $2^{n_q}$ CNOTs and $2^{n_q}$ $Z$-rotations per Trotter step (with additional small contribution from the kinetic part, and negligible overhead for the second-order Trotter approximation). To obtain a "directionally-controlled" time evolution we simply need to add $2$ CNOTs around each $Z$-rotation, as discussed above. This leads to about $3\cdot2^{n_q}$ CNOTs and basically no change in the number of $Z$-rotations. For $n_q=9$, and considering also the contribution from the kinetic part, $529$ $Z$-rotations and $528$ CNOTs become $530$ $Z$-rotations and $528+2\times 529=1586$ CNOTs per "directionally-controlled" Trotter step (where $1$ $Z$-rotation is added to account for the all-I Pauli string in the "directionally-controlled" time evolution, which is no longer a global phase).

In the approach of Section \ref{sec:evo_with_mcp}, $n_q-2$ ancilla qubits are used to obtain a shorter circuit with linear dependence on the number of qubits. The number of $Z$-rotations in applying $\exp(-iV dt)$ is fixed to $3$, independent of the number of qubits. This is important, because now we only need to add $6$ CNOTs to transform it to "directionally-controlled" time evolution. For the kinetic term with $n_q=9$, we need $18$ CNOTs and $18$ $Z$-rotations to implement $\exp(-iT dt)$. This becomes $54$ CNOTs and $19$ $Z$-rotations for the "directionally-controlled" $\exp(-iT dt)$ (where, again, $1$ $Z$-rotation is added to account for the all-I Pauli string). In total, for $n_q=9$ and considering two-qubit and non-Clifford one-qubit gates, we have $83$ CNOTs, $7$ conditioned $cZ$, $28$ $T$ gates, and $22$ $Z$ rotations per Trotter step "directionally controlled". We see here the significant advantage over the previous approach, where $530$ $Z$-rotations and $1586$ CNOTs are needed.

\begin{figure} \begin{center}
\includegraphics[width=\linewidth]{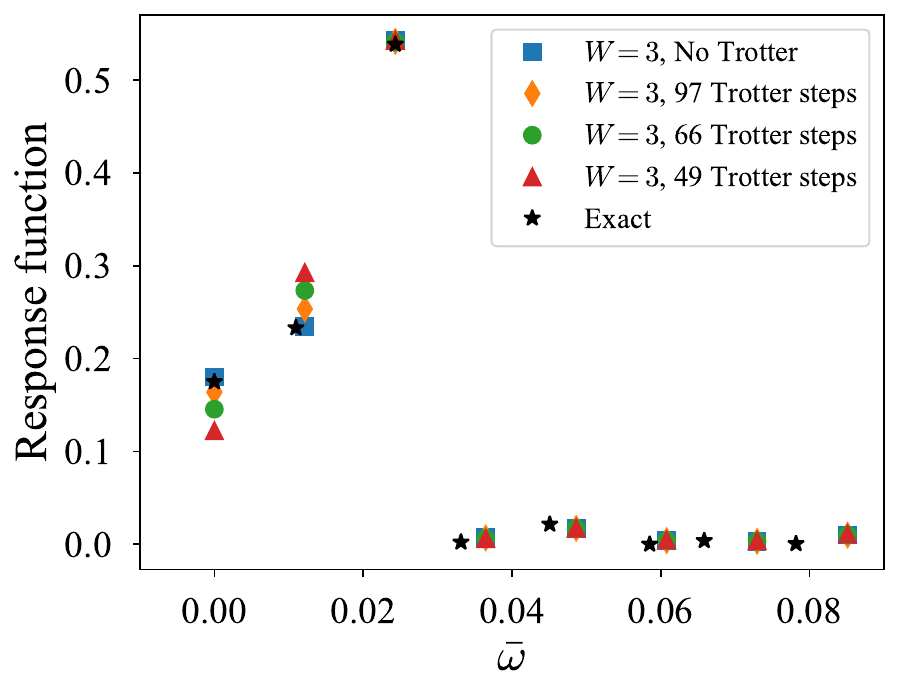}
\caption{\label{fig:response_Trotter}
Similar to Fig. \ref{fig:response_exact}, but including also calculations with Trotter approximation using the "directionally-controlled" time evolution formulation of QPE. The black stars correspond to the exact response function. The blue squares represent $W=3$ calculations with $\alpha>1$ assuming exact implementation, while orange diamonds, green circles and red triangles include Trotter approximation for the time evolution, with different number of steps using the second order Trotter formula ('T+V+T'). Exact ground state and exact application of the transition operator $\hat{O}$ are used here. }
\end{center}\end{figure}

To get a full gate count, we need to estimate the number of required Trotter steps.
The results of a calculation with Trotter approximation are shown in Fig. \ref{fig:response_Trotter}. We focus on the $W=3$ case. We can see that calculations with about $70$ Trotter steps lead to reasonable agreement with the exact $W=3$ results, describing well the width of the QE peak of the response. In that case, we can provide the gate count for the total circuit required for performing the relevant "directionally-controlled" time evolutions using the approach of Section \ref{sec:evo_with_mcp} with $n_q=9$. For $70$ Trotter steps, it includes about $5800$ CNOTs, $500$ conditioned $CZ$, $2000$ $T$ gates, and $1500$ $Z$-rotations. As the gate count is of the order of $10^3-10^4$ gates, it could be possible to run such a circuit on current or near-future hardware. 

We note that the energy resolution of these calculations is $15$ MeV ($\Delta {\cal H} = 1232$ MeV).
For heavier nuclei and larger momentum transfer, a coarser energy grid should be sufficient to describe the QE peak, which can reduce the total circuit depth. The calculations presented in this section were done assuming exact ground state preparation, but we note that similar results are obtained with approximated states.

Different improvements could possibly allow us to obtain a shorter circuit. Using symmetries like parity could further decrease the size of the relevant Hilbert space and thus reduce the number of qubits and possibly the circuit depth. Specific finite-range potentials, perhaps combined with coordinate-space mapping instead of momentum-space mapping, might reduce the number of Pauli strings in the Hamiltonian or the number of required Trotter steps. Other available algorithms for the calculation of response functions or time propagation could also lead to improvements.

\section{Noise}
\label{sec:noise}

\begin{figure} \begin{center}  
    \begin{tikzpicture}
        \node[anchor=south west,inner sep=0] (image) at (0,0) {\includegraphics[width=\linewidth]
{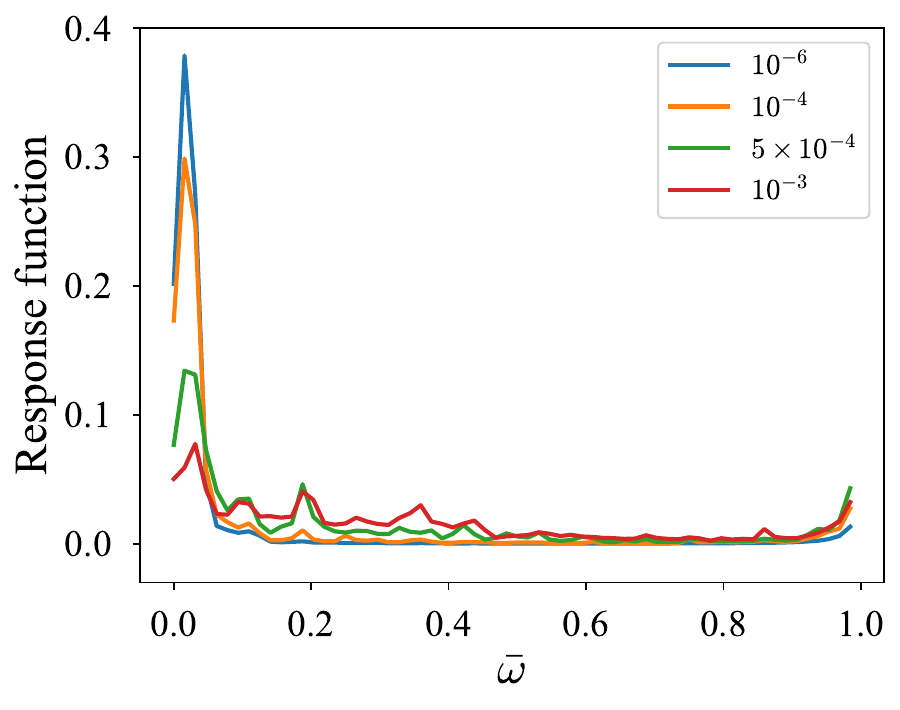}};
        \begin{scope}[x={(image.south east)},y={(image.north west)}]
            \node[anchor=south west,inner sep=0] (image) at (0.25,0.47) {\includegraphics[width=4.0cm]{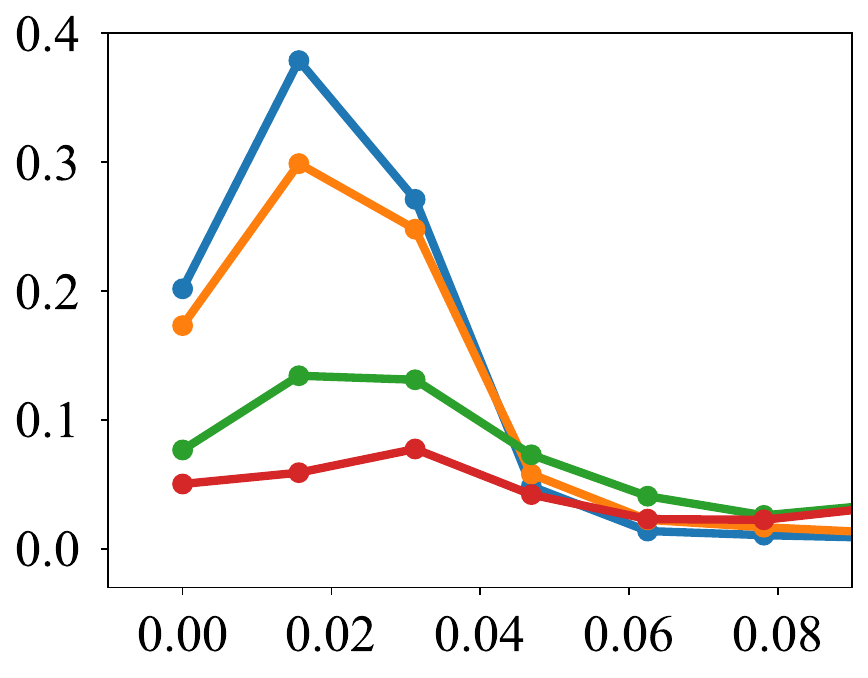}};
            \end{scope}
    \end{tikzpicture}
    \caption{\label{fig:noise_res_v0} 
    Results for the response function including depolarizing error channel for each of the two qubit gates in the circuit, and using $W=6$, $\alpha=1$, $\beta=0$ and second-order Trotter formula ('T+V+T'). Depolarizing error values are reported in the legend.
    The inset shows the same results with a focus around the QE peak.
}
\end{center} \end{figure}

In this section, we present an analysis performed on the quantum circuit through both statevector and noisy simulations. The circuit implements quantum phase estimation using second-order Trotter method to execute directionally-controlled time evolutions. It comprises a total of 22 qubits, nine are allocated for the target state, seven are used as auxiliary qubits for time evolution, and six are used as ancillary qubits for QPE. The total gate count of such circuit is $6\times 10^{3}$ CNOTs and $3.8\times 10^{3}$ Z-rotations, corresponding to a total of $63$ Trotter steps. We initialize the target register to the classically computed exact ground state using the IBM Qiskit~\cite{Qiskit} initialize function. The function implements the state preparation algorithm developed in Ref.~\cite{Shende2006}, resulting in a circuit having $502$ CNOTs and $1536$ Z-rotations.
To assess the impact of errors of two qubit gates (which have the largest errors in current hardware) we introduce a depolarizing error channel for each of the two-qubit gates (CNOT) within the circuit. The intensity of the error is systematically varied to evaluate its effects on the quality of the response function, utilizing the IBM Qiskit error modeling framework. For each value of the depolarizing error channel, we performed one hundred simulations to compute the average and variance estimators for the corresponding results. These simulations were conducted using the U.S. OLCF Andes HPC cluster, allocating one node per depolarizing error value. Each node's thousand simulations were distributed across its 32 cores.  The results are shown in Fig.~\ref{fig:noise_res_v0}. 
We see that depolarizing error of $10^{-4}$ has a only a small impact on the results of the calculations.
A marked degradation of the quality of the response function is observed for larger depolarizing error values, although the location of the peak is unchanged.
The signal vanishes when the depolarizing error is above $10^{-3}$.

\section{Summary}
\label{sec:summary}

To summarize, we have studied here the description of reactions, and specifically response functions relevant for nuclear systems, using quantum algorithms on a lattice. We focused on a qubit-efficient mapping due to its relevance for near-future devices.  

Focusing on the contact-interaction model, we studied this approach in detail, investigating the optimal integration of such a mapping with quantum algorithms relevant for reaction dynamics. We have constructed efficient circuits for the time evolution of the system using the Trotter approximation, showing that the large entanglement between the qubits in the qubit-efficient mapping does not necessarily translate to deep circuits. For the two-body case, we considered two different methods in momentum-space representation. For the first method, the time evolution is based on the evolution of each of the Pauli strings of the Hamiltonian, together with significant cancellation of CNOT gates. The gate complexity scales linearly with the lattice size. Second, utilizing the specific structure of the interaction and additional ancilla qubits, we presented circuit construction based on a multi-controlled gate implementation. The gate complexity of this method is significantly better, scaling only logarithmically with the system size. The number of ancilla qubits also scales logarithmically with the lattice size. Specifically, the number of arbitrary $Z$-rotations is significantly reduced in this approach, important for fault-tolerant implementations. The number of CNOT gates, which are the main source of physical noise in most of current hardware, is much smaller as well. We have presented an extension of the latter approach to the many-body case using coordinate-space first-quantization mapping.

For ground state preparation, we have investigated two possible methods, involving either an energy filter approach or a direct state initialization assuming the ground-state wave function is known (which can be the case if we consider a small number of particles). In the first approach, we can utilize the efficient time evolution with the Hamiltonian, but it requires multiple Trotter steps. The second approach requires no Trotter approximation, but its scaling with the lattice size is less favorable. The choice between these two options depends on the lattice size.

Finally, we have focused on the calculation of inclusive response functions using a QPE-based algorithm with modifications relevant for quasi-elastic reactions. Studying the number of Trotter steps required and utilizing the most efficient time evolution method, we have concluded that a circuit with an order of $10^3-10^4$ gates is required to obtain a reliable calculation for the two-body case. Therefore, this problem is a good candidate for implementation on current or near-future hardware. We have performed an end-to-end numerical calculation of the response function, including noise simulation. We also discussed possible approaches to further reduce the circuit depth.

Future studies should extend this work to heavier systems and other interaction models. Finite range interactions and spin-isospin dependence (for nuclear systems) should be considered, as well as three-body forces. We plan to study different approaches for anti-symmetrization as discussed before. 
Additional mappings can also be considered, studying the different relevant trade-offs.
Other available algorithms for inclusive response functions should also be considered, together with developing algorithms for exclusive reactions, where, e.g., a particle is knocked out of the system and its momentum is measured.

Making such progress, alongside advancements in quantum hardware, will allow us to study the possibility of performing exclusive response function calculations on quantum hardware. Such calculations are highly needed for analyzing, e.g., lepton-nucleus reactions, which are relevant for various experimental efforts but are not possible on classical computers for few-body systems. Specifically, the study of qubit-efficient approaches is essential for utilizing expected quantum devices with a few tens of logical qubits for this purpose.

\section*{Acknowledgments}

We are thankful for useful discussions with the QuEra team and with Alessandro Roggero.
This work was carried out under the auspices of the National Nuclear Security Administration of the U.S. Department of Energy at Los Alamos National Laboratory under Contract No. 89233218CNA000001, and Oak Ridge Leadership Computing Facility at the Oak Ridge National Laboratory, which is supported by the office of Science of the U. S. Department of Energy under contract No. DE-AC05-00OR22725. IS and JC gratefully acknowledge support by the Los Alamos Information Science and Technology Institute Rapid Response program and by the Advanced Simulation and Computing (ASC) program, under Physics and Engineering Models and Beyond Moore’s Law projects. AB’s work is supported by the U.S. Department of Energy, Office of Science, Nuclear Physics Quantum Horizons initiative, as is part of JC's work. This work was partially funded by the U. S. Department of Energy, Office of Science, Advanced Scientific Computing Program Office under FWP ERKJ382. 
 JC and AB also acknowledge the Quantum Science Center
 for partial support of their work on this project.
 The work of RW was supported by the Laboratory Directed Research and Development program of Los Alamos National Laboratory under project number 20210763PRD1. RW also acknowledges support by the Edwin Thompson Jaynes Postdoctoral Fellowship of the Washington University Physics Department. \\

 The data produced for this paper are available upon request. Please find the most up-to-date contact information here \cite{DataAvailability}.

\appendix

\section{Details on anti-symmetrization}
\label{sec:app_anti_symmetrization}

Following Ref. \cite{PhysRevResearch.4.023154}, fermionic anti-symmetrization can be enforced within a qubit-efficient mapping. In the case of nuclear systems, we can build Slater determinants for each specie (neutrons or protons) using single-particle states, similar to a configuration interaction in quantum chemistry or shell-model approach in nuclear physics. The many-body states will be product of neutron and proton Slater determinants. Working with momentum basis states, each single-particle state consists of the momentum and spin of the particle. There are $2N$ such single-particle states, where $N$ is the number of lattice sites. For $Z$ protons, there are $2N \choose Z$ possible proton Slater determinants. Similarly, there are $2N \choose A-Z$ possible neutron Slater determinants. The total number of antisymmetric many-body basis states is ${2N \choose Z}{2N \choose A-Z}$, and, thus, the number of required qubits is $\lceil \log_2{\left({2N \choose Z}{2N \choose A-Z}\right)} \rceil$. It can be further reduced, if we only keep many-body states that have zero total CM momentum. The mapping to qubits is defined by mapping each of the many-body basis states to a different bit-string (with number of digits as the number of qubits). The Hamiltonian is mapped to Pauli strings following Eqs. \eqref{eq:H_with_basis} and \eqref{eq:mapping}. Alternatively, one can use the first quantization mapping and follow available algorithms for antisymmetrization \cite{AbramsLloyd1997,Berry2018a}.

\section{Order of basis states}
\label{sec:app_order_states}

In the qubit-efficient mapping discussed in this work, each basis state is mapped to a bit-string of $n_q$ digits. We order the basis states in a specific order and map each state to the corresponding string, organized in binary order. The order of the basis states can impact the resulting number of Pauli strings in the Hamiltonian. 

In this work, we organize the momentum-space two-body basis states $|\bs{k},-\bs{k}\rangle$ in the following way. First, the allowed $k_x$ values are organized in this order: 
$\{0, \frac{2\pi}{L_x},2\frac{2\pi}{L_x},...,(\frac{N_x}{2}-1) \frac{2\pi}{L_x}, -\frac{N_x}{2} \frac{2\pi}{L_x} ,-(\frac{N_x}{2}-1)\frac{2\pi}{L_x},...,-\frac{2\pi}{L_x}\}$, and similarly for $k_y$ and $k_z$. Then, the three-dimensional vectors $|\bs{k}\rangle$ are organized in the following order. We choose the first value of $k_x$, $k_y$ and $k_z$, i.e., the zero-momentum state. Then, the values of $k_x$ and $k_y$ are fixed, and the value of $k_z$ is changed according to the above order. Once we go aver all values of $k_z$, the value of $k_y$ is changed to the next one, and we go again over all $k_z$ values. Once we go over all $k_y$ values in this way, we change the value of $k_x$ to the next one, and go again over all values of $k_y$ and $k_z$ in this way, etc. This defines the order of the basis states $|\bs{k},-\bs{k}\rangle$ used in this work.

\section{Energy filter - technical details}
\label{sec:app_energy_filter}

To apply the algorithm discussed in Section \ref{subsec:energy_filter} on a quantum computer, a Trotter approximation can be used for the time evolution operator $\exp[-i({\cal H}-E_0 I_N) t_\Delta \otimes Y_a]$. Similarly to the discussion in Section \ref{sec:time_evo}, we can separate the operator $({\cal H}-E_0 I_N) \otimes Y_a$ into two operators: $(T-E_0 I_N)\otimes Y_a$ and $V \otimes Y_a$. Each of these two operators translates to a sum of commuting Pauli strings in the mapping we employ in this work. We can use the Trotter approximation, similar to Eqs. \eqref{eq:1st_trotter} and \eqref{eq:2nd_trotter}. 

To build an appropriate circuit, we can follow the same idea of the Gray-Code order discussed in Section \ref{subsec:evo_with_Pauli_strings} with slight modifications. The operator $V \otimes Y_a$ can be diagonalized, 
\be \label{eq:VY_diag}
V \otimes Y_a = \left[ H^{\otimes n_q}\otimes S_a H_{a} \right] \left[ D\otimes Z_a \right] \left[ H^{\otimes n_q} \otimes H_{a} S_a^\dagger \right],
\ee
where $D$ was defined in Eq. \eqref{eq:D_def}, the subscript $a$ denotes the ancillary qubit, and $S$ is the $S$ phase gate. Therefore, in a single Trotter step, the operator $\exp[-i V\otimes Y_a \delta t]$, can be implemented with the appropriate diagonalizing gates at the beginning and at the end of the circuit, while the time propagation involves the operator $D$, mapped to all Pauli strings with $I$ and $Z$ matrices, and a fixed $Z$ for the ancillary qubit. Using the Gray-Code order for the Pauli strings and considering the cancellations of CNOTs we obtain a circuit with $512$ CNOTs and $512$ Z-rotations (including the all-$I$ string). For the $(T-E_0 I_N)\otimes Y_a$ operator, only $Y_a$ should be diagonalized and an optimized circuit involves $24$ CNOTs and 19 $Z$-rotations. Therefore, if we use first-order approximation, Eq. \eqref{eq:1st_trotter}, each Trotter step involves $536$ CNOTs and $531$ $Z$-rotations. For $40$ step, this is $21440$ CNOTs and $21240$ $Z$-rotations. Second-order approximations, Eq. \eqref{eq:2nd_trotter}, results in only $24$ more CNOTs and 19 more $Z$-rotations for the entire time evolution.

Alternatively, we can also use the ideas of Section \ref{sec:evo_with_mcp} to obtain a shorter circuit with the help of ancilla qubits. The time evolution with the potential term is given by
\be 
e^{-iV \otimes Y_a dt} = \left[ H^{\otimes n_q}\otimes S_a H_{a} \right]  e^{-iD\otimes Z_a dt} \left[ H^{\otimes n_q} \otimes H_{a} S_a^\dagger \right].
\ee
Due to the $Z_a$ in the exponent, if the ancilla qubit is in state $\ket{0}$, we should apply $e^{-iD dt}$, which can be done following the circuit in Eq. \eqref{eq:exp_V_with_mcp} (and the discussion around it). If the ancilla qubit is in state $\ket{1}$ we should apply $e^{+iD dt}$. This can be applied by following the same circuit construction, only replacing the angle $\theta$ with $-\theta$. 
The two cases can be dealt with by putting a pair of CNOTs around each explicit $\theta$-dependent rotation, where each CNOT is controlled on the ancilla qubit and targeted on the rotated qubit, leading to inverse rotation if the ancilla qubit is in state $\ket{1}$. Since there are only $3$ such rotations (Eq. \eqref{eq:cU_circuit}), this adds only $6$ CNOTs compared to the circuit for $\exp(-iV dt)$, with a total of $29$ CNOTs, $7$ conditioned $cZ$, $28$ $T$ gates, and $3$ $Z$-rotations for $n_q=9$ (together with additional single-qubit Clifford gates as discussed in Section \ref{sec:evo_with_mcp}, and the additional diagonilizing gates $S_a H_a$ on the ancilla qubit $a$). 
For the $(T-E_0 I_N)\otimes Y_a$ operator, we can follow the same approach as above involving $24$ CNOTs and 19 $Z$-rotations. In total, we have here $53$ CNOTs, $7$ conditioned $cZ$, $28$ $T$ gates, and $22$ $Z$-rotations per first-order Trotter step. For $40$ steps we get $2120$ CNOTS, $280$ conditioned $cZ$, $1120$ T gates, and $880$ $Z$-rotations.
As before, second-order Trotter approximation, Eq. \eqref{eq:2nd_trotter}, results in only $24$ more CNOTs and 19 more $Z$-rotations for the full time evolution.

\section{State initialization algorithm - technical details}
\label{sec:app_state_ini}

We provide here more technical details regarding the algorithm presented in Section \ref{subsec:initialization}.
As discussed, we assume to know the expansion of the desired state $|\Psi_0 \rangle$ using the basis states $\{|v_j\rangle \}$ and coefficients $\{g_j\}$ (Eq. \eqref{eq:gs_expan}).
The computer is initialized to the state $|\psi_i\rangle$, given in Eq. \eqref{eq:ini_state_bj}, with  {\it real} non-zero coefficients $\{b_j\}$.
In our mapping, this corresponds to a state that includes all $2^{n_q}$ bit-string combinations. If the computer is initialized to the state where all qubits are in the $|0\rangle$ state, $|\psi_i\rangle$ can be created, for example, by applying the Hadamard gate on each qubit, resulting in $b_j=1/\sqrt{2^{n_q}}$ for all $j$. Other states can also be created by applying a $Y$-rotation on each qubit, for example. 

Next, the operator $\exp[-i Q \otimes Y_a]$ is applied to the state $|\psi_i\rangle \otimes |0\rangle$, using an ancillary qubit $a$. $Q$ is defined in Eq. \eqref{eq:Q_op} as the diagonal Hermitian operator $Q = \sum_j d_j |v_j\rangle \langle v_j|$. It includes only Pauli strings with $I$ and $Z$ matrices when mapped to qubits, see Eq. \eqref{eq:mapping}.
If we measure the ancillary qubit in state $|0\rangle$, we obtain the state $|\psi_Q\rangle \equiv {\cal N}\cos(Q)|\psi_i \rangle$, where ${\cal N}$ is an appropriate normalization factor.
Since $Q$ is diagonal, this state can be written as
\be
|\psi_Q \rangle = {\cal N} \sum_j \cos(d_j) b_j |v_j \rangle.
\ee
We can now choose the values of $d_j$ such that 
\be \label{eq:d_values}
\cos(d_j) = \gamma \frac{|g_j|}{b_j},
\ee
for all $j$. $\gamma$ is a $j$-independent real and positive number. Notice that we must use the absolute value of $g_j$, because $\cos(d_j)$ is a real number. With this choice, we get ${\cal N} = 1/\gamma$, because $\sum_j |g_j|^2 =1$, and 
\be
|\psi_Q \rangle = \sum_j |g_j| |v_j \rangle.
\ee
This is very close to the desired state $|\Psi_0\rangle$. We are only missing the phases of the coefficients. Before accounting for the phases, notice that, in order to satisfy Eq. \eqref{eq:d_values}, we must have 
\be
\gamma \left|\frac{g_j}{b_j} \right| \leq 1
\ee
for all $j$.
The success probability, i.e., the probability of measuring the ancillary qubit in state $|0\rangle$ and obtaining the state $|\psi_Q\rangle$, is given by 
\be
P = \langle \psi_i | \cos^2(Q)|\psi_i \rangle = \sum_j cos^2(d_j) b_j^2 = \gamma^2.
\ee
Therefore, to maximize the success probability, we need to maximize the value of $\gamma$. The largest possible value of $\gamma$ is
\be
\gamma_{max} \equiv \min\left\{ \left| \frac{b_j}{g_j}\right| \right\}.
\ee
Therefore, when creating the state $|\psi_i\rangle$ with coefficients $\{b_j\}$, one should try to maximize the value of $\gamma_{max}$. Notice that $\gamma_{max}\leq 1$ because there must be at least one $j$ for which $|b_j|\leq|g_j|$ (because both $|\Psi_0\rangle$ and $|\psi_i\rangle$ are normalized to 1). If we choose to create $|\psi_i\rangle$ using the Hadamard gates, then we get $P \geq 1/2^{n_q}$ (because $|g_j|\leq 1$). 

Now, in order to obtain $|\Psi_0\rangle$ from $|\psi_Q\rangle$, we define the Hermitian diagonal operator
\be
\Theta = \sum_j \theta_j |v_j\rangle \langle v_j|, 
\ee
where $g_j=|g_j|\exp(i\theta_j)$. Applying the operator $\exp(i\Theta)$ on $|\psi_Q\rangle$, we obtain
\be
\exp(i\Theta) |\psi_Q\rangle
= \sum_j e^{i\theta_j} |g_j| |v_j\rangle = |\Psi_0\rangle.
\ee
Indeed, we obtained the state $|\Psi_0\rangle$. If all $\{g_j\}$ are real numbers, this step is not necessary (and $g_j$ should be used instead of $|g_j|$ in Eq. \eqref{eq:d_values}).
As discussed in Section \ref{subsec:initialization}, exact implementation of this algorithm requires at most $2^{n_q}$ CNOTs and $2^{n_q}$ $Z$-rotations if the coefficients $\{g_j\}$ are all real (similar to Eq. \eqref{eq:VY_diag} and the discussion around it). Generally, it requires at most 
at most $2^{n_q+1}-2$ CNOTs and $2^{n_q+1}-1$ $Z$-rotations in total. As also discussed in Section \ref{subsec:initialization}, a reduced number of gates is obtained if an approximate implementation is sufficient.

We can examine the gate count and success probability of this algorithm for our $9$-qubit two-body case. 
If the initial state $|\psi_i\rangle$ is created with Hadamard gates as described above, i.e. $b_j=1/\sqrt{2^{n_q}}$,
the success probability is guaranteed to be at least $1/2^9\approx 0.002$. For the two-body ground state, the success probability is actually $\approx 0.0026$, i.e. successful result once in about every $385$ tries in average. But, this can be significantly improved. 
Starting with the all-zero $9$-bit state, we apply single qubit $Y$-rotation gates on all qubits, with the same rotation angle $\theta$. As a result, all bit strings are created. The amplitude of each string is equal to $\cos^{9-p}(\theta/2) \sin^p(\theta/2)$, where $p$ is the number of $1$'s in the string.  $\theta$ can be chosen to maximize success probability. With the same mapping used in the sections above, we obtain $P\approx 0.03$, i.e. success about every 33 tries in average. In fact, this is much larger than $1/2^{n_q}$.
We can increase the success probability even more by using a slightly different mapping. 
We start by organizing the basis momentum states according to their total momentum magnitude, from smallest to largest. Then, they can be mapped to bit strings, such that the first state, i.e. the zero momentum state, is mapped to the all-zero $9$-bit string. The next 9 states are mapped to strings with a single $1$ digit and $8$ zeros. The following states are mapped to strings with two $1$'s, etc. 
Applying single qubit $Y$-rotations on the all-zero $9$-bit state, and optimizing the angle gives in this case $P\approx 0.14$, i.e. success every 7 tries in average. We do note that using this latter mapping results in more Pauli strings in the mapping of the kinetic energy. 

As mentioned above, following Ref. \cite{Welch_2014}, an approximated state can be created with a shorter quantum circuit. This is obtained by omitting Pauli strings with small coefficients in the operator $Q$. For example, if we use the last mapping described above, and omit all Pauli strings with coefficients smaller than $0.002$, we are left with $300$ strings out of the initial $512$ strings. Under this approximation, applying $\exp[-i Q \otimes Y_a]$ will involve only $300$ $Z$-rotations and a similar number of CNOTs \cite{Welch_2014}, compared to the $512$ $Z$-rotations and CNOTs for the exact case. Under this approximation, the energy of the state differs by only $10\%$ from  the exact ground state.

\section{Modified response-function algorithm}
\label{sec:app_modified_QPE}

In the modified response function algorithm discussed in Section \ref{sec:res_func}, the relevant operators are given in Eq. \eqref{eq:QPE_mod_op}. We consider here the case of $\beta=0$ for simplicity. The total "time" evolution, assuming $W_\alpha$ ancillary qubits, is given by
\be
t_{tot} =
2\pi\alpha \sum_{k=0}^{W_\alpha-1} 2^k = 2\pi \alpha(2^{W_\alpha}-1).
\ee
We can consider $\alpha=2^n$ for integer $n\geq 0$. The relevant energy domain in which the response function is evaluated is then $[0,1/2^n]$, compared to the range $[0,1]$ in the original algorithm, i.e., $\alpha=1$. Therefore, to obtain the same energy resolution as the case of $W_1$ ancillary qubits with $\alpha=1$, we can use $W_{2^n} = W_1-n$ anciallry qubits with $\alpha=2^n$ (because the number of energy grid points is $2^W$). This is the reduction in the number of ancillary qubits that we get in the modified approach (keeping the same resolution), if we know that the response function is negligible outside some energy range. Substituting it to the total time evolution, we get (for $\alpha=2^n$ and $W_{2^n} = W_1-n$)
\be
t_{tot} =
2\pi  (2^{W_1}-2^n) =
2\pi  (2^{W_1}-\alpha).
\ee
This can be compared to the total time evolution for the original case, i.e. for $\alpha=1$ and $W_1$ ancillary qubits
\be
t_{tot} = 2\pi(2^{W_1}-1).
\ee
We can see that we get a reduction in the total time (because $\alpha>1$), on top of the reduction in the number of ancillary qubits. However, notice the reduction in time by a factor smaller than $2$ (because $n\leq W_1 -1$, to keep $W_{2^n} \geq 1$).

\section{Implications of global phase in cU}
\label{sec:app_global_phase}.

We have provided an explicit circuit for the $cU$ gate in Eq. \eqref{eq:cU_circuit}. As mentioned, it comes with a global phase of $\theta/4$. $\theta$ appears in the definition of $U$, Eq. \eqref{eq:U_def}. This circuit for $cU$ enters into a larger circuit to implement $c^{n_q-1}U$. As a result, the global phase in $cU$ can become a relative phase in the circuit for $c^{n_q-1}U$, which should be corrected. Indeed, if we consider for example the gate $c(cU)$, $cU$ is applied on the first two qubits only if the third qubit is in state $\ket{1}$. Then, the global phase in $cU$ becomes an undesired relative phase, depending on the state of the third qubit. However, in the way we implement the multi-control $U$ gate, e.g. in Figs. \ref{fig:mcU_with_almost_Toffoli} and \ref{fig:c3U_mid_circuit}, the situation is different. We use ancilla qubits, and the gate $cU$, controlled on an ancilla, is applied regardless if this ancillary qubit is in state $\ket{0}$ or $\ket{1}$, using the circuit of Eq. \eqref{eq:cU_circuit} for $cU$. Therefore, the global phase is applied on all states and remains a global phase also for the final circuit of $c^{n_q-1}U$, as well as for the circuit of $\exp(-iV dt)$, Eq. \eqref{eq:exp_V_with_mcp}. 

This global phase has some implications for other circuits discussed in this work. When we follow the circuit in Eq. \eqref{eq:exp_V_with_mcp}, as discussed here, we eventually apply $\exp\left[-i(V dt-\theta/4)\right]$ due to the global phase. $\theta$ was defined as $\theta =  V_0 dt$. Thus, we apply
\be
\exp\left[-i\left(V-\frac{V_0}{4}\right) dt\right].
\ee
In other words, the global phase is equivalent to a shift of the potential term by $-V_0/{4}$. Since we only consider time evolution with the full Hamiltonian in this work, this shift of the potential can be corrected by shifting the kinetic part in the opposite direction
\be
T \to T + \frac{V_0}{4}.
\ee
These shifts have no impact on the accuracy of the Trotter approximations used in this work, as the constant terms used for shifting are commuting with all terms and can be freely moved between the potential and kinetic terms. It also does not change the gate count of the circuits in this work as the kinetic term already includes the all-$I$ string, and this shift only changes its numerical coefficient. 

This shift of $T$ is relevant for the application discussed in Section \ref{subsec:energy_filter}, where the time evolution operator $\exp[-i({\cal H}-E_0 I_N) t \otimes Y_a]$ is applied. The global phase becomes a relative phase due to the entanglement with the ancillary qubit $a$. The shift in T corrects it. 
This is also needed in Section \ref{sec:res_func}, for the "directionally-controlled" time evolutions in the application of QPE.

\bibliography{references}

\begin{thebibliography}{92}%
\makeatletter
\providecommand \@ifxundefined [1]{%
 \@ifx{#1\undefined}
}%
\providecommand \@ifnum [1]{%
 \ifnum #1\expandafter \@firstoftwo
 \else \expandafter \@secondoftwo
 \fi
}%
\providecommand \@ifx [1]{%
 \ifx #1\expandafter \@firstoftwo
 \else \expandafter \@secondoftwo
 \fi
}%
\providecommand \natexlab [1]{#1}%
\providecommand \enquote  [1]{``#1''}%
\providecommand \bibnamefont  [1]{#1}%
\providecommand \bibfnamefont [1]{#1}%
\providecommand \citenamefont [1]{#1}%
\providecommand \href@noop [0]{\@secondoftwo}%
\providecommand \href [0]{\begingroup \@sanitize@url \@href}%
\providecommand \@href[1]{\@@startlink{#1}\@@href}%
\providecommand \@@href[1]{\endgroup#1\@@endlink}%
\providecommand \@sanitize@url [0]{\catcode `\\12\catcode `\$12\catcode
  `\&12\catcode `\#12\catcode `\^12\catcode `\_12\catcode `\%12\relax}%
\providecommand \@@startlink[1]{}%
\providecommand \@@endlink[0]{}%
\providecommand \url  [0]{\begingroup\@sanitize@url \@url }%
\providecommand \@url [1]{\endgroup\@href {#1}{\urlprefix }}%
\providecommand \urlprefix  [0]{URL }%
\providecommand \Eprint [0]{\href }%
\providecommand \doibase [0]{https://doi.org/}%
\providecommand \selectlanguage [0]{\@gobble}%
\providecommand \bibinfo  [0]{\@secondoftwo}%
\providecommand \bibfield  [0]{\@secondoftwo}%
\providecommand \translation [1]{[#1]}%
\providecommand \BibitemOpen [0]{}%
\providecommand \bibitemStop [0]{}%
\providecommand \bibitemNoStop [0]{.\EOS\space}%
\providecommand \EOS [0]{\spacefactor3000\relax}%
\providecommand \BibitemShut  [1]{\csname bibitem#1\endcsname}%
\let\auto@bib@innerbib\@empty
\bibitem [{\citenamefont {Jordan}\ and\ \citenamefont {Wigner}(1928)}]{JW}%
  \BibitemOpen
  \bibfield  {author} {\bibinfo {author} {\bibfnamefont {P.}~\bibnamefont
  {Jordan}}\ and\ \bibinfo {author} {\bibfnamefont {E.}~\bibnamefont
  {Wigner}},\ }\bibfield  {title} {\bibinfo {title} {{\"U}ber das paulische
  {\"a}quivalenzverbot},\ }\href {https://doi.org/10.1007/BF01331938}
  {\bibfield  {journal} {\bibinfo  {journal} {Zeitschrift f{\"u}r Physik}\
  }\textbf {\bibinfo {volume} {47}},\ \bibinfo {pages} {631} (\bibinfo {year}
  {1928})}\BibitemShut {NoStop}%
\bibitem [{\citenamefont {Bravyi}\ and\ \citenamefont
  {Kitaev}(2002)}]{BRAVYI2002210}%
  \BibitemOpen
  \bibfield  {author} {\bibinfo {author} {\bibfnamefont {S.~B.}\ \bibnamefont
  {Bravyi}}\ and\ \bibinfo {author} {\bibfnamefont {A.~Y.}\ \bibnamefont
  {Kitaev}},\ }\bibfield  {title} {\bibinfo {title} {Fermionic quantum
  computation},\ }\href
  {https://doi.org/https://doi.org/10.1006/aphy.2002.6254} {\bibfield
  {journal} {\bibinfo  {journal} {Annals of Physics}\ }\textbf {\bibinfo
  {volume} {298}},\ \bibinfo {pages} {210} (\bibinfo {year}
  {2002})}\BibitemShut {NoStop}%
\bibitem [{\citenamefont {Tavernelli}\ \emph {et~al.}(2016)\citenamefont
  {Tavernelli}, \citenamefont {Staar}, \citenamefont {Fuhrer},\ and\
  \citenamefont {Moll}}]{Tavernelli:2016ckh}%
  \BibitemOpen
  \bibfield  {author} {\bibinfo {author} {\bibfnamefont {I.}~\bibnamefont
  {Tavernelli}}, \bibinfo {author} {\bibfnamefont {P.}~\bibnamefont {Staar}},
  \bibinfo {author} {\bibfnamefont {A.}~\bibnamefont {Fuhrer}},\ and\ \bibinfo
  {author} {\bibfnamefont {N.}~\bibnamefont {Moll}},\ }\bibfield  {title}
  {\bibinfo {title} {{Optimizing qubit resources for quantum chemistry
  simulations in second quantization on a quantum computer}},\ }\href
  {https://doi.org/10.1088/1751-8113/49/29/295301} {\bibfield  {journal}
  {\bibinfo  {journal} {J. Phys. A}\ }\textbf {\bibinfo {volume} {49}},\
  \bibinfo {pages} {295301} (\bibinfo {year} {2016})}\BibitemShut {NoStop}%
\bibitem [{\citenamefont {Bravyi}\ \emph {et~al.}(2017)\citenamefont {Bravyi},
  \citenamefont {Gambetta}, \citenamefont {Mezzacapo},\ and\ \citenamefont
  {Temme}}]{Bravyi:2017eoo}%
  \BibitemOpen
  \bibfield  {author} {\bibinfo {author} {\bibfnamefont {S.}~\bibnamefont
  {Bravyi}}, \bibinfo {author} {\bibfnamefont {J.~M.}\ \bibnamefont
  {Gambetta}}, \bibinfo {author} {\bibfnamefont {A.}~\bibnamefont
  {Mezzacapo}},\ and\ \bibinfo {author} {\bibfnamefont {K.}~\bibnamefont
  {Temme}},\ }\bibfield  {title} {\bibinfo {title} {{Tapering off qubits to
  simulate fermionic Hamiltonians}},\ }\href@noop {} {\  (\bibinfo {year}
  {2017})},\ \Eprint {https://arxiv.org/abs/1701.08213} {arXiv:1701.08213
  [quant-ph]} \BibitemShut {NoStop}%
\bibitem [{\citenamefont {Babbush}\ \emph {et~al.}(2017)\citenamefont
  {Babbush}, \citenamefont {Berry}, \citenamefont {Sanders}, \citenamefont
  {Kivlichan}, \citenamefont {Scherer}, \citenamefont {Wei}, \citenamefont
  {Love},\ and\ \citenamefont {Aspuru-Guzik}}]{Babbush:2017oum}%
  \BibitemOpen
  \bibfield  {author} {\bibinfo {author} {\bibfnamefont {R.}~\bibnamefont
  {Babbush}}, \bibinfo {author} {\bibfnamefont {D.~W.}\ \bibnamefont {Berry}},
  \bibinfo {author} {\bibfnamefont {Y.~R.}\ \bibnamefont {Sanders}}, \bibinfo
  {author} {\bibfnamefont {I.~D.}\ \bibnamefont {Kivlichan}}, \bibinfo {author}
  {\bibfnamefont {A.}~\bibnamefont {Scherer}}, \bibinfo {author} {\bibfnamefont
  {A.~Y.}\ \bibnamefont {Wei}}, \bibinfo {author} {\bibfnamefont {P.~J.}\
  \bibnamefont {Love}},\ and\ \bibinfo {author} {\bibfnamefont
  {A.}~\bibnamefont {Aspuru-Guzik}},\ }\bibfield  {title} {\bibinfo {title}
  {{Exponentially more precise quantum simulation of fermions in the
  configuration interaction representation}},\ }\href
  {https://doi.org/10.1088/2058-9565/aa9463} {\bibfield  {journal} {\bibinfo
  {journal} {Quantum Sci. Technol.}\ }\textbf {\bibinfo {volume} {3}},\
  \bibinfo {pages} {015006} (\bibinfo {year} {2017})}\BibitemShut {NoStop}%
\bibitem [{\citenamefont {Steudtner}\ and\ \citenamefont
  {Wehner}(2018)}]{Steudtner:2018ujo}%
  \BibitemOpen
  \bibfield  {author} {\bibinfo {author} {\bibfnamefont {M.}~\bibnamefont
  {Steudtner}}\ and\ \bibinfo {author} {\bibfnamefont {S.}~\bibnamefont
  {Wehner}},\ }\bibfield  {title} {\bibinfo {title} {{Fermion-to-qubit mappings
  with varying resource requirements for quantum simulation}},\ }\href
  {https://doi.org/10.1088/1367-2630/aac54f} {\bibfield  {journal} {\bibinfo
  {journal} {New J. Phys.}\ }\textbf {\bibinfo {volume} {20}},\ \bibinfo
  {pages} {063010} (\bibinfo {year} {2018})}\BibitemShut {NoStop}%
\bibitem [{\citenamefont {Di~Matteo}\ \emph {et~al.}(2021)\citenamefont
  {Di~Matteo}, \citenamefont {McCoy}, \citenamefont {Gysbers}, \citenamefont
  {Miyagi}, \citenamefont {Woloshyn},\ and\ \citenamefont
  {Navr\'atil}}]{DiMatteo:2020dhe}%
  \BibitemOpen
  \bibfield  {author} {\bibinfo {author} {\bibfnamefont {O.}~\bibnamefont
  {Di~Matteo}}, \bibinfo {author} {\bibfnamefont {A.}~\bibnamefont {McCoy}},
  \bibinfo {author} {\bibfnamefont {P.}~\bibnamefont {Gysbers}}, \bibinfo
  {author} {\bibfnamefont {T.}~\bibnamefont {Miyagi}}, \bibinfo {author}
  {\bibfnamefont {R.~M.}\ \bibnamefont {Woloshyn}},\ and\ \bibinfo {author}
  {\bibfnamefont {P.}~\bibnamefont {Navr\'atil}},\ }\bibfield  {title}
  {\bibinfo {title} {{Improving Hamiltonian encodings with the Gray code}},\
  }\href {https://doi.org/10.1103/PhysRevA.103.042405} {\bibfield  {journal}
  {\bibinfo  {journal} {Phys. Rev. A}\ }\textbf {\bibinfo {volume} {103}},\
  \bibinfo {pages} {042405} (\bibinfo {year} {2021})},\ \Eprint
  {https://arxiv.org/abs/2008.05012} {arXiv:2008.05012 [quant-ph]} \BibitemShut
  {NoStop}%
\bibitem [{\citenamefont {Kirby}\ \emph {et~al.}(2022)\citenamefont {Kirby},
  \citenamefont {Fuller}, \citenamefont {Hadfield},\ and\ \citenamefont
  {Mezzacapo}}]{Kirby:2021vkt}%
  \BibitemOpen
  \bibfield  {author} {\bibinfo {author} {\bibfnamefont {W.}~\bibnamefont
  {Kirby}}, \bibinfo {author} {\bibfnamefont {B.}~\bibnamefont {Fuller}},
  \bibinfo {author} {\bibfnamefont {C.}~\bibnamefont {Hadfield}},\ and\
  \bibinfo {author} {\bibfnamefont {A.}~\bibnamefont {Mezzacapo}},\ }\bibfield
  {title} {\bibinfo {title} {{Second-Quantized Fermionic Operators with
  Polylogarithmic Qubit and Gate Complexity}},\ }\href
  {https://doi.org/10.1103/PRXQuantum.3.020351} {\bibfield  {journal} {\bibinfo
   {journal} {PRX Quantum}\ }\textbf {\bibinfo {volume} {3}},\ \bibinfo {pages}
  {020351} (\bibinfo {year} {2022})},\ \Eprint
  {https://arxiv.org/abs/2109.14465} {arXiv:2109.14465 [quant-ph]} \BibitemShut
  {NoStop}%
\bibitem [{\citenamefont {Shee}\ \emph {et~al.}(2022)\citenamefont {Shee},
  \citenamefont {Tsai}, \citenamefont {Hong}, \citenamefont {Cheng},\ and\
  \citenamefont {Goan}}]{PhysRevResearch.4.023154}%
  \BibitemOpen
  \bibfield  {author} {\bibinfo {author} {\bibfnamefont {Y.}~\bibnamefont
  {Shee}}, \bibinfo {author} {\bibfnamefont {P.-K.}\ \bibnamefont {Tsai}},
  \bibinfo {author} {\bibfnamefont {C.-L.}\ \bibnamefont {Hong}}, \bibinfo
  {author} {\bibfnamefont {H.-C.}\ \bibnamefont {Cheng}},\ and\ \bibinfo
  {author} {\bibfnamefont {H.-S.}\ \bibnamefont {Goan}},\ }\bibfield  {title}
  {\bibinfo {title} {Qubit-efficient encoding scheme for quantum simulations of
  electronic structure},\ }\href
  {https://doi.org/10.1103/PhysRevResearch.4.023154} {\bibfield  {journal}
  {\bibinfo  {journal} {Phys. Rev. Res.}\ }\textbf {\bibinfo {volume} {4}},\
  \bibinfo {pages} {023154} (\bibinfo {year} {2022})}\BibitemShut {NoStop}%
\bibitem [{\citenamefont {{Kandala}}\ \emph {et~al.}(2017)\citenamefont
  {{Kandala}}, \citenamefont {{Mezzacapo}}, \citenamefont {{Temme}},
  \citenamefont {{Takita}}, \citenamefont {{Brink}}, \citenamefont {{Chow}},\
  and\ \citenamefont {{Gambetta}}}]{Kandala:2017HE}%
  \BibitemOpen
  \bibfield  {author} {\bibinfo {author} {\bibfnamefont {A.}~\bibnamefont
  {{Kandala}}}, \bibinfo {author} {\bibfnamefont {A.}~\bibnamefont
  {{Mezzacapo}}}, \bibinfo {author} {\bibfnamefont {K.}~\bibnamefont
  {{Temme}}}, \bibinfo {author} {\bibfnamefont {M.}~\bibnamefont {{Takita}}},
  \bibinfo {author} {\bibfnamefont {M.}~\bibnamefont {{Brink}}}, \bibinfo
  {author} {\bibfnamefont {J.~M.}\ \bibnamefont {{Chow}}},\ and\ \bibinfo
  {author} {\bibfnamefont {J.~M.}\ \bibnamefont {{Gambetta}}},\ }\bibfield
  {title} {\bibinfo {title} {{Hardware-efficient variational quantum
  eigensolver for small molecules and quantum magnets}},\ }\href
  {https://doi.org/10.1038/nature23879} {\bibfield  {journal} {\bibinfo
  {journal} {\nat}\ }\textbf {\bibinfo {volume} {549}},\ \bibinfo {pages} {242}
  (\bibinfo {year} {2017})},\ \Eprint {https://arxiv.org/abs/1704.05018}
  {arXiv:1704.05018 [quant-ph]} \BibitemShut {NoStop}%
\bibitem [{\citenamefont {Tilly}\ \emph {et~al.}(2022)\citenamefont {Tilly}
  \emph {et~al.}}]{Tilly:2021}%
  \BibitemOpen
  \bibfield  {author} {\bibinfo {author} {\bibfnamefont {J.}~\bibnamefont
  {Tilly}} \emph {et~al.},\ }\bibfield  {title} {\bibinfo {title} {{The
  Variational Quantum Eigensolver: A review of methods and best practices}},\
  }\href {https://doi.org/10.1016/j.physrep.2022.08.003} {\bibfield  {journal}
  {\bibinfo  {journal} {Phys. Rept.}\ }\textbf {\bibinfo {volume} {986}},\
  \bibinfo {pages} {1} (\bibinfo {year} {2022})},\ \Eprint
  {https://arxiv.org/abs/2111.05176} {arXiv:2111.05176 [quant-ph]} \BibitemShut
  {NoStop}%
\bibitem [{\citenamefont {{Farhi}}\ \emph {et~al.}(2000)\citenamefont
  {{Farhi}}, \citenamefont {{Goldstone}}, \citenamefont {{Gutmann}},\ and\
  \citenamefont {{Sipser}}}]{Farhi2000}%
  \BibitemOpen
  \bibfield  {author} {\bibinfo {author} {\bibfnamefont {E.}~\bibnamefont
  {{Farhi}}}, \bibinfo {author} {\bibfnamefont {J.}~\bibnamefont
  {{Goldstone}}}, \bibinfo {author} {\bibfnamefont {S.}~\bibnamefont
  {{Gutmann}}},\ and\ \bibinfo {author} {\bibfnamefont {M.}~\bibnamefont
  {{Sipser}}},\ }\bibfield  {title} {\bibinfo {title} {{Quantum Computation by
  Adiabatic Evolution}},\ }\href@noop {} {\bibfield  {journal} {\bibinfo
  {journal} {arXiv e-prints}\ ,\ \bibinfo {eid} {quant-ph/0001106}} (\bibinfo
  {year} {2000})},\ \Eprint {https://arxiv.org/abs/quant-ph/0001106}
  {arXiv:quant-ph/0001106 [quant-ph]} \BibitemShut {NoStop}%
\bibitem [{\citenamefont {Albash}\ and\ \citenamefont
  {Lidar}(2018)}]{Albash2018}%
  \BibitemOpen
  \bibfield  {author} {\bibinfo {author} {\bibfnamefont {T.}~\bibnamefont
  {Albash}}\ and\ \bibinfo {author} {\bibfnamefont {D.~A.}\ \bibnamefont
  {Lidar}},\ }\bibfield  {title} {\bibinfo {title} {Adiabatic quantum
  computation},\ }\href {https://doi.org/10.1103/RevModPhys.90.015002}
  {\bibfield  {journal} {\bibinfo  {journal} {Rev. Mod. Phys.}\ }\textbf
  {\bibinfo {volume} {90}},\ \bibinfo {pages} {015002} (\bibinfo {year}
  {2018})}\BibitemShut {NoStop}%
\bibitem [{\citenamefont {{Motta}}\ \emph {et~al.}(2020)\citenamefont
  {{Motta}}, \citenamefont {{Sun}}, \citenamefont {{Tan}}, \citenamefont
  {{O'Rourke}}, \citenamefont {{Ye}}, \citenamefont {{Minnich}}, \citenamefont
  {{Brand{\~a}o}},\ and\ \citenamefont {{Chan}}}]{Motta:2020}%
  \BibitemOpen
  \bibfield  {author} {\bibinfo {author} {\bibfnamefont {M.}~\bibnamefont
  {{Motta}}}, \bibinfo {author} {\bibfnamefont {C.}~\bibnamefont {{Sun}}},
  \bibinfo {author} {\bibfnamefont {A.~T.~K.}\ \bibnamefont {{Tan}}}, \bibinfo
  {author} {\bibfnamefont {M.~J.}\ \bibnamefont {{O'Rourke}}}, \bibinfo
  {author} {\bibfnamefont {E.}~\bibnamefont {{Ye}}}, \bibinfo {author}
  {\bibfnamefont {A.~J.}\ \bibnamefont {{Minnich}}}, \bibinfo {author}
  {\bibfnamefont {F.~G.~S.~L.}\ \bibnamefont {{Brand{\~a}o}}},\ and\ \bibinfo
  {author} {\bibfnamefont {G.~K.-L.}\ \bibnamefont {{Chan}}},\ }\bibfield
  {title} {\bibinfo {title} {{Determining eigenstates and thermal states on a
  quantum computer using quantum imaginary time evolution}},\ }\href
  {https://doi.org/10.1038/s41567-019-0704-4} {\bibfield  {journal} {\bibinfo
  {journal} {Nature Physics}\ }\textbf {\bibinfo {volume} {16}},\ \bibinfo
  {pages} {205} (\bibinfo {year} {2020})},\ \Eprint
  {https://arxiv.org/abs/1901.07653} {arXiv:1901.07653 [quant-ph]} \BibitemShut
  {NoStop}%
\bibitem [{\citenamefont {Kosugi}\ \emph {et~al.}(2022)\citenamefont {Kosugi},
  \citenamefont {Nishiya}, \citenamefont {Nishi},\ and\ \citenamefont
  {Matsushita}}]{PhysRevResearch.4.033121}%
  \BibitemOpen
  \bibfield  {author} {\bibinfo {author} {\bibfnamefont {T.}~\bibnamefont
  {Kosugi}}, \bibinfo {author} {\bibfnamefont {Y.}~\bibnamefont {Nishiya}},
  \bibinfo {author} {\bibfnamefont {H.}~\bibnamefont {Nishi}},\ and\ \bibinfo
  {author} {\bibfnamefont {Y.-i.}\ \bibnamefont {Matsushita}},\ }\bibfield
  {title} {\bibinfo {title} {Imaginary-time evolution using forward and
  backward real-time evolution with a single ancilla: First-quantized
  eigensolver algorithm for quantum chemistry},\ }\href
  {https://doi.org/10.1103/PhysRevResearch.4.033121} {\bibfield  {journal}
  {\bibinfo  {journal} {Phys. Rev. Research}\ }\textbf {\bibinfo {volume}
  {4}},\ \bibinfo {pages} {033121} (\bibinfo {year} {2022})}\BibitemShut
  {NoStop}%
\bibitem [{\citenamefont {Turro}\ \emph {et~al.}(2022)\citenamefont {Turro},
  \citenamefont {Roggero}, \citenamefont {Amitrano}, \citenamefont {Luchi},
  \citenamefont {Wendt}, \citenamefont {Dubois}, \citenamefont {Quaglioni},\
  and\ \citenamefont {Pederiva}}]{Turro2022}%
  \BibitemOpen
  \bibfield  {author} {\bibinfo {author} {\bibfnamefont {F.}~\bibnamefont
  {Turro}}, \bibinfo {author} {\bibfnamefont {A.}~\bibnamefont {Roggero}},
  \bibinfo {author} {\bibfnamefont {V.}~\bibnamefont {Amitrano}}, \bibinfo
  {author} {\bibfnamefont {P.}~\bibnamefont {Luchi}}, \bibinfo {author}
  {\bibfnamefont {K.~A.}\ \bibnamefont {Wendt}}, \bibinfo {author}
  {\bibfnamefont {J.~L.}\ \bibnamefont {Dubois}}, \bibinfo {author}
  {\bibfnamefont {S.}~\bibnamefont {Quaglioni}},\ and\ \bibinfo {author}
  {\bibfnamefont {F.}~\bibnamefont {Pederiva}},\ }\bibfield  {title} {\bibinfo
  {title} {{Imaginary-time propagation on a quantum chip}},\ }\href
  {https://doi.org/10.1103/PhysRevA.105.022440} {\bibfield  {journal} {\bibinfo
   {journal} {Physical Review A}\ }\textbf {\bibinfo {volume} {105}},\ \bibinfo
  {pages} {1} (\bibinfo {year} {2022})}\BibitemShut {NoStop}%
\bibitem [{\citenamefont {Jouzdani}\ \emph {et~al.}(2022)\citenamefont
  {Jouzdani}, \citenamefont {Johnson}, \citenamefont {Mucciolo},\ and\
  \citenamefont {Stetcu}}]{Jouzdani2022}%
  \BibitemOpen
  \bibfield  {author} {\bibinfo {author} {\bibfnamefont {P.}~\bibnamefont
  {Jouzdani}}, \bibinfo {author} {\bibfnamefont {C.~W.}\ \bibnamefont
  {Johnson}}, \bibinfo {author} {\bibfnamefont {E.~R.}\ \bibnamefont
  {Mucciolo}},\ and\ \bibinfo {author} {\bibfnamefont {I.}~\bibnamefont
  {Stetcu}},\ }\bibfield  {title} {\bibinfo {title} {Alternative approach to
  quantum imaginary time evolution},\ }\href
  {https://doi.org/10.1103/PhysRevA.106.062435} {\bibfield  {journal} {\bibinfo
   {journal} {Phys. Rev. A}\ }\textbf {\bibinfo {volume} {106}},\ \bibinfo
  {pages} {062435} (\bibinfo {year} {2022})}\BibitemShut {NoStop}%
\bibitem [{\citenamefont {Ge}\ \emph {et~al.}(2019)\citenamefont {Ge},
  \citenamefont {Tura},\ and\ \citenamefont {Cirac}}]{Ge2018}%
  \BibitemOpen
  \bibfield  {author} {\bibinfo {author} {\bibfnamefont {Y.}~\bibnamefont
  {Ge}}, \bibinfo {author} {\bibfnamefont {J.}~\bibnamefont {Tura}},\ and\
  \bibinfo {author} {\bibfnamefont {J.~I.}\ \bibnamefont {Cirac}},\ }\bibfield
  {title} {\bibinfo {title} {Faster ground state preparation and high-precision
  ground energy estimation with fewer qubits},\ }\href
  {https://doi.org/https://doi.org/10.1063/1.5027484} {\bibfield  {journal}
  {\bibinfo  {journal} {J. Math. Phys.}\ }\textbf {\bibinfo {volume} {60}},\
  \bibinfo {pages} {022202} (\bibinfo {year} {2019})}\BibitemShut {NoStop}%
\bibitem [{\citenamefont {Dong}\ \emph {et~al.}(2022)\citenamefont {Dong},
  \citenamefont {Lin},\ and\ \citenamefont {Tong}}]{Dong2022}%
  \BibitemOpen
  \bibfield  {author} {\bibinfo {author} {\bibfnamefont {Y.}~\bibnamefont
  {Dong}}, \bibinfo {author} {\bibfnamefont {L.}~\bibnamefont {Lin}},\ and\
  \bibinfo {author} {\bibfnamefont {Y.}~\bibnamefont {Tong}},\ }\bibfield
  {title} {\bibinfo {title} {Ground-state preparation and energy estimation on
  early fault-tolerant quantum computers via quantum eigenvalue transformation
  of unitary matrices},\ }\href {https://doi.org/10.1103/PRXQuantum.3.040305}
  {\bibfield  {journal} {\bibinfo  {journal} {PRX Quantum}\ }\textbf {\bibinfo
  {volume} {3}},\ \bibinfo {pages} {040305} (\bibinfo {year}
  {2022})}\BibitemShut {NoStop}%
\bibitem [{\citenamefont {Keen}\ \emph {et~al.}(2021)\citenamefont {Keen},
  \citenamefont {Dumitrescu},\ and\ \citenamefont {Wang}}]{Keen2021}%
  \BibitemOpen
  \bibfield  {author} {\bibinfo {author} {\bibfnamefont {T.}~\bibnamefont
  {Keen}}, \bibinfo {author} {\bibfnamefont {E.}~\bibnamefont {Dumitrescu}},\
  and\ \bibinfo {author} {\bibfnamefont {Y.}~\bibnamefont {Wang}},\ }\bibfield
  {title} {\bibinfo {title} {{Quantum Algorithms for Ground-State Preparation
  and Green's Function Calculation}},\ }\href@noop {} {\bibfield  {journal}
  {\bibinfo  {journal} {arXiv e-prints}\ ,\ \bibinfo {eid} {arXiv:2112.05731}}
  (\bibinfo {year} {2021})},\ \Eprint {https://arxiv.org/abs/2112.05731}
  {arXiv:2112.05731 [quant-ph]} \BibitemShut {NoStop}%
\bibitem [{\citenamefont {Choi}\ \emph {et~al.}(2021)\citenamefont {Choi},
  \citenamefont {Lee}, \citenamefont {Bonitati}, \citenamefont {Qian},\ and\
  \citenamefont {Watkins}}]{Choi2021}%
  \BibitemOpen
  \bibfield  {author} {\bibinfo {author} {\bibfnamefont {K.}~\bibnamefont
  {Choi}}, \bibinfo {author} {\bibfnamefont {D.}~\bibnamefont {Lee}}, \bibinfo
  {author} {\bibfnamefont {J.}~\bibnamefont {Bonitati}}, \bibinfo {author}
  {\bibfnamefont {Z.}~\bibnamefont {Qian}},\ and\ \bibinfo {author}
  {\bibfnamefont {J.}~\bibnamefont {Watkins}},\ }\bibfield  {title} {\bibinfo
  {title} {Rodeo algorithm for quantum computing},\ }\href
  {https://doi.org/10.1103/PhysRevLett.127.040505} {\bibfield  {journal}
  {\bibinfo  {journal} {Phys. Rev. Lett.}\ }\textbf {\bibinfo {volume} {127}},\
  \bibinfo {pages} {040505} (\bibinfo {year} {2021})}\BibitemShut {NoStop}%
\bibitem [{\citenamefont {Stetcu}\ \emph {et~al.}(2023)\citenamefont {Stetcu},
  \citenamefont {Baroni},\ and\ \citenamefont {Carlson}}]{Stetcu:2022nhy}%
  \BibitemOpen
  \bibfield  {author} {\bibinfo {author} {\bibfnamefont {I.}~\bibnamefont
  {Stetcu}}, \bibinfo {author} {\bibfnamefont {A.}~\bibnamefont {Baroni}},\
  and\ \bibinfo {author} {\bibfnamefont {J.}~\bibnamefont {Carlson}},\
  }\bibfield  {title} {\bibinfo {title} {{Projection algorithm for state
  preparation on quantum computers}},\ }\href
  {https://doi.org/10.1103/PhysRevC.108.L031306} {\bibfield  {journal}
  {\bibinfo  {journal} {Phys. Rev. C}\ }\textbf {\bibinfo {volume} {108}},\
  \bibinfo {pages} {L031306} (\bibinfo {year} {2023})},\ \Eprint
  {https://arxiv.org/abs/2211.10545} {arXiv:2211.10545 [quant-ph]} \BibitemShut
  {NoStop}%
\bibitem [{\citenamefont {Bergholm}\ \emph {et~al.}(2005)\citenamefont
  {Bergholm}, \citenamefont {Vartiainen}, \citenamefont {M\"ott\"onen},\ and\
  \citenamefont {Salomaa}}]{Bergholm:2005ibb}%
  \BibitemOpen
  \bibfield  {author} {\bibinfo {author} {\bibfnamefont {V.}~\bibnamefont
  {Bergholm}}, \bibinfo {author} {\bibfnamefont {J.~J.}\ \bibnamefont
  {Vartiainen}}, \bibinfo {author} {\bibfnamefont {M.}~\bibnamefont
  {M\"ott\"onen}},\ and\ \bibinfo {author} {\bibfnamefont {M.~M.}\ \bibnamefont
  {Salomaa}},\ }\bibfield  {title} {\bibinfo {title} {{Quantum circuits with
  uniformly controlled one-qubit gates}},\ }\href
  {https://doi.org/10.1103/PhysRevA.71.052330} {\bibfield  {journal} {\bibinfo
  {journal} {Phys. Rev. A}\ }\textbf {\bibinfo {volume} {71}},\ \bibinfo
  {pages} {052330} (\bibinfo {year} {2005})}\BibitemShut {NoStop}%
\bibitem [{\citenamefont {Plesch}\ and\ \citenamefont
  {Brukner}(2011)}]{Plesch:2011vwn}%
  \BibitemOpen
  \bibfield  {author} {\bibinfo {author} {\bibfnamefont {M.}~\bibnamefont
  {Plesch}}\ and\ \bibinfo {author} {\bibfnamefont {v.}~\bibnamefont
  {Brukner}},\ }\bibfield  {title} {\bibinfo {title} {{Quantum-state
  preparation with universal gate decompositions}},\ }\href
  {https://doi.org/10.1103/PhysRevA.83.032302} {\bibfield  {journal} {\bibinfo
  {journal} {Phys. Rev. A}\ }\textbf {\bibinfo {volume} {83}},\ \bibinfo
  {pages} {032302} (\bibinfo {year} {2011})}\BibitemShut {NoStop}%
\bibitem [{\citenamefont {Zhang}\ \emph
  {et~al.}(2021{\natexlab{a}})\citenamefont {Zhang}, \citenamefont {Yung},\
  and\ \citenamefont {Yuan}}]{Zhang:2021uwi}%
  \BibitemOpen
  \bibfield  {author} {\bibinfo {author} {\bibfnamefont {X.-M.}\ \bibnamefont
  {Zhang}}, \bibinfo {author} {\bibfnamefont {M.-H.}\ \bibnamefont {Yung}},\
  and\ \bibinfo {author} {\bibfnamefont {X.}~\bibnamefont {Yuan}},\ }\bibfield
  {title} {\bibinfo {title} {{Low-depth quantum state preparation}},\ }\href
  {https://doi.org/10.1103/PhysRevResearch.3.043200} {\bibfield  {journal}
  {\bibinfo  {journal} {Phys. Rev. Res.}\ }\textbf {\bibinfo {volume} {3}},\
  \bibinfo {pages} {043200} (\bibinfo {year} {2021}{\natexlab{a}})},\ \Eprint
  {https://arxiv.org/abs/2102.07533} {arXiv:2102.07533 [quant-ph]} \BibitemShut
  {NoStop}%
\bibitem [{\citenamefont {Zhang}\ \emph
  {et~al.}(2021{\natexlab{b}})\citenamefont {Zhang}, \citenamefont {Wang},\
  and\ \citenamefont {Ying}}]{Zhang:2021bue}%
  \BibitemOpen
  \bibfield  {author} {\bibinfo {author} {\bibfnamefont {Z.}~\bibnamefont
  {Zhang}}, \bibinfo {author} {\bibfnamefont {Q.}~\bibnamefont {Wang}},\ and\
  \bibinfo {author} {\bibfnamefont {M.}~\bibnamefont {Ying}},\ }\bibfield
  {title} {\bibinfo {title} {{Parallel Quantum Algorithm for Hamiltonian
  Simulation}},\ }\href@noop {} {\  (\bibinfo {year} {2021}{\natexlab{b}})},\
  \Eprint {https://arxiv.org/abs/2105.11889} {arXiv:2105.11889 [quant-ph]}
  \BibitemShut {NoStop}%
\bibitem [{\citenamefont {Rosenthal}(2021)}]{Rosenthal:2021rcb}%
  \BibitemOpen
  \bibfield  {author} {\bibinfo {author} {\bibfnamefont {G.}~\bibnamefont
  {Rosenthal}},\ }\bibfield  {title} {\bibinfo {title} {{Query and Depth Upper
  Bounds for Quantum Unitaries via Grover Search}},\ }\href@noop {} {\
  (\bibinfo {year} {2021})},\ \Eprint {https://arxiv.org/abs/2111.07992}
  {arXiv:2111.07992 [quant-ph]} \BibitemShut {NoStop}%
\bibitem [{\citenamefont {Zhang}\ \emph {et~al.}(2022)\citenamefont {Zhang},
  \citenamefont {Li},\ and\ \citenamefont {Yuan}}]{Zhang:2022pue}%
  \BibitemOpen
  \bibfield  {author} {\bibinfo {author} {\bibfnamefont {X.-M.}\ \bibnamefont
  {Zhang}}, \bibinfo {author} {\bibfnamefont {T.}~\bibnamefont {Li}},\ and\
  \bibinfo {author} {\bibfnamefont {X.}~\bibnamefont {Yuan}},\ }\bibfield
  {title} {\bibinfo {title} {{Quantum State Preparation with Optimal Circuit
  Depth: Implementations and Applications}},\ }\href
  {https://doi.org/10.1103/PhysRevLett.129.230504} {\bibfield  {journal}
  {\bibinfo  {journal} {Phys. Rev. Lett.}\ }\textbf {\bibinfo {volume} {129}},\
  \bibinfo {pages} {230504} (\bibinfo {year} {2022})},\ \Eprint
  {https://arxiv.org/abs/2201.11495} {arXiv:2201.11495 [quant-ph]} \BibitemShut
  {NoStop}%
\bibitem [{\citenamefont {Sun}\ \emph {et~al.}(2023)\citenamefont {Sun},
  \citenamefont {Tian}, \citenamefont {Yang}, \citenamefont {Yuan},\ and\
  \citenamefont {Zhang}}]{Sun:2023bwt}%
  \BibitemOpen
  \bibfield  {author} {\bibinfo {author} {\bibfnamefont {X.}~\bibnamefont
  {Sun}}, \bibinfo {author} {\bibfnamefont {G.}~\bibnamefont {Tian}}, \bibinfo
  {author} {\bibfnamefont {S.}~\bibnamefont {Yang}}, \bibinfo {author}
  {\bibfnamefont {P.}~\bibnamefont {Yuan}},\ and\ \bibinfo {author}
  {\bibfnamefont {S.}~\bibnamefont {Zhang}},\ }\bibfield  {title} {\bibinfo
  {title} {{Asymptotically Optimal Circuit Depth for Quantum State Preparation
  and General Unitary Synthesis}},\ }\href
  {https://doi.org/10.1109/TCAD.2023.3244885} {\bibfield  {journal} {\bibinfo
  {journal} {IEEE Trans. Comput. Aided Design Integr. Circuits Syst.}\ }\textbf
  {\bibinfo {volume} {42}},\ \bibinfo {pages} {3301} (\bibinfo {year}
  {2023})}\BibitemShut {NoStop}%
\bibitem [{\citenamefont {Terhal}\ and\ \citenamefont
  {DiVincenzo}(2000)}]{Terhal:1998yh}%
  \BibitemOpen
  \bibfield  {author} {\bibinfo {author} {\bibfnamefont {B.~M.}\ \bibnamefont
  {Terhal}}\ and\ \bibinfo {author} {\bibfnamefont {D.~P.}\ \bibnamefont
  {DiVincenzo}},\ }\bibfield  {title} {\bibinfo {title} {{On the problem of
  equilibration and the computation of correlation functions on a quantum
  computer}},\ }\href {https://doi.org/10.1103/PhysRevA.61.22301} {\bibfield
  {journal} {\bibinfo  {journal} {Phys. Rev. A}\ }\textbf {\bibinfo {volume}
  {61}},\ \bibinfo {pages} {22301} (\bibinfo {year} {2000})},\ \Eprint
  {https://arxiv.org/abs/quant-ph/9810063} {arXiv:quant-ph/9810063}
  \BibitemShut {NoStop}%
\bibitem [{\citenamefont {Lidar}\ and\ \citenamefont
  {Wang}(1999)}]{Lidar:1998mf}%
  \BibitemOpen
  \bibfield  {author} {\bibinfo {author} {\bibfnamefont {D.~A.}\ \bibnamefont
  {Lidar}}\ and\ \bibinfo {author} {\bibfnamefont {H.}~\bibnamefont {Wang}},\
  }\bibfield  {title} {\bibinfo {title} {{Calculating the thermal rate constant
  with exponential speed up on a quantum computer}},\ }\href
  {https://doi.org/10.1103/PhysRevE.59.2429} {\bibfield  {journal} {\bibinfo
  {journal} {Phys. Rev. E}\ }\textbf {\bibinfo {volume} {59}},\ \bibinfo
  {pages} {2429} (\bibinfo {year} {1999})},\ \Eprint
  {https://arxiv.org/abs/quant-ph/9807009} {arXiv:quant-ph/9807009}
  \BibitemShut {NoStop}%
\bibitem [{\citenamefont {Ortiz}\ \emph {et~al.}(2001)\citenamefont {Ortiz},
  \citenamefont {Gubernatis}, \citenamefont {Knill},\ and\ \citenamefont
  {Laflamme}}]{Ortiz:2000gc}%
  \BibitemOpen
  \bibfield  {author} {\bibinfo {author} {\bibfnamefont {G.}~\bibnamefont
  {Ortiz}}, \bibinfo {author} {\bibfnamefont {J.~E.}\ \bibnamefont
  {Gubernatis}}, \bibinfo {author} {\bibfnamefont {E.}~\bibnamefont {Knill}},\
  and\ \bibinfo {author} {\bibfnamefont {R.}~\bibnamefont {Laflamme}},\
  }\bibfield  {title} {\bibinfo {title} {{Quantum algorithms for fermionic
  simulations}},\ }\href {https://doi.org/10.1103/PhysRevA.64.022319}
  {\bibfield  {journal} {\bibinfo  {journal} {Phys. Rev. A}\ }\textbf {\bibinfo
  {volume} {64}},\ \bibinfo {pages} {022319} (\bibinfo {year} {2001})},\
  \bibinfo {note} {[Erratum: Phys.Rev.A 65, 029902 (2002)]},\ \Eprint
  {https://arxiv.org/abs/cond-mat/0012334} {arXiv:cond-mat/0012334}
  \BibitemShut {NoStop}%
\bibitem [{\citenamefont {Somma}\ \emph {et~al.}(2002)\citenamefont {Somma},
  \citenamefont {Ortiz}, \citenamefont {Gubernatis}, \citenamefont {Knill},\
  and\ \citenamefont {Laflamme}}]{Soma-SpectrumState}%
  \BibitemOpen
  \bibfield  {author} {\bibinfo {author} {\bibfnamefont {R.}~\bibnamefont
  {Somma}}, \bibinfo {author} {\bibfnamefont {G.}~\bibnamefont {Ortiz}},
  \bibinfo {author} {\bibfnamefont {J.~E.}\ \bibnamefont {Gubernatis}},
  \bibinfo {author} {\bibfnamefont {E.}~\bibnamefont {Knill}},\ and\ \bibinfo
  {author} {\bibfnamefont {R.}~\bibnamefont {Laflamme}},\ }\bibfield  {title}
  {\bibinfo {title} {Simulating physical phenomena by quantum networks},\
  }\href {https://doi.org/10.1103/PhysRevA.65.042323} {\bibfield  {journal}
  {\bibinfo  {journal} {Phys. Rev. A}\ }\textbf {\bibinfo {volume} {65}},\
  \bibinfo {pages} {042323} (\bibinfo {year} {2002})}\BibitemShut {NoStop}%
\bibitem [{\citenamefont {Kassal}\ \emph {et~al.}(2008)\citenamefont {Kassal},
  \citenamefont {Jordan}, \citenamefont {Love}, \citenamefont {Mohseni},\ and\
  \citenamefont {Aspuru-Guzik}}]{Kassal_2008}%
  \BibitemOpen
  \bibfield  {author} {\bibinfo {author} {\bibfnamefont {I.}~\bibnamefont
  {Kassal}}, \bibinfo {author} {\bibfnamefont {S.~P.}\ \bibnamefont {Jordan}},
  \bibinfo {author} {\bibfnamefont {P.~J.}\ \bibnamefont {Love}}, \bibinfo
  {author} {\bibfnamefont {M.}~\bibnamefont {Mohseni}},\ and\ \bibinfo {author}
  {\bibfnamefont {A.}~\bibnamefont {Aspuru-Guzik}},\ }\bibfield  {title}
  {\bibinfo {title} {Polynomial-time quantum algorithm for the simulation of
  chemical dynamics},\ }\href {https://doi.org/10.1073/pnas.0808245105}
  {\bibfield  {journal} {\bibinfo  {journal} {Proceedings of the National
  Academy of Sciences}\ }\textbf {\bibinfo {volume} {105}},\ \bibinfo {pages}
  {18681–18686} (\bibinfo {year} {2008})}\BibitemShut {NoStop}%
\bibitem [{\citenamefont {Roggero}\ and\ \citenamefont
  {Carlson}(2019)}]{Rogerro2019}%
  \BibitemOpen
  \bibfield  {author} {\bibinfo {author} {\bibfnamefont {A.}~\bibnamefont
  {Roggero}}\ and\ \bibinfo {author} {\bibfnamefont {J.}~\bibnamefont
  {Carlson}},\ }\bibfield  {title} {\bibinfo {title} {Dynamic linear response
  quantum algorithm},\ }\href {https://doi.org/10.1103/PhysRevC.100.034610}
  {\bibfield  {journal} {\bibinfo  {journal} {Phys. Rev. C}\ }\textbf {\bibinfo
  {volume} {100}},\ \bibinfo {pages} {034610} (\bibinfo {year}
  {2019})}\BibitemShut {NoStop}%
\bibitem [{\citenamefont {Somma}(2019)}]{Somma:2019rmm}%
  \BibitemOpen
  \bibfield  {author} {\bibinfo {author} {\bibfnamefont {R.~D.}\ \bibnamefont
  {Somma}},\ }\bibfield  {title} {\bibinfo {title} {{Quantum eigenvalue
  estimation via time series analysis}},\ }\href
  {https://doi.org/10.1088/1367-2630/ab5c60} {\bibfield  {journal} {\bibinfo
  {journal} {New J. Phys.}\ }\textbf {\bibinfo {volume} {21}},\ \bibinfo
  {pages} {123025} (\bibinfo {year} {2019})}\BibitemShut {NoStop}%
\bibitem [{\citenamefont {Roggero}(2020)}]{Roggero:2020qoz}%
  \BibitemOpen
  \bibfield  {author} {\bibinfo {author} {\bibfnamefont {A.}~\bibnamefont
  {Roggero}},\ }\bibfield  {title} {\bibinfo {title} {{Spectral density
  estimation with the Gaussian Integral Transform}},\ }\href
  {https://doi.org/10.1103/PhysRevA.102.022409} {\bibfield  {journal} {\bibinfo
   {journal} {Phys. Rev. A}\ }\textbf {\bibinfo {volume} {102}},\ \bibinfo
  {pages} {022409} (\bibinfo {year} {2020})},\ \Eprint
  {https://arxiv.org/abs/2004.04889} {arXiv:2004.04889 [quant-ph]} \BibitemShut
  {NoStop}%
\bibitem [{\citenamefont {Rall}(2020)}]{PhysRevA.102.022408}%
  \BibitemOpen
  \bibfield  {author} {\bibinfo {author} {\bibfnamefont {P.}~\bibnamefont
  {Rall}},\ }\bibfield  {title} {\bibinfo {title} {Quantum algorithms for
  estimating physical quantities using block encodings},\ }\href
  {https://doi.org/10.1103/PhysRevA.102.022408} {\bibfield  {journal} {\bibinfo
   {journal} {Phys. Rev. A}\ }\textbf {\bibinfo {volume} {102}},\ \bibinfo
  {pages} {022408} (\bibinfo {year} {2020})}\BibitemShut {NoStop}%
\bibitem [{\citenamefont {Roggero}\ \emph {et~al.}(2020)\citenamefont
  {Roggero}, \citenamefont {Li}, \citenamefont {Carlson}, \citenamefont
  {Gupta},\ and\ \citenamefont {Perdue}}]{Roggero:2019myu}%
  \BibitemOpen
  \bibfield  {author} {\bibinfo {author} {\bibfnamefont {A.}~\bibnamefont
  {Roggero}}, \bibinfo {author} {\bibfnamefont {A.~C.~Y.}\ \bibnamefont {Li}},
  \bibinfo {author} {\bibfnamefont {J.}~\bibnamefont {Carlson}}, \bibinfo
  {author} {\bibfnamefont {R.}~\bibnamefont {Gupta}},\ and\ \bibinfo {author}
  {\bibfnamefont {G.~N.}\ \bibnamefont {Perdue}},\ }\bibfield  {title}
  {\bibinfo {title} {{Quantum Computing for Neutrino-Nucleus Scattering}},\
  }\href {https://doi.org/10.1103/PhysRevD.101.074038} {\bibfield  {journal}
  {\bibinfo  {journal} {Phys. Rev. D}\ }\textbf {\bibinfo {volume} {101}},\
  \bibinfo {pages} {074038} (\bibinfo {year} {2020})},\ \Eprint
  {https://arxiv.org/abs/1911.06368} {arXiv:1911.06368 [quant-ph]} \BibitemShut
  {NoStop}%
\bibitem [{\citenamefont {Trotter}(1959)}]{Trotter1959}%
  \BibitemOpen
  \bibfield  {author} {\bibinfo {author} {\bibfnamefont {H.~F.}\ \bibnamefont
  {Trotter}},\ }\bibfield  {title} {\bibinfo {title} {On the product of
  semi-groups of operators},\ }\href {http://www.jstor.org/stable/2033649}
  {\bibfield  {journal} {\bibinfo  {journal} {Proceedings of the American
  Mathematical Society}\ }\textbf {\bibinfo {volume} {10}},\ \bibinfo {pages}
  {545} (\bibinfo {year} {1959})}\BibitemShut {NoStop}%
\bibitem [{\citenamefont {Suzuki}(1985)}]{Suzuki:1985wzj}%
  \BibitemOpen
  \bibfield  {author} {\bibinfo {author} {\bibfnamefont {M.}~\bibnamefont
  {Suzuki}},\ }\bibfield  {title} {\bibinfo {title} {{Decomposition formulas of
  exponential operators and Lie exponentials with some applications to quantum
  mechanics and statistical physics}},\ }\href
  {https://doi.org/10.1063/1.526596} {\bibfield  {journal} {\bibinfo  {journal}
  {J. Math. Phys.}\ }\textbf {\bibinfo {volume} {26}},\ \bibinfo {pages} {601}
  (\bibinfo {year} {1985})}\BibitemShut {NoStop}%
\bibitem [{\citenamefont {Lloyd}(1996)}]{Lloyd1996}%
  \BibitemOpen
  \bibfield  {author} {\bibinfo {author} {\bibfnamefont {S.}~\bibnamefont
  {Lloyd}},\ }\bibfield  {title} {\bibinfo {title} {Universal quantum
  simulators},\ }\href {https://doi.org/10.1126/science.273.5278.1073}
  {\bibfield  {journal} {\bibinfo  {journal} {Science}\ }\textbf {\bibinfo
  {volume} {273}},\ \bibinfo {pages} {1073} (\bibinfo {year}
  {1996})}\BibitemShut {NoStop}%
\bibitem [{\citenamefont {Heyl}\ \emph {et~al.}(2019)\citenamefont {Heyl},
  \citenamefont {Hauke},\ and\ \citenamefont {Zoller}}]{Heyl_2019}%
  \BibitemOpen
  \bibfield  {author} {\bibinfo {author} {\bibfnamefont {M.}~\bibnamefont
  {Heyl}}, \bibinfo {author} {\bibfnamefont {P.}~\bibnamefont {Hauke}},\ and\
  \bibinfo {author} {\bibfnamefont {P.}~\bibnamefont {Zoller}},\ }\bibfield
  {title} {\bibinfo {title} {Quantum localization bounds trotter errors in
  digital quantum simulation},\ }\bibfield  {journal} {\bibinfo  {journal}
  {Science Advances}\ }\textbf {\bibinfo {volume} {5}},\ \href
  {https://doi.org/10.1126/sciadv.aau8342} {10.1126/sciadv.aau8342} (\bibinfo
  {year} {2019})\BibitemShut {NoStop}%
\bibitem [{\citenamefont {Kivlichan}\ \emph {et~al.}(2020)\citenamefont
  {Kivlichan}, \citenamefont {Gidney}, \citenamefont {Berry}, \citenamefont
  {Wiebe}, \citenamefont {McClean}, \citenamefont {Sun}, \citenamefont {Jiang},
  \citenamefont {Rubin}, \citenamefont {Fowler}, \citenamefont {Aspuru-Guzik},
  \citenamefont {Neven},\ and\ \citenamefont {Babbush}}]{Kivlichan_2020}%
  \BibitemOpen
  \bibfield  {author} {\bibinfo {author} {\bibfnamefont {I.~D.}\ \bibnamefont
  {Kivlichan}}, \bibinfo {author} {\bibfnamefont {C.}~\bibnamefont {Gidney}},
  \bibinfo {author} {\bibfnamefont {D.~W.}\ \bibnamefont {Berry}}, \bibinfo
  {author} {\bibfnamefont {N.}~\bibnamefont {Wiebe}}, \bibinfo {author}
  {\bibfnamefont {J.}~\bibnamefont {McClean}}, \bibinfo {author} {\bibfnamefont
  {W.}~\bibnamefont {Sun}}, \bibinfo {author} {\bibfnamefont {Z.}~\bibnamefont
  {Jiang}}, \bibinfo {author} {\bibfnamefont {N.}~\bibnamefont {Rubin}},
  \bibinfo {author} {\bibfnamefont {A.}~\bibnamefont {Fowler}}, \bibinfo
  {author} {\bibfnamefont {A.}~\bibnamefont {Aspuru-Guzik}}, \bibinfo {author}
  {\bibfnamefont {H.}~\bibnamefont {Neven}},\ and\ \bibinfo {author}
  {\bibfnamefont {R.}~\bibnamefont {Babbush}},\ }\bibfield  {title} {\bibinfo
  {title} {Improved fault-tolerant quantum simulation of condensed-phase
  correlated electrons via trotterization},\ }\href
  {https://doi.org/10.22331/q-2020-07-16-296} {\bibfield  {journal} {\bibinfo
  {journal} {Quantum}\ }\textbf {\bibinfo {volume} {4}},\ \bibinfo {pages}
  {296} (\bibinfo {year} {2020})}\BibitemShut {NoStop}%
\bibitem [{\citenamefont {Childs}\ \emph {et~al.}(2021)\citenamefont {Childs},
  \citenamefont {Su}, \citenamefont {Tran}, \citenamefont {Wiebe},\ and\
  \citenamefont {Zhu}}]{Childs:2019hts}%
  \BibitemOpen
  \bibfield  {author} {\bibinfo {author} {\bibfnamefont {A.~M.}\ \bibnamefont
  {Childs}}, \bibinfo {author} {\bibfnamefont {Y.}~\bibnamefont {Su}}, \bibinfo
  {author} {\bibfnamefont {M.~C.}\ \bibnamefont {Tran}}, \bibinfo {author}
  {\bibfnamefont {N.}~\bibnamefont {Wiebe}},\ and\ \bibinfo {author}
  {\bibfnamefont {S.}~\bibnamefont {Zhu}},\ }\bibfield  {title} {\bibinfo
  {title} {{Theory of Trotter Error with Commutator Scaling}},\ }\href
  {https://doi.org/10.1103/physrevx.11.011020} {\bibfield  {journal} {\bibinfo
  {journal} {Phys. Rev. X}\ }\textbf {\bibinfo {volume} {11}},\ \bibinfo
  {pages} {011020} (\bibinfo {year} {2021})},\ \Eprint
  {https://arxiv.org/abs/1912.08854} {arXiv:1912.08854 [quant-ph]} \BibitemShut
  {NoStop}%
\bibitem [{\citenamefont {Tran}\ \emph {et~al.}(2020)\citenamefont {Tran},
  \citenamefont {Chu}, \citenamefont {Su}, \citenamefont {Childs},\ and\
  \citenamefont {Gorshkov}}]{Tran_2020}%
  \BibitemOpen
  \bibfield  {author} {\bibinfo {author} {\bibfnamefont {M.~C.}\ \bibnamefont
  {Tran}}, \bibinfo {author} {\bibfnamefont {S.-K.}\ \bibnamefont {Chu}},
  \bibinfo {author} {\bibfnamefont {Y.}~\bibnamefont {Su}}, \bibinfo {author}
  {\bibfnamefont {A.~M.}\ \bibnamefont {Childs}},\ and\ \bibinfo {author}
  {\bibfnamefont {A.~V.}\ \bibnamefont {Gorshkov}},\ }\bibfield  {title}
  {\bibinfo {title} {Destructive error interference in product-formula lattice
  simulation},\ }\bibfield  {journal} {\bibinfo  {journal} {Physical Review
  Letters}\ }\textbf {\bibinfo {volume} {124}},\ \href
  {https://doi.org/10.1103/physrevlett.124.220502}
  {10.1103/physrevlett.124.220502} (\bibinfo {year} {2020})\BibitemShut
  {NoStop}%
\bibitem [{\citenamefont {Sieberer}\ \emph {et~al.}(2019)\citenamefont
  {Sieberer}, \citenamefont {Olsacher}, \citenamefont {Elben}, \citenamefont
  {Heyl}, \citenamefont {Hauke}, \citenamefont {Haake},\ and\ \citenamefont
  {Zoller}}]{Sieberer:2019htd}%
  \BibitemOpen
  \bibfield  {author} {\bibinfo {author} {\bibfnamefont {L.~M.}\ \bibnamefont
  {Sieberer}}, \bibinfo {author} {\bibfnamefont {T.}~\bibnamefont {Olsacher}},
  \bibinfo {author} {\bibfnamefont {A.}~\bibnamefont {Elben}}, \bibinfo
  {author} {\bibfnamefont {M.}~\bibnamefont {Heyl}}, \bibinfo {author}
  {\bibfnamefont {P.}~\bibnamefont {Hauke}}, \bibinfo {author} {\bibfnamefont
  {F.}~\bibnamefont {Haake}},\ and\ \bibinfo {author} {\bibfnamefont
  {P.}~\bibnamefont {Zoller}},\ }\bibfield  {title} {\bibinfo {title} {{Digital
  quantum simulation, Trotter errors, and quantum chaos of the kicked top}},\
  }\href {https://doi.org/10.1038/s41534-019-0192-5} {\bibfield  {journal}
  {\bibinfo  {journal} {npj Quantum Inf.}\ }\textbf {\bibinfo {volume} {5}},\
  \bibinfo {pages} {78} (\bibinfo {year} {2019})}\BibitemShut {NoStop}%
\bibitem [{\citenamefont {Layden}(2022)}]{Layden:2021ols}%
  \BibitemOpen
  \bibfield  {author} {\bibinfo {author} {\bibfnamefont {D.}~\bibnamefont
  {Layden}},\ }\bibfield  {title} {\bibinfo {title} {{First-Order Trotter Error
  from a Second-Order Perspective}},\ }\href
  {https://doi.org/10.1103/PhysRevLett.128.210501} {\bibfield  {journal}
  {\bibinfo  {journal} {Phys. Rev. Lett.}\ }\textbf {\bibinfo {volume} {128}},\
  \bibinfo {pages} {210501} (\bibinfo {year} {2022})},\ \Eprint
  {https://arxiv.org/abs/2107.08032} {arXiv:2107.08032 [quant-ph]} \BibitemShut
  {NoStop}%
\bibitem [{\citenamefont {Berry}\ \emph
  {et~al.}(2015{\natexlab{a}})\citenamefont {Berry}, \citenamefont {Childs},\
  and\ \citenamefont {Kothari}}]{Berry_2015}%
  \BibitemOpen
  \bibfield  {author} {\bibinfo {author} {\bibfnamefont {D.~W.}\ \bibnamefont
  {Berry}}, \bibinfo {author} {\bibfnamefont {A.~M.}\ \bibnamefont {Childs}},\
  and\ \bibinfo {author} {\bibfnamefont {R.}~\bibnamefont {Kothari}},\
  }\bibfield  {title} {\bibinfo {title} {Hamiltonian simulation with nearly
  optimal dependence on all parameters},\ }in\ \href
  {https://doi.org/10.1109/focs.2015.54} {\emph {\bibinfo {booktitle} {2015
  IEEE 56th Annual Symposium on Foundations of Computer Science}}}\ (\bibinfo
  {publisher} {IEEE},\ \bibinfo {year} {2015})\BibitemShut {NoStop}%
\bibitem [{\citenamefont {Berry}\ \emph
  {et~al.}(2015{\natexlab{b}})\citenamefont {Berry}, \citenamefont {Childs},
  \citenamefont {Cleve}, \citenamefont {Kothari},\ and\ \citenamefont
  {Somma}}]{Berry:2014ivo}%
  \BibitemOpen
  \bibfield  {author} {\bibinfo {author} {\bibfnamefont {D.~W.}\ \bibnamefont
  {Berry}}, \bibinfo {author} {\bibfnamefont {A.~M.}\ \bibnamefont {Childs}},
  \bibinfo {author} {\bibfnamefont {R.}~\bibnamefont {Cleve}}, \bibinfo
  {author} {\bibfnamefont {R.}~\bibnamefont {Kothari}},\ and\ \bibinfo {author}
  {\bibfnamefont {R.~D.}\ \bibnamefont {Somma}},\ }\bibfield  {title} {\bibinfo
  {title} {{Simulating Hamiltonian Dynamics with a Truncated Taylor Series}},\
  }\href {https://doi.org/10.1103/PhysRevLett.114.090502} {\bibfield  {journal}
  {\bibinfo  {journal} {Phys. Rev. Lett.}\ }\textbf {\bibinfo {volume} {114}},\
  \bibinfo {pages} {090502} (\bibinfo {year} {2015}{\natexlab{b}})},\ \Eprint
  {https://arxiv.org/abs/1412.4687} {arXiv:1412.4687 [quant-ph]} \BibitemShut
  {NoStop}%
\bibitem [{\citenamefont {Low}\ and\ \citenamefont
  {Chuang}(2017)}]{Low:2016sck}%
  \BibitemOpen
  \bibfield  {author} {\bibinfo {author} {\bibfnamefont {G.~H.}\ \bibnamefont
  {Low}}\ and\ \bibinfo {author} {\bibfnamefont {I.~L.}\ \bibnamefont
  {Chuang}},\ }\bibfield  {title} {\bibinfo {title} {{Optimal Hamiltonian
  Simulation by Quantum Signal Processing}},\ }\href
  {https://doi.org/10.1103/PhysRevLett.118.010501} {\bibfield  {journal}
  {\bibinfo  {journal} {Phys. Rev. Lett.}\ }\textbf {\bibinfo {volume} {118}},\
  \bibinfo {pages} {010501} (\bibinfo {year} {2017})},\ \Eprint
  {https://arxiv.org/abs/1606.02685} {arXiv:1606.02685 [quant-ph]} \BibitemShut
  {NoStop}%
\bibitem [{\citenamefont {Low}\ and\ \citenamefont
  {Chuang}(2019)}]{Low:2016znh}%
  \BibitemOpen
  \bibfield  {author} {\bibinfo {author} {\bibfnamefont {G.~H.}\ \bibnamefont
  {Low}}\ and\ \bibinfo {author} {\bibfnamefont {I.~L.}\ \bibnamefont
  {Chuang}},\ }\bibfield  {title} {\bibinfo {title} {{Hamiltonian Simulation by
  Qubitization}},\ }\href {https://doi.org/10.22331/q-2019-07-12-163}
  {\bibfield  {journal} {\bibinfo  {journal} {Quantum}\ }\textbf {\bibinfo
  {volume} {3}},\ \bibinfo {pages} {163} (\bibinfo {year} {2019})},\ \Eprint
  {https://arxiv.org/abs/1610.06546} {arXiv:1610.06546 [quant-ph]} \BibitemShut
  {NoStop}%
\bibitem [{\citenamefont {Mei\ss{}ner}\ \emph {et~al.}(2024)\citenamefont
  {Mei\ss{}ner}, \citenamefont {Shen}, \citenamefont {Elhatisari},\ and\
  \citenamefont {Lee}}]{Meissner:2023cvo}%
  \BibitemOpen
  \bibfield  {author} {\bibinfo {author} {\bibfnamefont {U.-G.}\ \bibnamefont
  {Mei\ss{}ner}}, \bibinfo {author} {\bibfnamefont {S.}~\bibnamefont {Shen}},
  \bibinfo {author} {\bibfnamefont {S.}~\bibnamefont {Elhatisari}},\ and\
  \bibinfo {author} {\bibfnamefont {D.}~\bibnamefont {Lee}},\ }\bibfield
  {title} {\bibinfo {title} {{Ab~Initio Calculation of the Alpha-Particle
  Monopole Transition Form Factor}},\ }\href
  {https://doi.org/10.1103/PhysRevLett.132.062501} {\bibfield  {journal}
  {\bibinfo  {journal} {Phys. Rev. Lett.}\ }\textbf {\bibinfo {volume} {132}},\
  \bibinfo {pages} {062501} (\bibinfo {year} {2024})},\ \Eprint
  {https://arxiv.org/abs/2309.01558} {arXiv:2309.01558 [nucl-th]} \BibitemShut
  {NoStop}%
\bibitem [{\citenamefont {Lee}(2009)}]{Lee:2008fa}%
  \BibitemOpen
  \bibfield  {author} {\bibinfo {author} {\bibfnamefont {D.}~\bibnamefont
  {Lee}},\ }\bibfield  {title} {\bibinfo {title} {{Lattice simulations for few-
  and many-body systems}},\ }\href {https://doi.org/10.1016/j.ppnp.2008.12.001}
  {\bibfield  {journal} {\bibinfo  {journal} {Prog. Part. Nucl. Phys.}\
  }\textbf {\bibinfo {volume} {63}},\ \bibinfo {pages} {117} (\bibinfo {year}
  {2009})},\ \Eprint {https://arxiv.org/abs/0804.3501} {arXiv:0804.3501
  [nucl-th]} \BibitemShut {NoStop}%
\bibitem [{\citenamefont {Drut}\ and\ \citenamefont
  {Nicholson}(2013)}]{Drut:2012md}%
  \BibitemOpen
  \bibfield  {author} {\bibinfo {author} {\bibfnamefont {J.~E.}\ \bibnamefont
  {Drut}}\ and\ \bibinfo {author} {\bibfnamefont {A.~N.}\ \bibnamefont
  {Nicholson}},\ }\bibfield  {title} {\bibinfo {title} {{Lattice methods for
  strongly interacting many-body systems}},\ }\href
  {https://doi.org/10.1088/0954-3899/40/4/043101} {\bibfield  {journal}
  {\bibinfo  {journal} {J. Phys. G}\ }\textbf {\bibinfo {volume} {40}},\
  \bibinfo {pages} {043101} (\bibinfo {year} {2013})},\ \Eprint
  {https://arxiv.org/abs/1208.6556} {arXiv:1208.6556 [cond-mat.stat-mech]}
  \BibitemShut {NoStop}%
\bibitem [{\citenamefont {Lee}(2017)}]{Lee:2016fhn}%
  \BibitemOpen
  \bibfield  {author} {\bibinfo {author} {\bibfnamefont {D.}~\bibnamefont
  {Lee}},\ }\bibfield  {title} {\bibinfo {title} {{Lattice methods and the
  nuclear few- and many-body problem}},\ }\href
  {https://doi.org/10.1007/978-3-319-53336-0_6} {\bibfield  {journal} {\bibinfo
   {journal} {Lect. Notes Phys.}\ }\textbf {\bibinfo {volume} {936}},\ \bibinfo
  {pages} {237} (\bibinfo {year} {2017})},\ \Eprint
  {https://arxiv.org/abs/1609.00421} {arXiv:1609.00421 [nucl-th]} \BibitemShut
  {NoStop}%
\bibitem [{\citenamefont {L\"ahde}\ and\ \citenamefont
  {Mei\ss{}ner}(2019)}]{Lahde:2019npb}%
  \BibitemOpen
  \bibfield  {author} {\bibinfo {author} {\bibfnamefont {T.~A.}\ \bibnamefont
  {L\"ahde}}\ and\ \bibinfo {author} {\bibfnamefont {U.-G.}\ \bibnamefont
  {Mei\ss{}ner}},\ }\href {https://doi.org/10.1007/978-3-030-14189-9} {\emph
  {\bibinfo {title} {{Nuclear Lattice Effective Field Theory}: {An
  introduction}}}},\ Vol.\ \bibinfo {volume} {957}\ (\bibinfo  {publisher}
  {Springer},\ \bibinfo {year} {2019})\BibitemShut {NoStop}%
\bibitem [{\citenamefont {Elhatisari}\ \emph {et~al.}(2024)\citenamefont
  {Elhatisari} \emph {et~al.}}]{Elhatisari:2022zrb}%
  \BibitemOpen
  \bibfield  {author} {\bibinfo {author} {\bibfnamefont {S.}~\bibnamefont
  {Elhatisari}} \emph {et~al.},\ }\bibfield  {title} {\bibinfo {title}
  {{Wavefunction matching for solving quantum many-body problems}},\ }\href
  {https://doi.org/10.1038/s41586-024-07422-z} {\bibfield  {journal} {\bibinfo
  {journal} {Nature}\ }\textbf {\bibinfo {volume} {630}},\ \bibinfo {pages}
  {59} (\bibinfo {year} {2024})},\ \Eprint {https://arxiv.org/abs/2210.17488}
  {arXiv:2210.17488 [nucl-th]} \BibitemShut {NoStop}%
\bibitem [{\citenamefont {Abrams}\ and\ \citenamefont
  {Lloyd}(1997)}]{AbramsLloyd1997}%
  \BibitemOpen
  \bibfield  {author} {\bibinfo {author} {\bibfnamefont {D.~S.}\ \bibnamefont
  {Abrams}}\ and\ \bibinfo {author} {\bibfnamefont {S.}~\bibnamefont {Lloyd}},\
  }\bibfield  {title} {\bibinfo {title} {Simulation of many-body fermi systems
  on a universal quantum computer},\ }\href
  {https://doi.org/10.1103/PhysRevLett.79.2586} {\bibfield  {journal} {\bibinfo
   {journal} {Phys. Rev. Lett.}\ }\textbf {\bibinfo {volume} {79}},\ \bibinfo
  {pages} {2586} (\bibinfo {year} {1997})}\BibitemShut {NoStop}%
\bibitem [{\citenamefont {Berry}\ \emph {et~al.}(2018)\citenamefont {Berry},
  \citenamefont {Kieferov{\'a}}, \citenamefont {Scherer}, \citenamefont
  {Sanders}, \citenamefont {Low}, \citenamefont {Wiebe}, \citenamefont
  {Gidney},\ and\ \citenamefont {Babbush}}]{Berry2018a}%
  \BibitemOpen
  \bibfield  {author} {\bibinfo {author} {\bibfnamefont {D.~W.}\ \bibnamefont
  {Berry}}, \bibinfo {author} {\bibfnamefont {M.}~\bibnamefont
  {Kieferov{\'a}}}, \bibinfo {author} {\bibfnamefont {A.}~\bibnamefont
  {Scherer}}, \bibinfo {author} {\bibfnamefont {Y.~R.}\ \bibnamefont
  {Sanders}}, \bibinfo {author} {\bibfnamefont {G.~H.}\ \bibnamefont {Low}},
  \bibinfo {author} {\bibfnamefont {N.}~\bibnamefont {Wiebe}}, \bibinfo
  {author} {\bibfnamefont {C.}~\bibnamefont {Gidney}},\ and\ \bibinfo {author}
  {\bibfnamefont {R.}~\bibnamefont {Babbush}},\ }\bibfield  {title} {\bibinfo
  {title} {Improved techniques for preparing eigenstates of fermionic
  hamiltonians},\ }\href {https://doi.org/10.1038/s41534-018-0071-5} {\bibfield
   {journal} {\bibinfo  {journal} {npj Quantum Information}\ }\textbf {\bibinfo
  {volume} {4}},\ \bibinfo {pages} {22} (\bibinfo {year} {2018})}\BibitemShut
  {NoStop}%
\bibitem [{\citenamefont {Bedaque}\ and\ \citenamefont {van
  Kolck}(2002)}]{Bedaque:2002mn}%
  \BibitemOpen
  \bibfield  {author} {\bibinfo {author} {\bibfnamefont {P.~F.}\ \bibnamefont
  {Bedaque}}\ and\ \bibinfo {author} {\bibfnamefont {U.}~\bibnamefont {van
  Kolck}},\ }\bibfield  {title} {\bibinfo {title} {{Effective field theory for
  few nucleon systems}},\ }\href
  {https://doi.org/10.1146/annurev.nucl.52.050102.090637} {\bibfield  {journal}
  {\bibinfo  {journal} {Ann. Rev. Nucl. Part. Sci.}\ }\textbf {\bibinfo
  {volume} {52}},\ \bibinfo {pages} {339} (\bibinfo {year} {2002})},\ \Eprint
  {https://arxiv.org/abs/nucl-th/0203055} {arXiv:nucl-th/0203055} \BibitemShut
  {NoStop}%
\bibitem [{\citenamefont {Hammer}\ \emph {et~al.}(2020)\citenamefont {Hammer},
  \citenamefont {K\"onig},\ and\ \citenamefont {van Kolck}}]{Hammer:2019poc}%
  \BibitemOpen
  \bibfield  {author} {\bibinfo {author} {\bibfnamefont {H.~W.}\ \bibnamefont
  {Hammer}}, \bibinfo {author} {\bibfnamefont {S.}~\bibnamefont {K\"onig}},\
  and\ \bibinfo {author} {\bibfnamefont {U.}~\bibnamefont {van Kolck}},\
  }\bibfield  {title} {\bibinfo {title} {{Nuclear effective field theory:
  status and perspectives}},\ }\href
  {https://doi.org/10.1103/RevModPhys.92.025004} {\bibfield  {journal}
  {\bibinfo  {journal} {Rev. Mod. Phys.}\ }\textbf {\bibinfo {volume} {92}},\
  \bibinfo {pages} {025004} (\bibinfo {year} {2020})},\ \Eprint
  {https://arxiv.org/abs/1906.12122} {arXiv:1906.12122 [nucl-th]} \BibitemShut
  {NoStop}%
\bibitem [{\citenamefont {{van Kolck}}(1999)}]{VANKOLCK1999273}%
  \BibitemOpen
  \bibfield  {author} {\bibinfo {author} {\bibfnamefont {U.}~\bibnamefont {{van
  Kolck}}},\ }\bibfield  {title} {\bibinfo {title} {Effective field theory of
  short-range forces},\ }\href
  {https://doi.org/https://doi.org/10.1016/S0375-9474(98)00612-5} {\bibfield
  {journal} {\bibinfo  {journal} {Nuclear Physics A}\ }\textbf {\bibinfo
  {volume} {645}},\ \bibinfo {pages} {273} (\bibinfo {year}
  {1999})}\BibitemShut {NoStop}%
\bibitem [{\citenamefont {Chen}\ \emph {et~al.}(1999)\citenamefont {Chen},
  \citenamefont {Rupak},\ and\ \citenamefont {Savage}}]{CHEN1999386}%
  \BibitemOpen
  \bibfield  {author} {\bibinfo {author} {\bibfnamefont {J.-W.}\ \bibnamefont
  {Chen}}, \bibinfo {author} {\bibfnamefont {G.}~\bibnamefont {Rupak}},\ and\
  \bibinfo {author} {\bibfnamefont {M.~J.}\ \bibnamefont {Savage}},\ }\bibfield
   {title} {\bibinfo {title} {Nucleon-nucleon effective field theory without
  pions},\ }\href
  {https://doi.org/https://doi.org/10.1016/S0375-9474(99)00298-5} {\bibfield
  {journal} {\bibinfo  {journal} {Nuclear Physics A}\ }\textbf {\bibinfo
  {volume} {653}},\ \bibinfo {pages} {386} (\bibinfo {year}
  {1999})}\BibitemShut {NoStop}%
\bibitem [{\citenamefont {Watson}\ \emph {et~al.}(2023)\citenamefont {Watson},
  \citenamefont {Bringewatt}, \citenamefont {Shaw}, \citenamefont {Childs},
  \citenamefont {Gorshkov},\ and\ \citenamefont {Davoudi}}]{Watson:2023oov}%
  \BibitemOpen
  \bibfield  {author} {\bibinfo {author} {\bibfnamefont {J.~D.}\ \bibnamefont
  {Watson}}, \bibinfo {author} {\bibfnamefont {J.}~\bibnamefont {Bringewatt}},
  \bibinfo {author} {\bibfnamefont {A.~F.}\ \bibnamefont {Shaw}}, \bibinfo
  {author} {\bibfnamefont {A.~M.}\ \bibnamefont {Childs}}, \bibinfo {author}
  {\bibfnamefont {A.~V.}\ \bibnamefont {Gorshkov}},\ and\ \bibinfo {author}
  {\bibfnamefont {Z.}~\bibnamefont {Davoudi}},\ }\href@noop {} {\bibinfo
  {title} {{Quantum Algorithms for Simulating Nuclear Effective Field
  Theories}}} (\bibinfo {year} {2023}),\ \Eprint
  {https://arxiv.org/abs/2312.05344} {arXiv:2312.05344 [quant-ph]} \BibitemShut
  {NoStop}%
\bibitem [{\citenamefont {Wecker}\ \emph {et~al.}(2014)\citenamefont {Wecker},
  \citenamefont {Bauer}, \citenamefont {Clark}, \citenamefont {Hastings},\ and\
  \citenamefont {Troyer}}]{Wecker:2014vsy}%
  \BibitemOpen
  \bibfield  {author} {\bibinfo {author} {\bibfnamefont {D.}~\bibnamefont
  {Wecker}}, \bibinfo {author} {\bibfnamefont {B.}~\bibnamefont {Bauer}},
  \bibinfo {author} {\bibfnamefont {B.~K.}\ \bibnamefont {Clark}}, \bibinfo
  {author} {\bibfnamefont {M.~B.}\ \bibnamefont {Hastings}},\ and\ \bibinfo
  {author} {\bibfnamefont {M.}~\bibnamefont {Troyer}},\ }\bibfield  {title}
  {\bibinfo {title} {{Gate-count estimates for performing quantum chemistry on
  small quantum computers}},\ }\href
  {https://doi.org/10.1103/PhysRevA.90.022305} {\bibfield  {journal} {\bibinfo
  {journal} {Phys. Rev. A}\ }\textbf {\bibinfo {volume} {90}},\ \bibinfo
  {pages} {022305} (\bibinfo {year} {2014})}\BibitemShut {NoStop}%
\bibitem [{\citenamefont {Gidney}(2018)}]{Gidney2018halvingcostof}%
  \BibitemOpen
  \bibfield  {author} {\bibinfo {author} {\bibfnamefont {C.}~\bibnamefont
  {Gidney}},\ }\bibfield  {title} {\bibinfo {title} {Halving the cost of
  quantum addition},\ }\href {https://doi.org/10.22331/q-2018-06-18-74}
  {\bibfield  {journal} {\bibinfo  {journal} {{Quantum}}\ }\textbf {\bibinfo
  {volume} {2}},\ \bibinfo {pages} {74} (\bibinfo {year} {2018})}\BibitemShut
  {NoStop}%
\bibitem [{\citenamefont {Bocharov}\ \emph {et~al.}(2015)\citenamefont
  {Bocharov}, \citenamefont {Roetteler},\ and\ \citenamefont
  {Svore}}]{Bocharov:2015tpl}%
  \BibitemOpen
  \bibfield  {author} {\bibinfo {author} {\bibfnamefont {A.}~\bibnamefont
  {Bocharov}}, \bibinfo {author} {\bibfnamefont {M.}~\bibnamefont
  {Roetteler}},\ and\ \bibinfo {author} {\bibfnamefont {K.~M.}\ \bibnamefont
  {Svore}},\ }\bibfield  {title} {\bibinfo {title} {{Efficient Synthesis of
  Universal Repeat-Until-Success Quantum Circuits}},\ }\href
  {https://doi.org/10.1103/PhysRevLett.114.080502} {\bibfield  {journal}
  {\bibinfo  {journal} {Phys. Rev. Lett.}\ }\textbf {\bibinfo {volume} {114}},\
  \bibinfo {pages} {080502} (\bibinfo {year} {2015})}\BibitemShut {NoStop}%
\bibitem [{\citenamefont {Akahoshi}\ \emph {et~al.}(2024)\citenamefont
  {Akahoshi}, \citenamefont {Maruyama}, \citenamefont {Oshima}, \citenamefont
  {Sato},\ and\ \citenamefont {Fujii}}]{Akahoshi:2023xck}%
  \BibitemOpen
  \bibfield  {author} {\bibinfo {author} {\bibfnamefont {Y.}~\bibnamefont
  {Akahoshi}}, \bibinfo {author} {\bibfnamefont {K.}~\bibnamefont {Maruyama}},
  \bibinfo {author} {\bibfnamefont {H.}~\bibnamefont {Oshima}}, \bibinfo
  {author} {\bibfnamefont {S.}~\bibnamefont {Sato}},\ and\ \bibinfo {author}
  {\bibfnamefont {K.}~\bibnamefont {Fujii}},\ }\bibfield  {title} {\bibinfo
  {title} {{Partially Fault-Tolerant Quantum Computing Architecture with
  Error-Corrected Clifford Gates and Space-Time Efficient Analog Rotations}},\
  }\href {https://doi.org/10.1103/PRXQuantum.5.010337} {\bibfield  {journal}
  {\bibinfo  {journal} {PRX Quantum}\ }\textbf {\bibinfo {volume} {5}},\
  \bibinfo {pages} {010337} (\bibinfo {year} {2024})},\ \Eprint
  {https://arxiv.org/abs/2303.13181} {arXiv:2303.13181 [quant-ph]} \BibitemShut
  {NoStop}%
\bibitem [{\citenamefont {Welch}\ \emph {et~al.}(2014)\citenamefont {Welch},
  \citenamefont {Greenbaum}, \citenamefont {Mostame},\ and\ \citenamefont
  {Aspuru-Guzik}}]{Welch_2014}%
  \BibitemOpen
  \bibfield  {author} {\bibinfo {author} {\bibfnamefont {J.}~\bibnamefont
  {Welch}}, \bibinfo {author} {\bibfnamefont {D.}~\bibnamefont {Greenbaum}},
  \bibinfo {author} {\bibfnamefont {S.}~\bibnamefont {Mostame}},\ and\ \bibinfo
  {author} {\bibfnamefont {A.}~\bibnamefont {Aspuru-Guzik}},\ }\bibfield
  {title} {\bibinfo {title} {Efficient quantum circuits for diagonal unitaries
  without ancillas},\ }\href {https://doi.org/10.1088/1367-2630/16/3/033040}
  {\bibfield  {journal} {\bibinfo  {journal} {New Journal of Physics}\ }\textbf
  {\bibinfo {volume} {16}},\ \bibinfo {pages} {033040} (\bibinfo {year}
  {2014})}\BibitemShut {NoStop}%
\bibitem [{\citenamefont {Nielsen}\ and\ \citenamefont
  {Chuang}(2010)}]{Nielsen_Chuang_2010}%
  \BibitemOpen
  \bibfield  {author} {\bibinfo {author} {\bibfnamefont {M.~A.}\ \bibnamefont
  {Nielsen}}\ and\ \bibinfo {author} {\bibfnamefont {I.~L.}\ \bibnamefont
  {Chuang}},\ }\href {https://doi.org/https://doi.org/10.1017/CBO9780511976667}
  {\emph {\bibinfo {title} {Quantum Computation and Quantum Information: 10th
  Anniversary Edition}}}\ (\bibinfo  {publisher} {Cambridge University Press},\
  \bibinfo {year} {2010})\BibitemShut {NoStop}%
\bibitem [{\citenamefont {Selinger}(2013)}]{Selinger:2013ksm}%
  \BibitemOpen
  \bibfield  {author} {\bibinfo {author} {\bibfnamefont {P.}~\bibnamefont
  {Selinger}},\ }\bibfield  {title} {\bibinfo {title} {{Quantum circuits of
  T-depth one}},\ }\href {https://doi.org/10.1103/PhysRevA.87.042302}
  {\bibfield  {journal} {\bibinfo  {journal} {Phys. Rev. A}\ }\textbf {\bibinfo
  {volume} {87}},\ \bibinfo {pages} {042302} (\bibinfo {year}
  {2013})}\BibitemShut {NoStop}%
\bibitem [{\citenamefont {He}\ \emph {et~al.}(2017)\citenamefont {He},
  \citenamefont {Luo}, \citenamefont {Zhang}, \citenamefont {Wang},\ and\
  \citenamefont {Wang}}]{He:2017wow}%
  \BibitemOpen
  \bibfield  {author} {\bibinfo {author} {\bibfnamefont {Y.}~\bibnamefont
  {He}}, \bibinfo {author} {\bibfnamefont {M.-X.}\ \bibnamefont {Luo}},
  \bibinfo {author} {\bibfnamefont {E.}~\bibnamefont {Zhang}}, \bibinfo
  {author} {\bibfnamefont {H.-K.}\ \bibnamefont {Wang}},\ and\ \bibinfo
  {author} {\bibfnamefont {X.-F.}\ \bibnamefont {Wang}},\ }\bibfield  {title}
  {\bibinfo {title} {{Decompositions of n-qubit Toffoli Gates with Linear
  Circuit Complexity}},\ }\href {https://doi.org/10.1007/s10773-017-3389-4}
  {\bibfield  {journal} {\bibinfo  {journal} {Int. J. Theor. Phys.}\ }\textbf
  {\bibinfo {volume} {56}},\ \bibinfo {pages} {2350} (\bibinfo {year}
  {2017})}\BibitemShut {NoStop}%
\bibitem [{\citenamefont {Jones}(2013)}]{Jones:2013gpb}%
  \BibitemOpen
  \bibfield  {author} {\bibinfo {author} {\bibfnamefont {C.}~\bibnamefont
  {Jones}},\ }\bibfield  {title} {\bibinfo {title} {{Low-overhead constructions
  for the fault-tolerant Toffoli gate}},\ }\href
  {https://doi.org/10.1103/PhysRevA.87.022328} {\bibfield  {journal} {\bibinfo
  {journal} {Phys. Rev. A}\ }\textbf {\bibinfo {volume} {87}},\ \bibinfo
  {pages} {022328} (\bibinfo {year} {2013})}\BibitemShut {NoStop}%
\bibitem [{\citenamefont {Carlson}\ \emph {et~al.}(2002)\citenamefont
  {Carlson}, \citenamefont {Jourdan}, \citenamefont {Schiavilla},\ and\
  \citenamefont {Sick}}]{Carlson:2001mp}%
  \BibitemOpen
  \bibfield  {author} {\bibinfo {author} {\bibfnamefont {J.}~\bibnamefont
  {Carlson}}, \bibinfo {author} {\bibfnamefont {J.}~\bibnamefont {Jourdan}},
  \bibinfo {author} {\bibfnamefont {R.}~\bibnamefont {Schiavilla}},\ and\
  \bibinfo {author} {\bibfnamefont {I.}~\bibnamefont {Sick}},\ }\bibfield
  {title} {\bibinfo {title} {{Longitudinal and transverse quasielastic response
  functions of light nuclei}},\ }\href
  {https://doi.org/10.1103/PhysRevC.65.024002} {\bibfield  {journal} {\bibinfo
  {journal} {Phys. Rev. C}\ }\textbf {\bibinfo {volume} {65}},\ \bibinfo
  {pages} {024002} (\bibinfo {year} {2002})},\ \Eprint
  {https://arxiv.org/abs/nucl-th/0106047} {arXiv:nucl-th/0106047} \BibitemShut
  {NoStop}%
\bibitem [{\citenamefont {Lovato}\ \emph {et~al.}(2015)\citenamefont {Lovato},
  \citenamefont {Gandolfi}, \citenamefont {Carlson}, \citenamefont {Pieper},\
  and\ \citenamefont {Schiavilla}}]{Lovato:2015qka}%
  \BibitemOpen
  \bibfield  {author} {\bibinfo {author} {\bibfnamefont {A.}~\bibnamefont
  {Lovato}}, \bibinfo {author} {\bibfnamefont {S.}~\bibnamefont {Gandolfi}},
  \bibinfo {author} {\bibfnamefont {J.}~\bibnamefont {Carlson}}, \bibinfo
  {author} {\bibfnamefont {S.~C.}\ \bibnamefont {Pieper}},\ and\ \bibinfo
  {author} {\bibfnamefont {R.}~\bibnamefont {Schiavilla}},\ }\bibfield  {title}
  {\bibinfo {title} {{Electromagnetic and neutral-weak response functions of
  $^4$He and $^{12}$C}},\ }\href {https://doi.org/10.1103/PhysRevC.91.062501}
  {\bibfield  {journal} {\bibinfo  {journal} {Phys. Rev. C}\ }\textbf {\bibinfo
  {volume} {91}},\ \bibinfo {pages} {062501} (\bibinfo {year} {2015})},\
  \Eprint {https://arxiv.org/abs/1501.01981} {arXiv:1501.01981 [nucl-th]}
  \BibitemShut {NoStop}%
\bibitem [{\citenamefont {Lovato}\ \emph {et~al.}(2016)\citenamefont {Lovato},
  \citenamefont {Gandolfi}, \citenamefont {Carlson}, \citenamefont {Pieper},\
  and\ \citenamefont {Schiavilla}}]{Lovato:2016gkq}%
  \BibitemOpen
  \bibfield  {author} {\bibinfo {author} {\bibfnamefont {A.}~\bibnamefont
  {Lovato}}, \bibinfo {author} {\bibfnamefont {S.}~\bibnamefont {Gandolfi}},
  \bibinfo {author} {\bibfnamefont {J.}~\bibnamefont {Carlson}}, \bibinfo
  {author} {\bibfnamefont {S.~C.}\ \bibnamefont {Pieper}},\ and\ \bibinfo
  {author} {\bibfnamefont {R.}~\bibnamefont {Schiavilla}},\ }\bibfield  {title}
  {\bibinfo {title} {{Electromagnetic response of $^{12}$C: A first-principles
  calculation}},\ }\href {https://doi.org/10.1103/PhysRevLett.117.082501}
  {\bibfield  {journal} {\bibinfo  {journal} {Phys. Rev. Lett.}\ }\textbf
  {\bibinfo {volume} {117}},\ \bibinfo {pages} {082501} (\bibinfo {year}
  {2016})},\ \Eprint {https://arxiv.org/abs/1605.00248} {arXiv:1605.00248
  [nucl-th]} \BibitemShut {NoStop}%
\bibitem [{\citenamefont {Lovato}\ \emph {et~al.}(2018)\citenamefont {Lovato},
  \citenamefont {Gandolfi}, \citenamefont {Carlson}, \citenamefont {Lusk},
  \citenamefont {Pieper},\ and\ \citenamefont {Schiavilla}}]{Lovato:2017cux}%
  \BibitemOpen
  \bibfield  {author} {\bibinfo {author} {\bibfnamefont {A.}~\bibnamefont
  {Lovato}}, \bibinfo {author} {\bibfnamefont {S.}~\bibnamefont {Gandolfi}},
  \bibinfo {author} {\bibfnamefont {J.}~\bibnamefont {Carlson}}, \bibinfo
  {author} {\bibfnamefont {E.}~\bibnamefont {Lusk}}, \bibinfo {author}
  {\bibfnamefont {S.~C.}\ \bibnamefont {Pieper}},\ and\ \bibinfo {author}
  {\bibfnamefont {R.}~\bibnamefont {Schiavilla}},\ }\bibfield  {title}
  {\bibinfo {title} {{Quantum Monte Carlo calculation of neutral-current
  $\nu-^{12}C$ inclusive quasielastic scattering}},\ }\href
  {https://doi.org/10.1103/PhysRevC.97.022502} {\bibfield  {journal} {\bibinfo
  {journal} {Phys. Rev. C}\ }\textbf {\bibinfo {volume} {97}},\ \bibinfo
  {pages} {022502} (\bibinfo {year} {2018})},\ \Eprint
  {https://arxiv.org/abs/1711.02047} {arXiv:1711.02047 [nucl-th]} \BibitemShut
  {NoStop}%
\bibitem [{\citenamefont {Sobczyk}\ \emph {et~al.}(2021)\citenamefont
  {Sobczyk}, \citenamefont {Acharya}, \citenamefont {Bacca},\ and\
  \citenamefont {Hagen}}]{Sobczyk:2021dwm}%
  \BibitemOpen
  \bibfield  {author} {\bibinfo {author} {\bibfnamefont {J.~E.}\ \bibnamefont
  {Sobczyk}}, \bibinfo {author} {\bibfnamefont {B.}~\bibnamefont {Acharya}},
  \bibinfo {author} {\bibfnamefont {S.}~\bibnamefont {Bacca}},\ and\ \bibinfo
  {author} {\bibfnamefont {G.}~\bibnamefont {Hagen}},\ }\bibfield  {title}
  {\bibinfo {title} {{Ab initio computation of the longitudinal response
  function in $^{40}$Ca}},\ }\href
  {https://doi.org/10.1103/PhysRevLett.127.072501} {\bibfield  {journal}
  {\bibinfo  {journal} {Phys. Rev. Lett.}\ }\textbf {\bibinfo {volume} {127}},\
  \bibinfo {pages} {072501} (\bibinfo {year} {2021})},\ \Eprint
  {https://arxiv.org/abs/2103.06786} {arXiv:2103.06786 [nucl-th]} \BibitemShut
  {NoStop}%
\bibitem [{\citenamefont {Sobczyk}\ \emph {et~al.}(2024)\citenamefont
  {Sobczyk}, \citenamefont {Acharya}, \citenamefont {Bacca},\ and\
  \citenamefont {Hagen}}]{Sobczyk:2023sxh}%
  \BibitemOpen
  \bibfield  {author} {\bibinfo {author} {\bibfnamefont {J.~E.}\ \bibnamefont
  {Sobczyk}}, \bibinfo {author} {\bibfnamefont {B.}~\bibnamefont {Acharya}},
  \bibinfo {author} {\bibfnamefont {S.}~\bibnamefont {Bacca}},\ and\ \bibinfo
  {author} {\bibfnamefont {G.}~\bibnamefont {Hagen}},\ }\bibfield  {title}
  {\bibinfo {title} {{Ca40 transverse response function from coupled-cluster
  theory}},\ }\href {https://doi.org/10.1103/PhysRevC.109.025502} {\bibfield
  {journal} {\bibinfo  {journal} {Phys. Rev. C}\ }\textbf {\bibinfo {volume}
  {109}},\ \bibinfo {pages} {025502} (\bibinfo {year} {2024})},\ \Eprint
  {https://arxiv.org/abs/2310.03109} {arXiv:2310.03109 [nucl-th]} \BibitemShut
  {NoStop}%
\bibitem [{\citenamefont {Pastore}\ \emph {et~al.}(2020)\citenamefont
  {Pastore}, \citenamefont {Carlson}, \citenamefont {Gandolfi}, \citenamefont
  {Schiavilla},\ and\ \citenamefont {Wiringa}}]{Pastore:2019urn}%
  \BibitemOpen
  \bibfield  {author} {\bibinfo {author} {\bibfnamefont {S.}~\bibnamefont
  {Pastore}}, \bibinfo {author} {\bibfnamefont {J.}~\bibnamefont {Carlson}},
  \bibinfo {author} {\bibfnamefont {S.}~\bibnamefont {Gandolfi}}, \bibinfo
  {author} {\bibfnamefont {R.}~\bibnamefont {Schiavilla}},\ and\ \bibinfo
  {author} {\bibfnamefont {R.~B.}\ \bibnamefont {Wiringa}},\ }\bibfield
  {title} {\bibinfo {title} {{Quasielastic lepton scattering and back-to-back
  nucleons in the short-time approximation}},\ }\href
  {https://doi.org/10.1103/PhysRevC.101.044612} {\bibfield  {journal} {\bibinfo
   {journal} {Phys. Rev. C}\ }\textbf {\bibinfo {volume} {101}},\ \bibinfo
  {pages} {044612} (\bibinfo {year} {2020})},\ \Eprint
  {https://arxiv.org/abs/1909.06400} {arXiv:1909.06400 [nucl-th]} \BibitemShut
  {NoStop}%
\bibitem [{\citenamefont {Andreoli}\ \emph {et~al.}(2022)\citenamefont
  {Andreoli}, \citenamefont {Carlson}, \citenamefont {Lovato}, \citenamefont
  {Pastore}, \citenamefont {Rocco},\ and\ \citenamefont
  {Wiringa}}]{Andreoli:2021cxo}%
  \BibitemOpen
  \bibfield  {author} {\bibinfo {author} {\bibfnamefont {L.}~\bibnamefont
  {Andreoli}}, \bibinfo {author} {\bibfnamefont {J.}~\bibnamefont {Carlson}},
  \bibinfo {author} {\bibfnamefont {A.}~\bibnamefont {Lovato}}, \bibinfo
  {author} {\bibfnamefont {S.}~\bibnamefont {Pastore}}, \bibinfo {author}
  {\bibfnamefont {N.}~\bibnamefont {Rocco}},\ and\ \bibinfo {author}
  {\bibfnamefont {R.~B.}\ \bibnamefont {Wiringa}},\ }\bibfield  {title}
  {\bibinfo {title} {{Electron scattering on A=3 nuclei from quantum Monte
  Carlo based approaches}},\ }\href
  {https://doi.org/10.1103/PhysRevC.105.014002} {\bibfield  {journal} {\bibinfo
   {journal} {Phys. Rev. C}\ }\textbf {\bibinfo {volume} {105}},\ \bibinfo
  {pages} {014002} (\bibinfo {year} {2022})},\ \Eprint
  {https://arxiv.org/abs/2108.10824} {arXiv:2108.10824 [nucl-th]} \BibitemShut
  {NoStop}%
\bibitem [{\citenamefont {Andreoli}\ \emph {et~al.}(2024)\citenamefont
  {Andreoli}, \citenamefont {King}, \citenamefont {Pastore}, \citenamefont
  {Piarulli}, \citenamefont {Carlson}, \citenamefont {Gandolfi},\ and\
  \citenamefont {Wiringa}}]{Andreoli:2024ovl}%
  \BibitemOpen
  \bibfield  {author} {\bibinfo {author} {\bibfnamefont {L.}~\bibnamefont
  {Andreoli}}, \bibinfo {author} {\bibfnamefont {G.~B.}\ \bibnamefont {King}},
  \bibinfo {author} {\bibfnamefont {S.}~\bibnamefont {Pastore}}, \bibinfo
  {author} {\bibfnamefont {M.}~\bibnamefont {Piarulli}}, \bibinfo {author}
  {\bibfnamefont {J.}~\bibnamefont {Carlson}}, \bibinfo {author} {\bibfnamefont
  {S.}~\bibnamefont {Gandolfi}},\ and\ \bibinfo {author} {\bibfnamefont
  {R.~B.}\ \bibnamefont {Wiringa}},\ }\bibfield  {title} {\bibinfo {title}
  {{Quantum Monte Carlo calculations of electron scattering from C12 in the
  short-time approximation}},\ }\href
  {https://doi.org/10.1103/PhysRevC.110.064004} {\bibfield  {journal} {\bibinfo
   {journal} {Phys. Rev. C}\ }\textbf {\bibinfo {volume} {110}},\ \bibinfo
  {pages} {064004} (\bibinfo {year} {2024})},\ \Eprint
  {https://arxiv.org/abs/2407.06986} {arXiv:2407.06986 [nucl-th]} \BibitemShut
  {NoStop}%
\bibitem [{\citenamefont {Benhar}\ \emph {et~al.}(1994)\citenamefont {Benhar},
  \citenamefont {Fabrocini}, \citenamefont {Fantoni},\ and\ \citenamefont
  {Sick}}]{Benhar:1994hw}%
  \BibitemOpen
  \bibfield  {author} {\bibinfo {author} {\bibfnamefont {O.}~\bibnamefont
  {Benhar}}, \bibinfo {author} {\bibfnamefont {A.}~\bibnamefont {Fabrocini}},
  \bibinfo {author} {\bibfnamefont {S.}~\bibnamefont {Fantoni}},\ and\ \bibinfo
  {author} {\bibfnamefont {I.}~\bibnamefont {Sick}},\ }\bibfield  {title}
  {\bibinfo {title} {{Spectral function of finite nuclei and scattering of GeV
  electrons}},\ }\href {https://doi.org/10.1016/0375-9474(94)90920-2}
  {\bibfield  {journal} {\bibinfo  {journal} {Nucl. Phys. A}\ }\textbf
  {\bibinfo {volume} {579}},\ \bibinfo {pages} {493} (\bibinfo {year}
  {1994})}\BibitemShut {NoStop}%
\bibitem [{\citenamefont {Rocco}\ \emph {et~al.}(2016)\citenamefont {Rocco},
  \citenamefont {Lovato},\ and\ \citenamefont {Benhar}}]{Rocco:2015cil}%
  \BibitemOpen
  \bibfield  {author} {\bibinfo {author} {\bibfnamefont {N.}~\bibnamefont
  {Rocco}}, \bibinfo {author} {\bibfnamefont {A.}~\bibnamefont {Lovato}},\ and\
  \bibinfo {author} {\bibfnamefont {O.}~\bibnamefont {Benhar}},\ }\bibfield
  {title} {\bibinfo {title} {{Unified description of electron-nucleus
  scattering within the spectral function formalism}},\ }\href
  {https://doi.org/10.1103/PhysRevLett.116.192501} {\bibfield  {journal}
  {\bibinfo  {journal} {Phys. Rev. Lett.}\ }\textbf {\bibinfo {volume} {116}},\
  \bibinfo {pages} {192501} (\bibinfo {year} {2016})},\ \Eprint
  {https://arxiv.org/abs/1512.07426} {arXiv:1512.07426 [nucl-th]} \BibitemShut
  {NoStop}%
\bibitem [{\citenamefont {Rocco}\ \emph {et~al.}(2019)\citenamefont {Rocco},
  \citenamefont {Barbieri}, \citenamefont {Benhar}, \citenamefont {De~Pace},\
  and\ \citenamefont {Lovato}}]{Rocco:2018mwt}%
  \BibitemOpen
  \bibfield  {author} {\bibinfo {author} {\bibfnamefont {N.}~\bibnamefont
  {Rocco}}, \bibinfo {author} {\bibfnamefont {C.}~\bibnamefont {Barbieri}},
  \bibinfo {author} {\bibfnamefont {O.}~\bibnamefont {Benhar}}, \bibinfo
  {author} {\bibfnamefont {A.}~\bibnamefont {De~Pace}},\ and\ \bibinfo {author}
  {\bibfnamefont {A.}~\bibnamefont {Lovato}},\ }\bibfield  {title} {\bibinfo
  {title} {{Neutrino-Nucleus Cross Section within the Extended Factorization
  Scheme}},\ }\href {https://doi.org/10.1103/PhysRevC.99.025502} {\bibfield
  {journal} {\bibinfo  {journal} {Phys. Rev. C}\ }\textbf {\bibinfo {volume}
  {99}},\ \bibinfo {pages} {025502} (\bibinfo {year} {2019})},\ \Eprint
  {https://arxiv.org/abs/1810.07647} {arXiv:1810.07647 [nucl-th]} \BibitemShut
  {NoStop}%
\bibitem [{\citenamefont {Abrams}\ and\ \citenamefont
  {Lloyd}(1999)}]{Abrams1999}%
  \BibitemOpen
  \bibfield  {author} {\bibinfo {author} {\bibfnamefont {D.~S.}\ \bibnamefont
  {Abrams}}\ and\ \bibinfo {author} {\bibfnamefont {S.}~\bibnamefont {Lloyd}},\
  }\bibfield  {title} {\bibinfo {title} {Quantum algorithm providing
  exponential speed increase for finding eigenvalues and eigenvectors},\ }\href
  {https://doi.org/10.1103/PhysRevLett.83.5162} {\bibfield  {journal} {\bibinfo
   {journal} {Phys. Rev. Lett.}\ }\textbf {\bibinfo {volume} {83}},\ \bibinfo
  {pages} {5162} (\bibinfo {year} {1999})}\BibitemShut {NoStop}%
\bibitem [{\citenamefont {Cleve}\ \emph {et~al.}(1998)\citenamefont {Cleve},
  \citenamefont {Ekert}, \citenamefont {Macchiavello},\ and\ \citenamefont
  {Mosca}}]{Cleve:1997dh}%
  \BibitemOpen
  \bibfield  {author} {\bibinfo {author} {\bibfnamefont {R.}~\bibnamefont
  {Cleve}}, \bibinfo {author} {\bibfnamefont {A.}~\bibnamefont {Ekert}},
  \bibinfo {author} {\bibfnamefont {C.}~\bibnamefont {Macchiavello}},\ and\
  \bibinfo {author} {\bibfnamefont {M.}~\bibnamefont {Mosca}},\ }\bibfield
  {title} {\bibinfo {title} {{Quantum algorithms revisited}},\ }\href
  {https://doi.org/10.1098/rspa.1998.0164} {\bibfield  {journal} {\bibinfo
  {journal} {Proc. Roy. Soc. Lond. A}\ }\textbf {\bibinfo {volume} {454}},\
  \bibinfo {pages} {339} (\bibinfo {year} {1998})},\ \Eprint
  {https://arxiv.org/abs/quant-ph/9708016} {arXiv:quant-ph/9708016}
  \BibitemShut {NoStop}%
\bibitem [{\citenamefont {Wecker}\ \emph {et~al.}(2015)\citenamefont {Wecker},
  \citenamefont {Hastings}, \citenamefont {Wiebe}, \citenamefont {Clark},
  \citenamefont {Nayak},\ and\ \citenamefont {Troyer}}]{Wecker:2015fib}%
  \BibitemOpen
  \bibfield  {author} {\bibinfo {author} {\bibfnamefont {D.}~\bibnamefont
  {Wecker}}, \bibinfo {author} {\bibfnamefont {M.~B.}\ \bibnamefont
  {Hastings}}, \bibinfo {author} {\bibfnamefont {N.}~\bibnamefont {Wiebe}},
  \bibinfo {author} {\bibfnamefont {B.~K.}\ \bibnamefont {Clark}}, \bibinfo
  {author} {\bibfnamefont {C.}~\bibnamefont {Nayak}},\ and\ \bibinfo {author}
  {\bibfnamefont {M.}~\bibnamefont {Troyer}},\ }\bibfield  {title} {\bibinfo
  {title} {{Solving strongly correlated electron models on a quantum
  computer}},\ }\href {https://doi.org/10.1103/PhysRevA.92.062318} {\bibfield
  {journal} {\bibinfo  {journal} {Phys. Rev. A}\ }\textbf {\bibinfo {volume}
  {92}},\ \bibinfo {pages} {062318} (\bibinfo {year} {2015})}\BibitemShut
  {NoStop}%
\bibitem [{\citenamefont {{Qiskit contributors}}(2023)}]{Qiskit}%
  \BibitemOpen
  \bibfield  {author} {\bibinfo {author} {\bibnamefont {{Qiskit
  contributors}}},\ }\href {https://doi.org/10.5281/zenodo.2573505} {\bibinfo
  {title} {Qiskit: An open-source framework for quantum computing}} (\bibinfo
  {year} {2023})\BibitemShut {NoStop}%
\bibitem [{\citenamefont {Shende}\ \emph {et~al.}(2006)\citenamefont {Shende},
  \citenamefont {Bullock},\ and\ \citenamefont {Markov}}]{Shende2006}%
  \BibitemOpen
  \bibfield  {author} {\bibinfo {author} {\bibfnamefont {V.}~\bibnamefont
  {Shende}}, \bibinfo {author} {\bibfnamefont {S.}~\bibnamefont {Bullock}},\
  and\ \bibinfo {author} {\bibfnamefont {I.}~\bibnamefont {Markov}},\
  }\bibfield  {title} {\bibinfo {title} {Synthesis of quantum-logic circuits},\
  }\href {https://doi.org/10.1109/TCAD.2005.855930} {\bibfield  {journal}
  {\bibinfo  {journal} {IEEE Transactions on Computer-Aided Design of
  Integrated Circuits and Systems}\ }\textbf {\bibinfo {volume} {25}},\
  \bibinfo {pages} {1000} (\bibinfo {year} {2006})}\BibitemShut {NoStop}%
\bibitem [{Dat(2025)}]{DataAvailability}%
  \BibitemOpen
  \href@noop {} {\bibinfo {title} {Data availability contact information}},\
  \bibinfo {howpublished}
  {\url{https://docs.google.com/document/d/1kPKNG9ZK3I5O2q3GNCWI8ED57yKRQ3jzlzgSh2xyZNs/edit?usp=share_link}}
  (\bibinfo {year} {2025})\BibitemShut {NoStop}%
\end{thebibliography}%

\end{document}